\documentstyle[12pt]{article}

\def\r#1{{\rm #1}}

\def\bm#1{{\mathord{\mbox{\boldmath $#1$}}}}
\def\sbm#1{{\mathord{\mbox{\boldmath $\scriptstyle #1$}}}}
\def\bg#1{\hbox{\rlap{\hbox{$#1$}}\kern0.6pt $#1$}}
\def\Frac(#1/#2){\left(\frac{#1}{#2}\right)}
\def\Kr(#1;#2#3){{\GAMMA^{#1}_{#2#3}}}
\def\utilde#1{\lower6pt\hbox{${\textstyle #1} \atop
\mkern-2mu\raise2pt\hbox{$\tilde{}$}$}}

\def\Tr{\r{Tr}}
\def\Tp{\,{}^t\!}
\def\RF{\bm{R}}
\def\CF{\bm{C}}
\def\Lie{\hbox{\rlap{$\cal L$}$-$}}

\def\RSCD{\;{}^3\!\nabla}
\def\chiral{{}^\pm\!}
\def\sharp{\mathord{\raise2pt\hbox{\tiny\#}}}

\def\therefore{\mbox{\setbox0=\hbox{X}\hbox{$\ldotp$}\raise0.7\ht0\hbox{$\ldotp$}\hbox{$\ldotp$}} }
\def\because{\mbox{\setbox0=\hbox{X}\raise0.7\ht0\hbox{$\ldotp$}\hbox{$\ldotp$}\raise0.7\ht0\hbox{$\ldotp$}}\kern0pt }

\def\A{{\cal A}}
\def\B{{\cal B}}
\def\C{{\cal C}}
\def\D{{\cal D}}
\def\E{{\cal E}}
\def\F{{\cal F}}
\def\G{{\cal G}}
\def\H{{\cal H}}

\def\L{{\cal L}}

\def\O{{\cal O}}
\def\P{{\cal P}}

\def\R{{\cal R}}

\def\U{{\cal U}}
\def\V{{\cal V}}

\def\GAMMA{{\mit \Gamma}}

\def\CITE#1{$^{\hbox{\small \cite{#1}}}$}

\catcode`\@=11
\@addtoreset{equation}{section}   % Makes \section reset 'equation' counter.
\@addtoreset{equation}{subsection}
\def\CAPTION#1{\def\@makecaption##1##2{} \caption{#1}}
\catcode`\@=12
\begin{document}

\def\KUCP{ \hrule height 0pt depth 0pt \par
\vskip -2.5cm \par
{\leftskip 11 cm \par
\noindent
KUCP-0048\hfil\break
October 1992 \par
}
\vskip 1.5cm \par
}

\begin{titlepage}

\KUCP

{\multiply\baselineskip by 3 \divide\baselineskip by 2

\begin{center}\LARGE
Quantum Gravity by the Complex Canonical Formulation
\end{center}

\medskip

\begin{center}\large
Hideo Kodama
\end{center}

\medskip
\begin{center}\sl
Department of Fundamental Sciences, FIHS \\
Kyoto University, Yoshida, Kyoto 606, Japan
\end{center}

\vspace{4cm}

The basic features of the complex canonical formulation of general
relativity and the recent developments in the quantum gravity program
based on it are reviewed. The exposition is intended to be
complementary to the review articles available already and some
original arguments are included.  In particular the conventional
treatment of the Hamiltonian constraint and quantum states in the
canonical approach to quantum gravity is criticized and a new
formulation is proposed.

 }

\vspace{1cm}

\begin{center}
To appear in Int. J. Theor. Phys. D.
\end{center}

\end{titlepage}

\tableofcontents \par \pagebreak \par

\section{Introduction}

The most important concepts introduced in fundamental physics in this
half century are renormalizability and gauge principle. Various
experimental verifications of the standard model of elementary
particles have marked their success. However, there remains one
important field which is not encompassed by these ideas: the
gravitational interaction.

Of course general relativity, the most successful classical
theory of gravity, is a gauge field theory in the sense that gravity is
described by a $SO(3,1)$ connection.  However, the requirement of the
general covariance has made it quite different from ordinary
gauge theories: the local gauge symmetry is intimately  connected with
space-time diffeomorphisms. Actually it is this feature that makes
general relativity work as a theory of gravity.

This difference gives rise to the serious problem of unrenormalizable
divergences when one tries to construct a quantum theory of general
relativity. It implies that the dynamics becomes more and more
intricate as one goes to smaller scales.  In the classical theory this
does not cause any trouble because one can suppress local excitations.
In contrast such a suppression is not possible in the quantum regime
due to the existence of uncontrollable quantum fluctuations.

Though various approaches have been tried to attack this problem
historically, they are now converging to three main streams.  The
first is to construct a new theory of gravity which does not suffer
from the above difficulty.  The most successful approach along these
lines is the superstring theory.  Second is the canonical approach in
which one tries to find a new framework to handle the nonperturbative
nature of quantum gravity by starting from the conventional canonical
quantization of general relativity.  The third is the path-integral
approach which differs from the second in that it formulates the
theory in terms of sums over histories.

Superstring theory appears to be quite elegant in its formulation and
fascinating in that it gives a unified treatment of all the
interactions.  In spite of these nice features it is yet at a
premature stage as a theory of quantum gravity since it can treat
gravity only perturbatively\CITE{GreenSchwarzWitten87}.  The spacetime
structure is built into the theory just as a classical object and its
quantum dynamics cannot be studied.  Two entities of completely
different nature coexist. Further, since no satisfactory mechanism of
dimensional reduction is found yet, there remains a significant
ambiguity in its prediction for low energy physics.

In contrast, the canonical approach is far from being elegant and
cannot restrict the structure of interactions other than gravity.
Further there exists no justification for assuming that gravity is
described by general relativity on small scales.  The path-integral
approach shares these features except that it utilizes the
path-integral which is conceptually powerful but technically
ill-defined.  The main reason why people follow these approaches in
spite of these limitations is in that they are currently the only
approaches in which one can address the problem of the quantum
dynamics of four-dimensional spacetimes directly without introducing
extra ambiguities. This point is reflected in the historical
development that the canonical approach was awakened from its long
sleep relatively recently by the interesting work of Hawking and
others\CITE{Hawking82a,HartleHawking83,Hawking84,Vilenkin83,Vilenkin84}
on quantum cosmology.

It is quite interesting that these approaches are now coming closer
and closer in spite of the differences in their starting points and
features.  It is the introduction of new canonical variables by
Ashtekar\CITE{Ashtekar86a,Ashtekar87} that has played a very important
role in this development.

Ashtekar's theory was proposed as a rewriting of the traditional
canonical theory of general relativity in terms of new variables and
has two fascinating but one embarrassing feature: all the fundamental
equations in the canonical theory are polynomial and the takes a form
of a $SO(3,\CF)$ gauge theory, but the new variables are complex.

Soon after its proposal Samuel\CITE{Samuel87} and Jacobson and
Smolin\CITE{JacobsonSmolin87,JacobsonSmolin88a} showed that Ashtekar's
complex canonical theory can be derived from the first-order Palatini
action by its (anti-)self-dual decomposition.  They also clarified
that the above three features are intimately related.  Further with
the help of this elegant reformulation gravity systems coupled with
matter were rewritten in the same form by several
authors\CITE{AshtekarRomanoTate89,Jacobson88b,GorobeyLukyanenko90}.

Almost at the same time Jacobson and Smolin\CITE{JacobsonSmolin88b}
revealed that Ashtekar's complex canonical theory when the above three
features are fully utilized may provide a new breakthrough for the
study of canonical quantum gravity.  They found an infinite family of
solutions to the quantum Hamiltonian constraint equation which is the
central equation in the canonical gravity. This was astonishing
because no exact solution had been found in the conventional ADM-WD
formulation since its first detailed study by DeWitt\CITE{DeWitt67a}.
The essential ingredients were the introduction of a holomorphic
connection representation\CITE{Ashtekar86b,Ashtekar86a} and loop
integral variables.  The former was a natural consequence of adopting
the complex connection as the fundamental variable and the latter of
the gauge theoretical structure of the formulation.

There were, however, a few unpleasant aspects in these solutions.
First they appeared to be unphysical since they represent spacetimes
with spatial metrics everywhere
degenerate\CITE{Husain89,BruegmannPullin91}. Second they do not
satisfy the diffeomorphism constraint.  Since the diffeomorphism
invariance is the most important feature of general relativity, the
latter was regarded as a crucial defect.

Resolution was brought about by the introduction of the loop space
representation by Rovelli and
Smolin\CITE{RovelliSmolin88,RovelliSmolin90}, which is in a sense a
natural development provoked by the introduction of loop variables but
is quite foreign to the conventional framework of canonical quantum
gravity.  There all the operators and the fundamental equations are
transferred into the space of loops in a three-dimensional base
manifold, and the problem of diffeomorphism invariance is reduced to
the task of finding the knot or link invariants.  Actually a solution
to all the constraint equations which is nondegenerate at a point is
found\CITE{BruegmannGambiniPullin92a} and is extended to a series of
solutions\CITE{BruegmannGambiniPullin92b,BruegmannGambiniPullin92c}
with the help of an exact solution of non-loop-integral type in the
connection representation\CITE{Kodama90} and the link between the
Jones Polynomial and the Chern-Simons topological field theory found
by Witten\CITE{Witten89}.  Stimulated by these successes, a program
has started to reformulate the other conventional gauge theories in
terms of the loop space language in order to transfer all the physics
onto the loop space and find physical interpretations of the loop
space
objects\CITE{AshtekarRovelliSmolin91,AshtekarRovelliSmolin92,Smolin92,AshtekarRovelliSmolin92}.

In spite of these exciting developments there remain lots of important
problems yet to be solved in this approach to quantum gravity based on
the complex canonical theory.  Some of them are common to the
traditional ADM-WD approach, such as the extraction of dynamics, the
construction of diffeomorphism invariant operators and the
interpretation. Though some investigations have been made on these
problems by specializing the formalism to the Bianchi minisuperspace
models\CITE{Kodama88a,Kodama90,AshtekarPullin89,AshtekarTateUggla92}
or the space-times with one or two Killing
vectors\CITE{HusainSmolin89,HusainPullin90}, all of them are of a
preliminary nature except for the recent construction of some finite
geometrical operators on the loop space\CITE{AshtekarRovelliSmolin92}.
The other problems are specific to the complex canonical theory.  In
particular the treatment of the reality condition, which is closely
connected with the definition of the inner product of quantum states,
is left as a quite difficult problem.

In the present paper I review the basic formulation of this complex
canonical theory and the achievements and the problems in its
application to quantum gravity.  Since several good
reviews\CITE{Ashtekar88a,Smolin89,Rovelli91a,Smolin92} including the
excellent book by Ashtekar himself\CITE{Ashtekar91a} already exist, I
have tried to make this paper complementary to them keeping the
exposition self-contained.  In particular I have inserted a long
section explaining the general structure of the canonical quantization
program and its difficulties in order to help the readers to look at
the present status of the theory objectively.  A large fraction of the
discussion in this section is original and in particular includes an
important criticism on the conventional treatment of the quantum
Hamiltonian constraint and its interpretation.  Further as a technical
point I avoid the use of the spinorial notation and describe the
theory in the vector language as far as possible because the
spinorial notation seems to be cumbersome for non-specialists.  Since
the main interest is in quantum gravity, I will not touch on some of
the topics which are not directly relevant to it.

The main body of the paper consists of three sections. We begin in the
next section by examining the basic structures and the characteristic
features of the three main formulations of the classical canonical
gravity, the metric approach, the real triad approach and the chiral
approach(the complex canonical formulation), in order to see the
similarities and the differences of them.  In particular I explain in
some detail how the latter two formulations are derived from the
covariant actions to make clear the origin of the new features of the
complex canonical theory and examine the correspondence among the
three formulations.

In \S3 I outline the generic structure of the canonical quantization
program of gravity, and discuss its difficulty associated with the
specification of dynamics and its origin, as a preliminary to the next
section.  In particular, by analyzing the physical meanings and roles
of the three invariances in general relativity, the ordinary gauge
invariance, the spatial diffeomorphism invariance and the time
coordinate transformation invariance, I criticize the conventional
treatment of the Hamiltonian constraint and quantum states, and
propose a new treatment which I call the probability amplitude
functional formalism.  Some simple examples are given to illustrate
how this formalism works.

In the light of the general framework given in \S3, I describe in \S4
what has been achieved and what to be done yet in the quantum gravity
program based on the complex canonical formulation.  The exposition is
limited to the basic aspects and technical details are often omitted
since they are described in the reviews cited above.  Section 5 is
devoted to summary and discussion.

Finally I comment on the notation used in this paper.  I adopt the
natural units $c=\hbar=1$ and use $\kappa^2=8\pi G$ in stead of $G$.
The signature of the spacetime metric $g_{\mu\nu}$ is $[-,+,+,+]$, and
the completely antisymmetric symbols always denote tensor densities
normalized by $\epsilon_{0123}=1$, $\epsilon^{0123}=-1$ and
$\epsilon_{123}=\epsilon^{123}=1$.  The spacetime coordinate and
spatial coordinate indices are denoted by the greek letters and the
Latin letters starting from $j$, respectively, and the four- and
three- dimensional internal indices by the Latin letters between $a$
and $h$, and the capital letters starting from $I$, respectively.  The
zeroth component of the spacetime coordinate indices is denoted by $t$
instead of $0$ where it is necessary to distinguish it from that of
the internal indices.  Of course these letters are sometimes used in
other senses due to the limited amount of symbols.

\section{Canonical Formulation of General Relativity}

In this first part of the paper we first summarize the basic
structures and the characteristic features of two conventional real
canonical formulations of the classical theory of general relativity,
the metric approach and the real triad approach.  Then we derive the
complex canonical formulation and look at its structures and relation
to the conventional ones.

\subsection{Metric Approach}

In the metric approach one starts from the action which is expressed
in terms of the space-time metric $g_{\mu\nu}$ and the material field
variables as
\begin{equation}
S=S_\r{G} + S_\r{M},
\end{equation}
where $S_\r{G}$ is the Einstein-Hilbert action
\begin{equation}
S_\r{G}=\int_M d^4x {1\over 2\kappa^2}\sqrt{-g}R
+ \int_{\partial M} d\Sigma_\mu {1\over\kappa^2}k^\mu,
\label{EinsteinHilbertAction}\end{equation}
and $S_\r{M}$ is the action for matter.  The second term in the
right-hand side of Eq.(\ref{EinsteinHilbertAction}) is the surface
term to cancel the second derivative terms in the first term, where
$k^\mu$ is the vector density defined in terms of an appropriate tetrad
$e_a^\mu$ as
\begin{equation}
k^\mu := -\sqrt{-g} \eta^{ab}(\nabla_{e_a}e_b)^\mu,
\end{equation}
and $d\Sigma_\mu$ is the three-dimensional volume element
\begin{equation}
d\Sigma_\mu = {1\over 3!}\epsilon_{\mu\nu\lambda\sigma}dx^\nu\wedge dx^\lambda
\wedge dx^\sigma.
\end{equation}

In the present subsection we only consider a three-component real
scalar field $\Phi$ coupled with a $SO(3)$ gauge field $\bm{A}_\mu$
for simplicity. For this system $S_\r{M}$ is given by
\begin{equation}
S_\r{M}=\int_M d^4x \sqrt{-g}[-{1\over4}\bm{F}_{\mu\nu}\cdot
\bm{F}^{\mu\nu} -{1\over2}\bm{D}_\mu \Phi\cdot \bm{D}^\mu \Phi -
V(\Phi)],
\end{equation}
where
\begin{eqnarray}
&&\bm{F}_{\mu\nu} = \partial_\mu\bm{A}_\nu - \partial_\nu\bm{A}_\mu -e
\bm{A}_\mu\times\bm{A}_\nu,\\ &&\bm{D}_\mu \Phi =(\partial_\mu - e
\bm{A}_\mu \times)\Phi.
\end{eqnarray}

In order to construct a canonical theory of this system, we must
foliate the spacetime into a family of space-like slices with constant
time $t$, and decompose the fundamental variables to three dimensional
tensors on each slice. In the present paper we assume that the spacetime
manifold $M$ has the structure $\RF\times \Sigma$ after this time slicing
where $\RF$ and $\Sigma$ correspond to the curves with constant spatial
coordinates and the time slices, respectively.

First, by expressing the future-directed unit
normal vector $n$ to each slice in terms of the lapse function $N$ and
the shift vector $N^j$ as
\begin{equation}
n = N^{-1}(\partial_t - N^j\partial_j)
\end{equation}
the space-time metric is written in terms of $N$, $N^j$ and the
intrinsic metric $q_{jk}$ of each slice as
\begin{equation}
ds^2 = -N^2dt^2 + q_{jk}(dx^j+N^jdt)(dx^k+N^kdt).
\end{equation}
This (3+1)-decomposition leads to the following expression for the
gravitational Lagrangian density:
\begin{equation}
\sqrt{-g}R=N\sqrt{q}({}^3\!R+K_{jk}K^{jk}-K^2)-\partial_0(2\sqrt{q}K)
+\partial_j[2\sqrt{q}(N^jK-\RSCD^jN)],
\label{4curvature:decomposition}\end{equation}
where $\RSCD_j$ is the three-dimensional Riemannian covariant
derivative with respect to $q_{jk}$, all the spatial-coordinate
indices are raised and lowered by $q_{jk}$, and $K_{jk}$ is the
extrinsic curvature of each slice defined by
\begin{eqnarray}
&&K_{jk}:=-\nabla_j n_k ={1\over2N}(-\dot q_{jk}+\RSCD_jN_k +
\RSCD_kN_j),
\label{K:def}\\
&&K:=K^j_j=q^{jk}K_{jk}.
\end{eqnarray}
In order to decompose the boundary term, we choose the tetrad such that
$e_0=n$.  Then, since $k^0=\sqrt{q}K$ on each time slice and
$k^j=-\sqrt{q}[KN^j-\RSCD^j N + N(\RSCD_{e_I}e_I)^j]$ on
$\RF\times\partial\Sigma$, the total derivative terms in
(\ref{4curvature:decomposition}) are
canceled out by the boundary term.

Under the (3+1)-decomposition the action is written in terms of
$q_{jk}$, $N$, $N^j$, $\bm{A}_\mu$, and $\Phi$.  However, since no
time derivatives of $N^\mu$($N^t:=N$) and $\bm{A}_0$ are contained in
the action as is seen from the expressions for $\sqrt{-g}R$ and
$\bm{F}_{\mu\nu}$, we can introduce the canonical momentums only for
$q_{jk}$, $\bm{A}_j$ and $\Phi$:
\begin{eqnarray}
&&p^{jk}:={\delta L\over \delta \dot q_{jk}} =-{\sqrt{q}\over
2\kappa^2} (K^{jk}-q^{jk}K),\label{p:def}\\ &&\bm{E}^j:={\delta L\over
\delta \dot{\bm{A}}_j}={\sqrt{q}\over
N}q^{jk}(\bm{F}_{0k}-N^l\bm{F}_{lk}),\label{E:def}\\ &&\Pi:={\delta L
\over \delta \dot \Phi}={\sqrt{q}\over N}(\bm{D}_0\Phi
-N^j\bm{D}_j\Phi).\label{Pi:def}
\end{eqnarray}
In terms of these variables the Lagrangian is written in the canonical
form as
\begin{eqnarray}
&& L=\int_\Sigma d^3x(\dot q_{jk}p^{jk} + \bm{E}^j\cdot\dot{\bm{A}}_j
+\Pi\cdot \dot\Phi) -H,\\
&& H=\int_\Sigma d^3x(N^\mu\H_\mu + \bm{A}_0\cdot
\bm{C}_{\sbm{A}}) + H_\infty,
\end{eqnarray}
where
\begin{eqnarray}
&\bm{C}_{\sbm{A}}&=-\bm{D}_j\bm{E}^j + e\Phi\times\Pi,\\ &\H_0
&={2\kappa^2\over\sqrt{q}}(p^{jk}p_{jk}-{1\over2}p^2)-{\sqrt{q}\over2\kappa^2}
{}^3\!R + \sqrt{q}T_{\mu\nu}n^\mu n^\nu,\\ &\H_j&=-2\RSCD_kp^k_j+
\sqrt{q}T_{j\mu}n^\mu,\\ &\sqrt{q}T_{\mu\nu}n^\mu n^\nu &={\Pi^2\over
2\sqrt{q}}+\sqrt{q}({1\over2}\bm{D}_j\Phi\cdot\bm{D}^j\Phi+V)
+{1\over2\sqrt{q}}\bm{E}_j\cdot\bm{E}^j \nonumber\\ && \qquad
+{\sqrt{q}\over2}q^{jk}q^{lm}\bm{F}_{jl}\bm{F}_{lm},\\
&\sqrt{q}T_{j\mu}n^\mu &=\Pi\cdot\bm{D}_j\Phi
+\bm{E}^k\cdot\bm{F}_{jk},\\
&H_\infty&=\int_{\partial\Sigma}dS_j[{1\over\kappa^2} \sqrt{q}N
(\RSCD_{e_I}e_I)^j+ \bm{A}\!_0\cdot \bm{E}^j].
\end{eqnarray}
In the last expression $dS_j=(1/2)\epsilon_{jkl}dx^k\wedge dx^l$.  For
asymptotically flat spacetimes $e_I$ should be taken so that it
approaches some fixed Descartian frame at infinity in order for
$H_\infty$ to be finite. Of course $H_\infty$ vanishes for the
spatially compact case.

Thus by setting the Poisson brackets among the fundamental canonical
variables as
\begin{eqnarray}
&&\{q_{jk}(\bm{x}),p^{lm}(\bm{y})\}=\delta^l_{(j}\delta^m_{k)}
\delta^3(\bm{x}-\bm{y}),\\
&&\{\bm{A}^I_j(\bm{x}),\bm{E}_J^k(\bm{y})\}=\delta^I_J\delta^k_j
\delta^3(\bm{x}-\bm{y}),\\
&&\{\Phi^I(\bm{x}),\Pi_J(\bm{y})\}=\delta^I_J
\delta^3(\bm{x}-\bm{y}),\\
&&\hbox{others are zero},
\end{eqnarray}
the variation of the action with respect to these variables yields the
canonical equation of motion for the canonical quantity
$f(q,p,\bm{A},\bm{E},\Phi,\Pi)$
\begin{equation}
\dot f = \{ f, H\}.
\end{equation}
On the other hand the variation with respect to the non-canonical
variables $N^\mu$ and $\bm{A}\!_0$ yields the constraint equations
\begin{eqnarray}
&& \H_\mu = 0,\\ && C_{\sbm{A}}=0.
\end{eqnarray}

These constraints are shown to be of the first-class, that is, weakly
closed under the Poisson bracket.  Actually the classical commutation
relations among them are given by
\begin{eqnarray}
&&\{C_{\sbm{A}}(\bg{\Lambda}_1),C_{\sbm{A}}(\bg{\Lambda}_2)\} =e
C_{\sbm{A}}(\bg{\Lambda}_1\times\bg{\Lambda}_2),\\
&&\{C_\r{M}(L_1),C_\r{M}(L_2)\}=C_\r{M}([L_1,L_2])+C_{\sbm{A}}(L_1^jL_2^k\bm{F}_{jk}),\\
&&\{C_\r{M}(L), C_\r{H}(T)\}=C_\r{H}(\Lie_LT)
-C_{\sbm{A}}(q^{-1/2}TL^j\bm{E}_j),\\
&&\{C_\r{H}(T_1),C_\r{H}(T_2)\}=C_\r{M}(T_1\RSCD T_2 - T_2\RSCD T_1),
\label{ADM:PoissonBracket:CHCH}\\
&&\hbox{others}=0,
\end{eqnarray}
where $C_\r{H}(T)$, $C_\r{M}(L)$ and $C_{\sbm{A}}(\bg{\Lambda})$ are
defined in terms of smooth functions or vectors $T$, $L^j$ and
$\bg{\Lambda}$ with compact supports as
\begin{eqnarray}
&&C_\r{H}(T):=\int_\Sigma d^3x T\H_0,\\
&&C_\r{M}(L):=\int_\Sigma d^3xL^j\H_j,\\
&&C_{\sbm{A}}(\bg{\Lambda}):=\int_\Sigma d^3x
\bg{\Lambda}\cdot\bm{C}_{\sbm{A}}.
\end{eqnarray}
Since the Hamiltonian is written as a linear combination of these
constraint functions apart from the term $H_\infty$ which does not
affect the local dynamics, this first-class nature guarantees the
consistency of the constraints with the canonical evolution equation.

Thus the general relativity theory can be consistently put in the
canonical form. In the classical region this canonical formalism works
well.  Actually it is utilized successfully in practical problems such
as numerically solving the Einstein equations.  However, when regarded
as the starting point of the quantum gravity program, it has some
difficulties.

First, though the momentum constraint functional $C_\r{M}$ has a
rather simple structure ( linear both in $q_{jk}$ and $p^{jk}$), the
Hamiltonian constraint functional $C_\r{H}$ is non-polynomial in
$q_{jk}$ and includes $\sqrt{q}$.  Of course, if one redefines $N$ to
$\bar N=N/(\sqrt{q}q^2)$, the new Hamiltonian constraint functional
becomes a polynomial in $q_{jk}$.  However, the resultant polynomial
is at least of 8th degree ( of 9th degree if the scalar field has a
non-vanishing potential), and its structure is quite complicated.
This complicated structure makes it difficult to find appropriate
operator orderings and regularization of operator products in
constructing the operator corresponding to the Hamiltonian constraint
functional as well as to solve the constraint equation.

The second difficulty is associated with the constraint algebra. In
general the appearance of first-class constraints is closely connected
with the gauge invariance of the original
Lagrangian\CITE{HalliwellHartle91}.  In the present case the
constraint $\H_\mu=0$ is related with the general covariance and the
constraint $\bm{C}_{\sbm{A}}=0$ with the $SO(3)$ gauge invariance.  In
fact the variation of the canonical variables under the infinitesimal
coordinate transformation
\begin{equation}
\delta t=T, \qquad \delta x^j=L^j,
\end{equation}
and the infinitesimal gauge transformation
\begin{eqnarray}
&&\delta\Phi = -e \bm{\Lambda}\times\Phi, \quad
\delta \Pi = -e \bm{\Lambda}\times\Pi, \\
&&\delta\bm{A}_\mu=-\bm{D}_\mu \bm{\Lambda},\quad
\delta \bm{E} = -e \bm{\Lambda}\times\bm{E}, \\
\end{eqnarray}
is expressed as the canonical transformation
\begin{equation}
\delta f=\{G, \delta f\} \ ;\quad f=f(q,p,\Phi,\Pi,\bm{A},\bm{E}),
\label{InfinitesimalCanonicalTrf}\end{equation}
with the generator
\begin{equation}
G=\int_\Sigma
d^3x[T(N^\mu\H_\mu+\bm{A}_0\cdot\bm{C}_{\sbm{A}})+L^j(\H_j+\bm{A}_j\cdot\bm{C}_{\sbm{A}})+\bg{\Lambda}\cdot\bm{C}_{\sbm{A}}].
\end{equation}
Here Eq.(\ref{InfinitesimalCanonicalTrf}) is valid for $T\not=0$ only
when the canonical variables satisfy the equation of motion.

Thus the constraint function $C_{\sbm{A}}$ is the generator of the
gauge transformation and the canonical quantity $C_\r{D}$ defined by a
linear combination of the constraint functions as
\begin{equation}
C_\r{D}(L):=C_\r{M}(L)+C_{\sbm{A}}(L^j\bm{A}_j),
\end{equation}
is the generator of the spatial coordinate transformation(or the
spatial diffeomorphism).  Further, the Poisson bracket algebra
generated by these constraint functions,
\begin{eqnarray}
&&\{C_{\sbm{A}}(\bg{\Lambda}_1),C_{\sbm{A}}(\bg{\Lambda}_2)\} =e
C_{\sbm{A}}(\bg{\Lambda}_1\times\bg{\Lambda}_2),\\
&&\{C_\r{D}(L),C_{\sbm{A}}(\bg{\Lambda})\}=C_{\sbm{A}}(\Lie_L\bg{\Lambda}),\\
&&\{C_\r{D}(L_1),C_\r{D}(L_2)\}=C_\r{D}([L_1,L_2]),
\end{eqnarray}
is completely isomorphic to the Lie algebra of the spatial coordinate
transformation group and the $SO(3)$ gauge transformation group
\begin{eqnarray}
&&[L(\xi_1),L(\xi_2)]=L([\xi_1,\xi_2]),\\
&&[L(\xi),L(\bg{\Lambda})]=L(\Lie_\xi\bg{\Lambda}),\\
&&[L(\bg{\Lambda}_1),L(\bg{\Lambda}_2)]=eL(\bg{\Lambda}_1\times\bg{\Lambda}_2),
\end{eqnarray}
Hence the constraints $C_{\sbm{A}}=0$ and $C_\r{D}=0$ are purely
kinematical ones.  This point is also confirmed from the structure of
their Poisson brackets with $C_\r{H}$ given by
\begin{eqnarray}
&&\{C_{\sbm{A}}, C_\r{H}(T)\}=0,\\ &&\{C_\r{D}(L),
C_\r{H}(T)\}=C_\r{H}(\Lie_LT).
\end{eqnarray}

In contrast, though the generator of the time-coordinate
transformation is written as a linear combination of the constraint
functions, it is not a canonical quantity since it contains the
noncanonical variables $N^\mu$ and $\bm{A}\!_0$.  This implies that
the group of canonical transformations generated by all the constraint
functions is not isomorphic to the group of the four-dimensional
diffeomorphisms and the gauge transformations.  This point is reflected
in the fact that the structure coefficients in
Eq.(\ref{ADM:PoissonBracket:CHCH}) depend on the canonical variables.
This peculiarity occurs because in contrast to the other
transformations the time-coordinate transformations can be represented
on the phase space only by eliminating the time derivatives appearing
in the transformation formula with the aid of the evolution equation.
This implies that the Hamiltonian constraint is of a dynamical nature
unlike the other constraints(cf. \cite{StoneKuchar92}).

Classically this Poisson bracket structure of the constraint functions
causes no problem.  However, if one tries to quantize the theory, it
introduces a nontrivial ambiguity in the operator ordering for the
constraint operators.  This ambiguity is common to all the canonical
approaches.  This problem will be discussed in more detail in \S4.

\subsection{Real Triad Approach}

If one would like to describe the interactions of spinor fields with
the gravitational field, one must introduce the local
pseud-orthonormal frame, the so-called tetrad.  The Einstein-Hilbert
action can be easily rewritten in terms of the tetrad.

Let $e_a=e_a^\mu \partial_\mu$ be a tetrad field and
$\theta^a=\theta^a_\mu dx^\mu$ be its dual 1-form basis:
\begin{equation}
\theta^a(e_b)=\theta^a_\mu e_b^\mu=\delta^a_b.
\end{equation}
The metric tensor $g$ is written in terms of $\theta^a$ as
\begin{equation}
g_{\mu\nu}=\eta_{ab}\theta^a_\mu\theta^b_\nu,
\label{MetricByTetrad}\end{equation}
where $\eta_{ab}$ is the flat metric.
Though the gravitational action for the tetrad is obtained simply by
substituting this expression into Eq.(\ref{EinsteinHilbertAction}), it
is better to introduce the connection form to make clear the structure
of the action.

In general the connection form for the covariant derivative $\nabla$
with respect to a given vector frame $e_a$ and its dual 1-form basis
$\theta^a$ is a set of 1-form $\omega^a{}_b$ defined
by\CITE{MisnerThorneWheeler73}
\begin{equation}
\nabla_X e_a=e_b\omega^b{}_a(X),\quad
\nabla_X\theta^a=-\omega^a{}_b(X)\theta^b.
\end{equation}
The curvature tensor $R^\lambda{}_{\sigma\mu\nu}$ of this covariant
derivative is related to the curvature 2-form defined by
\begin{equation}
\R^a{}_b=d\omega^a{}_b + \omega^a{}_c\wedge\omega^c{}_b
\end{equation}
as
\begin{equation}
\R^a{}_{b\mu\nu}=\theta^a_\lambda e_b^\sigma R^\lambda{}_{\sigma\mu\nu}.
\end{equation}
Hence the Lagrangian density of the Einstein-Hilbert action is written
in terms of the tetrad and the connection form as
\begin{equation}
\sqrt{-g}R=|\theta|e^{a\mu}e^{b\nu}\R_{ab\mu\nu},\label{EHactionByTetrad}
\end{equation}
where $|\theta|=\det(\theta^a_\mu)=\sqrt{-g}$ and the tetrad indices
$a, b, \ldots$ are lowered and raised by $\eta_{ab}$.

Thus in terms of the tetrad and the connection form the Einstein
theory can be put in a form similar to the gauge field theory.
Actually the action obtained from the Lagrangian density
(\ref{EHactionByTetrad}) is invariant under the local Lorentz
transformations as well as the general coordinate transformations.
The theory in the present form, however, cannot be regarded as a
genuine gauge theory since the connection form is not an independent
field but is assumed to be expressed in terms of the tetrad through
the condition that it corresponds to the Riemannian connection. This
condition is expressed by the following two equations:
\begin{eqnarray}
&&\Theta^a:=d\theta^a +
\omega^a{}_b\wedge\theta^b=0\quad(\hbox{Torsion free}),\\ &&(\nabla
g)_{ab}=\omega_{ab}+\omega_{ba}=0\quad(\hbox{Metricity}).
\end{eqnarray}
Apparently, if one would like to treat the connection form as an
independent field, one should impose these equations as the extra
constraints.  Interestingly, however, it is not the case: if we only
require the metricity condition, the torsion free condition is
obtained from the action.  Thus the Einstein theory can be formulated
as a gauge theory for the proper Lorentz group $SO_+(3,1)$.  This
point will play an important role in putting the theory into the
complex canonical form.

Before proving the above statement and its generalization to the case
in which the interactions with matter fields are included, we must
make some comments on the $SO_+(3,1)$ connection.

\subsubsection{$SO_+(3,1)$ connection}

{}From now on we denote the connection form by
$A^a{}_b=A^a{}_{b\mu}dx^\mu$ and reserve the symbol $\omega^a{}_b$ to
denote the Riemannian connection form expressed in terms of the
tetrad. Further the latin indices $a, b, \ldots$ are always raised or
lowered by $\eta_{ab}$ and $A^a{}_b$ is assumed to satisfy
\begin{equation}
A_{ab}=-A_{ba}, \label{MetricityOnA}
\end{equation}
which corresponds to the metricity condition.

In the above argument the connection form is regarded as defining an
linear connection in the tangent bundle $T(M)$.  From this standpoint,
for example, the covariant derivative of a vector field $V=V^ae_a$ is
expressed in terms of the connection form $A^a{}_b$ as
\begin{equation}
\nabla_X V = e_a(dV^a(X) + A^a{}_{b}(X)V^b).
\end{equation}
This expression consists of two parts: the part defining a derivative
of the component fields $V^a$ and the tetrad which maps the component
fields to a vector field.  Since $\nabla_X V$ is invariant under the
local Lorentz transformation $\Lambda\in SO_+(3,1)$ of the tetrad
\begin{eqnarray}
&&e_a \rightarrow e'_a=e_b(\Lambda^{-1})^b{}_a=\Lambda_a{}^b e_b, \\
&&\theta^a \rightarrow \theta'{}^a=\Lambda^a{}_b\theta^b
\end{eqnarray}
the connection form transforms under this transformation as
\begin{equation}
A \rightarrow A'=\Lambda A \Lambda^{-1} -d\Lambda \Lambda^{-1}.
\end{equation}
Thus the connection form can be regarded as a $SO_+(3,1)$ gauge
field.  Mathematically speaking this implies that the connection form
$A^a{}_b$ defines a connection in a principal fiber bundle
$P(M,SO_+(3,1))$ or vector bundles associated with
it\CITE{KobayashiNomizu63}.

Conversely, if a connection $A$ in the principal fiber bundle
$P(M,SO_+(3,1))$ is given, we can define an linear connection $\nabla$
in the tangent bundle $T(M)$ with the aid of a quantity $e_a^\mu$
which transforms as a covector under the local $SO_+(3,1)$
transformations and a contravariant vector under the general
coordinate transformation simultaneously.  If we define a metric by
Eq.(\ref{MetricByTetrad}), the linear connection satisfies the metricity
condition and $e_a=e_a^\mu\partial_\mu$ becomes a pseud-orthonormal
tetrad with respect to this metric.

Thus $A^a{}_b$ can be regarded either as the connection form defining
an linear connection in the tangent bundle $T(M)$ or as the
$SO_+(3,1)$ gauge field. Though both the view points are
mathematically equivalent, the second view point turns out to be more
natural and convenient if one would like to treat the connection form
as an independent gauge field.  Hence we adopt the second view point
and regard the connection form as defining a connection in a principal
fiber bundle $P(M,SO_+(3,1))$ throughout this paper.

Let the covariant derivative in this sense be denoted by $D$.  Then in
general the covariant derivative of a quantity $\phi$ which transforms
under the local $SO_+(3,1)$ transformation $\Lambda(x)$ as
\begin{equation}
\phi \rightarrow \rho(\Lambda)\phi \label{SOTensorTrf}
\end{equation}
with some representation $\rho:SO_+(3,1) \rightarrow GL(\RF^m)$ is
given by
\begin{equation}
D_X \phi = d\phi(X) + d\rho(A(x))\phi.
\end{equation}
For example, the covariant derivative of a vector field $V^a$ as a
section of $SO_+(3,1)$ vector bundle is given by
\begin{equation}
D_\mu V^a= \partial_\mu V^a + A^a{}_{b\mu}V^b.
\end{equation}
Similarly the covariant derivative of a 2-component spinor field $\xi$
as a section of a spinor bundle is given by
\begin{equation}
D_\mu \xi =\partial_\mu \xi + \chiral\A_{0I\mu}\sigma_I\xi,
\end{equation}
where $\sigma_I$ is the Pauli matrix and $\chiral\A$ is defined by
\begin{equation}
\chiral \A_{0I}:={1\over2}\left(A_{0I}\pm {i\over2}\epsilon_{IJK}A_{JK}\right).
\end{equation}
Here $+$ sign and $-$ sign corresponds to the left chiral and the
right chiral field spinors, respectively.  For a Dirac spinor field
$\psi$ these formulas are put together in the form
\begin{equation}
D_\mu \psi=\partial_\mu\psi
-{1\over8}A_{ab\mu}[\gamma^a,\gamma^b]\psi.
\end{equation}

$\eta_{ab}$ can be regarded as a natural metric of each fiber of the
vector bundle, and the metricity condition Eq.(\ref{MetricityOnA}) is
expressed as
\begin{equation}
D\eta_{ab}=0.
\end{equation}
{}From this condition it follows that the covariant derivative of
$\epsilon_{abcd}$ also vanishes:
\begin{equation}
D\epsilon_{abcd}=0
\end{equation}

One important point to be noted here is that tensor fields expressed
in the coordinate basis are regarded as scalar with respect to the
derivative $D$. For example
\begin{equation}
D_\mu V^\nu=\partial_\mu V^\nu.
\end{equation}
Thus $DT$ does not behave as a tensor under the coordinate
transformations even if $T$ is a tensor in general.  However,
restricted to the tensorial forms, that is, differential forms whose
values transform under the local $SO_+(3,1)$ transformation $\Lambda$
as in Eq.(\ref{SOTensorTrf}), we can define a covariant derivative
from $D$, so called the covariant exterior derivative, by
\begin{equation}
D\chi := d\chi + d\rho(A)\wedge\chi.
\end{equation}
Like the ordinary exterior derivative the equation
\begin{equation}
D(\chi\wedge\phi) = D\chi\wedge \phi + (-1)^p\chi\wedge
D\phi\quad(\chi:p\r{-form})
\end{equation}
holds, but $D^2$ does not vanish in general.

For example, the exterior covariant derivative of an ordinary 1-form,
$D_\mu V_\nu - D_\nu V_\mu$, behaves as a 2nd-rank covariant tensor
under the coordinate transformations.  In particular the dual tetrad
basis $\theta^a$ is a tensorial 1-form, and its covariant exterior
derivative coincides with the torsion form of the corresponding linear
connection:
\begin{equation}
\Theta^a:=D\theta^a=d\theta^a + A^a{}_b\wedge\theta^b.
\label{TorsionForm:def}\end{equation}
Taking the covariant exterior derivative of the torsion form, we get
the first Bianchi identity
\begin{equation}
D\Theta^a=F^a{}_b\wedge\theta^b, \label{1stBianchiIdentity}
\end{equation}
where $F^a{}_b$ is the curvature form defined by
\begin{equation}
F^a{}_b:=dA^a{}_b + A^a{}_c\wedge A^c{}_b.
\end{equation}
Further the covariant exterior derivative of the curvature form yields
the second Bianchi identity:
\begin{equation}
DF^a{}_b=0. \label{2ndBianchiIdentity}
\end{equation}
Finally since the identity
\begin{equation}
|\theta|e^\mu_a={1\over
4!}\epsilon^{\mu\nu\lambda\sigma}\epsilon_{abcd}\theta^c_\nu\theta^d_\lambda\theta^e_\sigma
\end{equation}
yields the equation
\begin{equation}
D(\epsilon_{abcd}\theta^c\wedge\theta^d\wedge\theta^e)=4!D_\mu(|\theta|e^\mu_a)d^4x,
\end{equation}
$D_\mu(|\theta|e^\mu_a)$ behaves as a scalar density under the
coordinate transformation.

\subsubsection{1st-order Palatini action}

Now we prove that the second-order action given by
Eq.(\ref{EHactionByTetrad}) is equivalent to the first-order Palatini
action obtained from it by treating the connection form as the
independent variable:
\begin{eqnarray}
&S_\r{G}(e,A) &:={1\over2\kappa^2}\int_M d^4x
|\theta|e^{a\mu}e^{b\nu}F_{ab\mu\nu}
-{1\over\kappa^2}\int_{\partial
M}d\Sigma_\mu|\theta|e^{a\mu}e^{b\nu}A_{ab\nu}\nonumber\\
&&={1\over2\kappa^2}\int_M [\Sigma^{ab}\wedge *F_{ab} - d(\Sigma^{ab}\wedge
*A_{ab})],
\end{eqnarray}
where
\begin{eqnarray}
&&\Sigma^{ab}:=\theta^a\wedge\theta^b,\label{Sigma:def}\\
&&*F_{ab}:={1\over2}\epsilon_{abcd}F^{cd}.
\end{eqnarray}

First varying $A$ in $S_\r{G}(e,A)$ we obtain
\begin{equation}
2\kappa^2\delta_AS_\r{G}(e,A)=-2 \int_M\Theta^a\wedge\theta^b\wedge *\delta
A_{ab}.
\end{equation}
Hence $\delta_AS_\r{G}(e,A)=0$ yields
\begin{equation}
\Theta^{[a}\wedge\theta^{b]}=0\quad \Leftrightarrow\quad \Theta^a=0.
\label{dSA:PureGravity}\end{equation}
Since the Riemannian connection is specified by the metricity and the
torsion free conditions, this equation determines $A^a{}_b$ to be
$A^a{}_b=\omega^a{}_b(e)$.  Therefore the total variational equation
$\delta S_\r{G}(e,A)=0$ is equivalent to the variational equation for
the 2nd-order action $\delta S_\r{G}(e)=0$.

This equivalence is easily extended to the case in which matter fields
are included.  For simplicity we consider as matter fields a
Yang-Mills field $\bm{A}=(\bm{A}^P)$, a real scalar field multiplet
$\Phi$ and a left-chiral spinor field multiplet $\xi$ coupled with the
gravitational field minimally. We do not lose any generality by
restricting to left chiral spinors since right chiral spinors can be
converted to left chiral spinors by the charge conjugation
\begin{equation}
\xi \rightarrow \xi_c = \pm i \sigma_2 \xi^*.
\end{equation}

 The action for the Yang-Mills field is given by
\begin{equation}
S_\r{YM}= -{1\over4} \int_M \Omega_4
g^{\mu\nu}g^{\lambda\sigma}\bm{F}_{\mu\lambda}\cdot\bm{F}_{\nu\sigma},
\end{equation}
where the field strength
$\bm{F}=(\bm{F}^P)=({1\over2}\bm{F}^P{}_{\mu\nu}dx^\mu\wedge dx^\nu)$
is expressed in terms of the gauge field $\bm{A}$ and the structure
constant $f^P_{QR}$ as
\begin{equation}
\bm{F}^P=d\bm{A}^P + f^P_{QR}\bm{A}^Q\wedge\bm{A}^R,
\end{equation}
and
\begin{equation}
\Omega_4=\sqrt{-g}d^4x=\theta^0\wedge\theta^1\wedge\theta^2\wedge\theta^3,
\end{equation}
The action for the scalar field is given by
\begin{equation}
S_\r{S}=-\int_M
\Omega_4[{1\over2}g^{\mu\nu}\bm{D}_\mu\Phi\cdot\bm{D}_\nu\Phi +
V(\Phi)],
\end{equation}
where $\bm{D}_\mu$ is the covariant derivative with respect to the
gauge field $\bm{A}$ and expressed in terms of the representation
matrix $\bm{T}=(T_P)$ satisfying
\begin{equation}
[T_P, T_Q]=f^R_{PQ}T_R,
\end{equation}
as
\begin{equation}
\bm{D}_\mu \Phi = \partial_\mu\Phi + e \bm{A}\!_\mu\cdot\bm{T}\Phi.
\end{equation}
Finally the action for the left-chiral spinor field is given by
\begin{equation}
S_\r{F}=\int_M \Omega_4[-{i\over2}\xi^\dagger\sigma^a\bm{D}_a\xi
+{i\over2}(\bm{D}_a\xi)^\dagger\sigma^a\xi-\xi_c^\dagger M(\Phi)\xi],
\end{equation}
where $\sigma_0=1$ and $\bm{D}_a$ is expressed as
\begin{equation}
\bm{D}_a\xi=e_a^\mu(\partial_\mu +
{}^+\!\A_{0I\mu}\sigma_I+e\bm{A}\cdot\bm{T})\xi.
\end{equation}
The total 1st-order action $S_1$ is given by the sum of these actions
and $S_\r{G}(e,A)$:
\begin{equation}
S_1=S_\r{G}(e,A)+S_\r{YM}(\bm{A},e)+
S_\r{S}(\Phi,\bm{A},e)+S_\r{F}(\xi,\bm{A},e,A).
\end{equation}

Since the connection form appears in $S_\r{F}$ as well as in
$S_\r{G}$, the variational equation $\delta_A S_1=0$ is modified from
Eq.(\ref{dSA:PureGravity}) to
\begin{equation}
\Theta^{[a}\wedge\theta^{b]} =
-{\kappa^2\over12}\epsilon^{[a}{}_{cde}S^{b]}\theta^c\wedge\theta^d\wedge\theta^e,
\end{equation}
where
\begin{equation}
S^a:=\xi^\dagger\sigma^a\xi.
\end{equation}
This equation can be solved with respect to $\Theta^a$ to yield
\begin{equation}
\Theta^a=-{\kappa^2\over4}\epsilon^a{}_{bcd}S^b\theta^c\wedge\theta^d.
\label{Torsion:spinor}\end{equation}

Thus the torsion does not vanish and the connection form $A$ is not
Riemannian when the gravitational field is coupled with spinor fields.
In spite of this we can prove that the first-order action $S_1$ is
equivalent to the second-order action obtained from $S_1$ by replacing
$A$ with the Riemannian connection form $\omega$.  First from the
definition of the torsion form (\ref{TorsionForm:def}) and
Eq.(\ref{Torsion:spinor}) the connection form is expressed in terms of
the Riemannian connection form and the spinor current $S^a$ as
\begin{equation}
A^a{}_b = \omega^a{}_b -{\kappa^2\over4}\epsilon^a{}_{bcd}S^c\theta^d.
\label{ConnectionForm:SpinorContribution}\end{equation}
Putting this expression into $S_\r{G}$ and $S_\r{F}$, we get
\begin{eqnarray}
&&S_\r{G}(e,A)=S_\r{G}(e,\omega) + {3\kappa^2\over16}\int_M \Omega_4
S_aS^a,\\
&&S_\r{F}(\xi,\bm{A},e,A)=S_\r{F}(\xi,\bm{A},e,\omega)-{3\kappa^2\over16}
\int_M \Omega_4 S_aS^a.
\end{eqnarray}
Thus the contributions of the spinor current to $S_\r{G}$ and
$S_\r{F}$ cancels:
\begin{equation}
S_1(e,A,\bm{A},\Phi,\xi)=S_1(e,\omega,\bm{A},\Phi,\xi)=S_2(e,\bm{A},\Phi,\xi).\end{equation}
This proves the equivalence of the first-order Palatini action and the
minimally coupled second-order action.

\subsubsection{$(3+1)$-decomposition}

Now we put the action into the canonical form.  It is a rather easy
job if we start from the first-order action.  For simplicity we
neglect the material fields here.  Further we restrict the freedom of
the local Lorentz transformation of the tetrad so that $e_0$ is
orthogonal to the $t$=const hypersurfaces.  We call this gauge {\it
the spatial gauge}. This partial gauge fixing does not affect the
physical content of the theory.

Under the spatial gauge $e_0$ is expressed in terms of the lapse
function $N$ and the shift vector $N^j$ as
\begin{equation}
e_0=N^{-1}(\partial_t-N^j\partial_j),
\end{equation}
and $e_I$ is tangential to the $t$=const hypersurfaces:
\begin{equation}
e_I=e_I^j\partial_j.
\end{equation}
Accordingly the dual basis is written as
\begin{equation}
\theta^0=Ndt, \qquad \theta^I=\theta^I_j(dx^j+N^jdt),
\end{equation}
and the intrinsic three dimensional metric $q_{jk}$ of the
constant-time hypersurfaces is expressed in terms of $\theta^I_j$ as
\begin{equation}
q_{jk}=\theta^I_j\theta^I_k.
\end{equation}

Since $\Sigma^{ab}$ is expressed as
\begin{eqnarray}
&&\Sigma^{0I}=N\theta^I_j dt\wedge dx^j,
\label{Sigma:decomposition1}\\
&&\Sigma^{IJ}=\theta^I_j\theta^J_k\left[dx^j\wedge
dx^k+dt\wedge(N^jdx^k-N^kdx^j)\right],\label{Sigma:decomposition2}
\end{eqnarray}
the first-order Lagrangian density is written as
\begin{eqnarray}
&\Sigma^{ab}\wedge *F_{ab} &=
\epsilon_{IJK}\left({1\over2}N\theta^I_jF_{JKkl}+
\theta^{I}_j\theta^{J}_mN^mF_{0Kkl}\right.\nonumber\\
&&\quad \left. - \theta^I_k\theta^J_lF_{0Ktj}\right)dt\wedge
dx^j\wedge dx^k\wedge dx^l.
\end{eqnarray}

In order to make this expression simpler, let us introduce the
following three dimensional vectors whose components are labeled by
the internal index $I, J, \ldots$:
\begin{eqnarray}
&&\tilde e^j:=(\tilde e^{jI}) := (\sqrt{q}e^{Ij}),\\
&&P_\mu:=(P_{I\mu}):=({1\over2\kappa^2}A_{0I\mu}),\\
&&Q_\mu:=(Q_{I\mu}):=({1\over2}\epsilon_{IJK}A_{JK\mu}).
\end{eqnarray}
The components of $F_{ab\mu\nu}$ appearing in the above equation are
expressed in terms of $P_\mu$ and $Q_\mu$ as
\begin{eqnarray}
&&{1\over2}\epsilon_{IJK}F_{JKjk}=(F_{jk}+4\kappa^4 P_j\times
P_k)_I,\\ &&F_{0Ijk}=2\kappa^2(D_jP_k-D_kP_j)_I,\\
&&F_{0Itj}=2\kappa^2(\partial_tP_j-D_jP_t-Q_t\times P_j)_I,
\end{eqnarray}
where
\begin{eqnarray}
&&F_{jk}:=\partial_j Q_k - \partial_kQ_j - Q_j\times Q_k,\\
&&D_jP_k:=\partial_jP_k - Q_j\times P_k.
\end{eqnarray}
Further from the identity
\begin{eqnarray}
&&\epsilon^{jkl}\theta^I_j\theta^J_k\theta^K_l=\epsilon^{IJK}\sqrt{q},
\label{TetradID:3D}\\
&&\epsilon_{IJK}\theta^I_j\theta^J_l\theta^K_l=\epsilon_{jkl}\sqrt{q}
\end{eqnarray}
we get
\begin{eqnarray}
&&\epsilon_{IJK}\theta^J_j\theta^K_k=\epsilon_{jkl}\tilde e^{Il},\\
&&\epsilon^{jkl}\theta^I_j={1\over\sqrt{q}}\epsilon^{IJK}\tilde
e^j_J\tilde e^k_K.
\end{eqnarray}

Putting these expressions into the above Lagrangian density, we
finally obtain the following gravitational Lagrangian in the canonical
form:
\begin{eqnarray}
&L_\r{G}&=\int_\Sigma d^3x \big[2\dot{\tilde e}{}^j\cdot P_j -(P_t\cdot C_\r{B}
+ Q_t\cdot C_\r{R} +N^jC_{\r{M}j} + \utilde N C_\r{H})\big] \nonumber\\
&& \quad + \int_{\partial\Sigma}dS_j\big[{1\over\kappa^2}\utilde N
(\tilde e^k\times\tilde e^j)\cdot Q_k + 2(\tilde e^j N^k -\tilde e^k N^j)
\cdot P_k \big],
\label{LG:Canonical:RealTetrad}\end{eqnarray}
where $\utilde N=N/\sqrt{q}$ and
\begin{eqnarray}
&&C_\r{B}:=2D_j\tilde e^j,\\ &&C_\r{R}:=2\tilde e^j\times P_j,\\
&&C_{\r{M}j}:=-2\tilde e^k\cdot (D_jP_k-D_kP_j),\\ &&C_\r{H}:=-(\tilde
e^j\times \tilde e^k)\cdot \left[{1\over2\kappa^2}F_{jk} + 2\kappa^2
P_j\times P_k\right].
\end{eqnarray}

{}From this expression we see that only the quantities $(\tilde e^j,
P_j)$ are dynamical canonical variables, and the others are
non-dynamical.  Among these non-dynamical variables, $P_t$, $Q_t$,
$N^j$ and $\utilde N$ play the role of Lagrange multipliers and the
variation of the action with respect to them yield the four sets of
constraints on the canonical variables,
\begin{eqnarray}
&&C_\r{B}=0 \label{BoostConstraint:RealTetrad},\\ &&C_\r{R}=0
\label{RotationConstraint:RealTetrad},\\ &&C_{\r{M}j}=0
\label{MomentumConstraint:RealTetrad},\\ &&C_\r{H}=0
\label{HamiltonianConstraint:RealTetrad}.
\end{eqnarray}
The third and the fourth of these correspond to the momentum and the
Hamiltonian constraints in the metric approach, respectively. On the
other hand the first and the second ones are new constraints arising
from the local Lorentz invariance of the theory, and represent the
generators of the Lorentz boost and the spatial rotation of the
tetrad, respectively.

In contrast to these variables, the variation of the action with
respect to $Q_j$ does not lead to a constraint but yields equations
determining $Q_j$ itself and $P_t$.  In fact the variation with
respect to $Q_j$ yields
\begin{eqnarray}
&0={\delta L_\r{G}\over\delta Q_j} &= 2\tilde
e^j\times(P_t-N^kP_k+{N\over4\kappa^2}C_\r{B}
+{1\over2\kappa^2}\partial_kN \tilde
e^k)\nonumber\\ &&\quad+N^jC_\r{R} + {N\over\kappa^2}D_ke^j\times\tilde e^k,
\end{eqnarray}
which is equivalent under the constraints $C_\r{B}=C_\r{R}=0$ to the
two equations
\begin{eqnarray}
&&\phi^{jk}:=\left(e^{(j}\times D_l e^{k)}\right)\cdot
e^l=0,\label{phi:def}\\
&&P_t=N^kP_k-{1\over2\kappa^2}\partial_kN\tilde e^k
-{1\over4\kappa^2}Ne^m\epsilon_{mjk}(e^j\times D_l e^k)\cdot \tilde
e^l.\label{Pt:SpatialGauge}
\end{eqnarray}

Since $Q_j$ is non-dynamical, we must eliminate them to obtain a
consistent canonical formalism.  This is achieved with the help of the
constraint $C_\r{B}=0$ and the equation $\phi^{jk}=0$.  To show this,
let us calculate the torsion form of the three-dimensional $SO(3)$
connection
\begin{equation}
D_j V_I=\partial_jV_I-\epsilon_{IJK}Q_{Jj}\times V_K.
\end{equation}
Since the torsion form is given by the covariant exterior derivative
of the three dimensional dual basis $\theta^I_j$ as
${}^3\!\Theta^I_{jk}=2D_{[j}\theta^I_{k]}$, we obtain the following
relation with the help of the identity equation (\ref{TetradID:3D}):
\begin{eqnarray}
&{}^3\!\Theta^{pq}&:={1\over2}e^p_I\epsilon^{qjk}\;{}^3\!\Theta^I_{jk}
\nonumber\\ && ={1\over4}(e^p\times e^q)\cdot C_\r{B} + \phi^{pq}.
\end{eqnarray}
Thus the equations $C_\r{B}=\phi^{jk}=0$ are equivalent to the torsion
free condition on the metric connection $D_j$, which implies that the
connection is the Riemannian connection with respect to the triad
$e^j_I$:
\begin{equation}
Q_{jI}={1\over2}\epsilon_{IJK}\omega_{JKj}(e).
\label{QbyOmega}\end{equation}

Thus, although the canonical Lagrangian
(\ref{LG:Canonical:RealTetrad}) appears to be a simple polynomial, it
is really a complicated rational function after the elimination of
$Q_j$.  In fact the momentum and the Hamiltonian constraint functions
coincide with those in the metric approach modulo the constraint
$C_\r{R}$.  To see this, let us examine the relation between $P_{Ij}$
and the momentum variable $p^{jk}$ in the metric approach.  First note
that for the three-dimensional Riemannian connection $D_je^k_I$ is
related to the Christoffel symbol $\Kr(j;kl)$ by
\begin{equation}
D_je^k_I=\partial_je^k_I+\omega^I{}_{Jj}e^k_J=\partial_je^k_I-(\RSCD_je_I)^k
=-\Kr(k;jl)e^l_I.
\label{DeByGamma}\end{equation}
{}From this it follows that
\begin{equation}
(D_jP_k)_I=(\RSCD_jP_{kl}+\Kr(m;jk)P_{ml})e^l_I,
\end{equation}
where
\begin{equation}
P_{jk}=P_{Ij}\theta^I_k.
\end{equation}
Hence the momentum constraint function is expressed as
\begin{equation}
C_{\r{M}j}=-2\sqrt{q}q^{kl}(\RSCD_jP_{kl}-\RSCD_kP_{jl}).
\end{equation}
Now let us define $p_{jk}$ by
\begin{equation}
p_{jk}:=-\sqrt{q}(P_{(jk)}-q_{jk}P_{lm}q^{lm}).
\end{equation}
Then from the relation
\begin{equation}
P_{[jk]}={1\over4}\epsilon_{jkl}(C_\r{R}\cdot e^l),
\end{equation}
we get
\begin{equation}
C_{\r{M}j}=\H_j + {1\over2}\sqrt{q}\epsilon_{jkl}\RSCD^k(C_\r{R}\cdot e^l).
\end{equation}
Similarly the Hamiltonian constraint function is written as
\begin{equation}
C_\r{H}=\sqrt{q}\H_0 -{\kappa^2\over4}C_\r{R}\cdot C_\r{R}.
\end{equation}
Since the Poisson brackets of $q^{jk}=e^j\cdot e^k$ and $p_{jk}$ with
$C_\r{R}$ vanishes, it follows from these equations that $q^{jk}$ and
$p_{jk}$ defined above satisfy the same evolution equations as in the
metric approach.  This implies that $p_{jk}$ here coincides
with the momentum variable defined in the metric approach.  In
particular from Eq.(\ref{p:def}) $P_{(jk)}$ is expressed in terms of
the extrinsic curvature $K_{jk}$ as
\begin{equation}
 P_{(jk)}={1\over2\kappa^2}K_{jk}, \qquad P_j\cdot \tilde e^j =
{1\over2\kappa^2}\sqrt{q}K.
\label{PbyK}\end{equation}

\subsection{Chiral 1st-Order Formalism}

As we have seen in the previous section, the Hamiltonian becomes a
complicated rational functions of the canonical variables in the
canonical theory obtained from the 1st-order Palatini action in spite
of its apparent simple structure at the start. The origin of this
complexity was that the spatial part $A_{IJ}$ of the connection form
is not dynamical and should be eliminated.  The main point of the
complex canonical theory is that this elimination procedure can be
avoided if the momentum variable is modified to a complex quantity to
include $A_{IJ}$ as its imaginary part.

\subsubsection{Chiral representation of the proper Lorentz group}

Let $X_{ab}$ be a quantity antisymmetric with respect to the indices,
$X_{ba}=-X_{ab}$, and $*X_{ab}$ be its dual defined by
\begin{equation}
*X_{ab}:={1\over2}\epsilon_{ab}{}^{cd}X_{cd}.
\end{equation}
This dual operation satisfies the identity equations
\begin{eqnarray}
&&**X_{ab}=-X_{ab},\\ &&*X_a{}^c\circ
*Y_{cb}={1\over2}\eta_{ab}X_{cd}\circ Y^{cd}+X_{bc}\circ Y^c{}_a,
\end{eqnarray}
where $\circ$ is any binary bilinear operation such as the exterior
product of differential forms.  From this it follows that
\begin{equation}
*X_{ab}\circ Y^{ab}=X_{ab}\circ *Y^{ab}.
\label{Star:Commutativity}\end{equation}

With the help of this operation let us define the pair of complex chiral
combinations of $X_{ab}$ by
\begin{equation}
\chiral X_{ab}:= {1\over2}(X_{ab}\pm i*X_{ab}).
\end{equation}
These quantities are eigen quantities of the dual operation:
\begin{equation}
*\chiral X_{ab}=\mp i\chiral X_{ab}.
\end{equation}
As we saw in \S2.2, the connection form $A_{ab}$ is coupled with the
left(right) chiral spinor in the chiral combination
${}^+\!\A_{ab}({}^-\!\A_{ab})$.  This is the reason why we call
$\chiral X$ the chiral combination.  Accordingly we call
${}^+\!X_{ab}$ and ${}^-\!X_{ab}$ the left chiral and the right chiral
combination, respectively.  In the literature these quantities are
often called the anti-dual and the dual variables as well,
respectively.

Owing to the self-duality of the chiral combination, only the half of
the components of $\chiral X_{ab}$ are independent:
\begin{equation}
\chiral X_{0I}=\pm {i\over2}\epsilon_{IJK}\chiral X_{JK}.
\end{equation}
This implies that the Lorentz group can be linearly represented on the
three-dimensional complex space $\CF^3$.  In fact the chiral
combinations yield natural isomorphisms between the proper Lorentz
group $SO_+(3,1)$ and the complex orthogonal group $SO(3,\CF)$.  To be
explicit, for $\Lambda\in SO_+(3,1)$ under which $X^{ab}$ transforms
\begin{equation}
X^{ab} \rightarrow \Lambda^a{}_c \Lambda^b{}_d X^{cd},
\end{equation}
$\chiral X_{0I}$ transforms as
\begin{equation}
\chiral X_{0I} \rightarrow \chiral X_{0J}\chiral O_{JI}
\end{equation}
where $\chiral O_{IJ}$ is matrix defined by
\begin{equation}
\chiral O_{IJ}=2\Lambda^0{}_{[0}\Lambda^J{}_{I]} \mp
i\epsilon_{JKL}\Lambda^K{}_{[0}\Lambda^L{}_{I]}.
\end{equation}
Since $\chiral O_{IJ}$ is shown to belong to $SO(3,\CF)$ and the
correspondence is one-to-one, it gives the isomorphism.

Another way to look at this isomorphism is to utilize the spinor
representation of $SO_+(3,1)$ to $SL(2,\CF)$.  For example the right
chiral representation is the double-valued correspondence $\Lambda\in
SO_+(3,1) \rightarrow V\in SL(2,\CF)$ determined by
\begin{equation}
V\sigma_a V^\dagger = \sigma_b\Lambda^b{}_a.
\label{RightChiralRepresentation:spinor}\end{equation}
If we parametrize $V$ as $V=z^a\sigma_a$ by a four-dimensional complex
time-like vector $z^a$ satisfying $\eta_{ab}z^az^b=-1$, $\Lambda$ is
expressed as
\begin{equation}
\Lambda^a{}_b=\eta^{ab}(|z^0|^2-z^I\bar z^I) +2z^{(a}\bar z^{b)}
+i\epsilon^{ab}{}_{cd}z^c\bar z^d.
\end{equation}
On the other hand the matrix $O$ defined by
\begin{equation}
V\sigma_I V^{-1} = O_{IJ}\sigma_J
\end{equation}
belongs to $SO(3,\CF)$ and gives a two-to-one representation of
$SL(2,\CF)$ to $SO(3,\CF)$ as is seen from its expression in terms of
$z^a$:
\begin{equation}
O_{IJ}=(1+2z^Iz^I)\delta_{IJ}-2z^Iz^J + 2i\epsilon_{IJK}z^0z^K.
\end{equation}
Combination of these representations yields the above isomorphism
based on the right chiral combination.  Similarly using the left chiral
representation which is obtained by assigning $(V^\dagger)^{-1}$ to
$\Lambda$ with $V$ defined by
Eq.(\ref{RightChiralRepresentation:spinor}) we obtain the isomorphism
based on the left chiral combination.

As we will see soon, the fact that the chiral combination makes it
possible to represent the Lorentz group on the three dimensional
complex space plays an important role in the complex canonical theory.

\subsubsection{Chiral action}

As shown by Jacobson\CITE{JacobsonSmolin87,JacobsonSmolin88a}, the
most elegant way to arrive at the complex canonical theory is to use
the chiral decomposition of the first-order Palatini action written in
terms of the tetrad and the connection form.

Let us define the chiral connection by
\begin{equation}
\chiral\A_{ab}:={1\over2}(A_{ab}\pm i*A_{ab}),
\end{equation}
and the chiral gravitational action $\chiral S_\r{G}$ in terms of the
first-order Palatini action by
\begin{equation}
\chiral S_\r{G}=S_\r{G}(e,\chiral \A).
\end{equation}
Then since the curvature form $F_{ab}[\chiral\A]$ constructed from
$\chiral\A_{ab}$ coincides with the chiral combination of the
curvature form $F_{ab}[A]$,
\begin{equation}
\chiral \F_{ab}:={1\over2}(F_{ab}\pm i*F_{ab})=F_{ab}[\chiral \A],
\end{equation}
the chiral Lagrangian density $\chiral \L_\r{G}$ corresponding to
$\chiral S_\r{G}$ is expressed as
\begin{eqnarray}
&\chiral\L_\r{G}&={1\over2\kappa^2}[\Sigma^{ab}\wedge *\chiral\F_{ab}
-d(\Sigma^{ab}\wedge *\chiral\A_{ab})] \nonumber\\
&&=\mp{i\over2\kappa^2}[\Sigma^{ab}\wedge \chiral\F_{ab}
-d(\Sigma^{ab}\wedge \chiral\A_{ab})] \nonumber\\
&&={1\over2}\L_\r{G} \mp {i\over4\kappa^2}[\Sigma^{ab}\wedge F_{ab}
-d(\Sigma^{ab}\wedge A_{ab})].
\label{ChiralLagrangianDensity:Gravity}\end{eqnarray}
{}From the definition of the torsion form the second term in the second
line of this equation is rewritten as
\begin{equation}
\Sigma^{ab}\wedge F_{ab} -d(\Sigma^{ab}\wedge A_{ab})
=-d(\theta^a\wedge d\theta_a) + \Theta_a\wedge\Theta^a.
\end{equation}

Next we define the left-chiral action ${}^+\!S_\r{F}$ for the left
chiral spinor multiplet by
\begin{equation}
{}^+\!S_\r{F}:=-\int_M \Omega_4[i\xi^\dagger\sigma^a\bm{D}_a\xi+
\xi_c^\dagger M(\Phi)\xi],
\label{ChiralAction:LeftSpinor}\end{equation}
and the right-chiral action ${}^-\!S_\r{F}$ by its complex conjugate.
Then the corresponding chiral Lagrangian density is written as
\begin{equation}
\chiral\L_\r{F}=\L_\r{F} \pm {i\over2}[-d({1\over2}\epsilon_{abcd}S^a
\theta^b\wedge\theta^c\wedge\theta^d) + S^a*\Sigma_{ab}\wedge\Theta^b].
\end{equation}

{}From these equations, if we define the total chiral action $\chiral S$
by
\begin{equation}
\chiral S:=2\chiral S_\r{G} + S_\r{YM} + S_\r{S} + \chiral S_\r{F},
\label{TotalChiralAction}\end{equation}
the real part of the total chiral Lagrangian density $\chiral\L$
coincides with the first-order Lagrangian density $\L$ written in
terms of the tetrad, the real connection form and the matter fields:
\begin{eqnarray}
&\chiral \L &=\L \pm
{i\over2}d({1\over\kappa^2}\theta^a\wedge
d\theta_a-{1\over2}\epsilon_{abcd}S^a\theta^b\wedge\theta^c\wedge\theta^d)\nonumber\\
&&\mp{i\over2\kappa^2}(\Theta_a+{\kappa^2\over2}*\Sigma_{ab}S^b)\wedge
(\Theta^a+{\kappa^2\over2}*\Sigma^a{}_cS^c),
\end{eqnarray}
where we have used the identity $\Sigma_{ac}\wedge\Sigma^c{}_b=0$.

The variational equations obtained from the chiral action splits into
two sets of equations corresponding to the real and the imaginary part
of the action.  Hence the chiral action appears to only admit a more
restricted class of solutions than the real first-order action.
However, it is not the case.  In fact, as was shown in \S2.2, the
variation of the real action with respect to the connection form
yields Eq.(\ref{Torsion:spinor}), but the variation of the imaginary
part of the action vanishes under this equation.  Therefore the chiral
action is equivalent to the original real first-order action.

\subsubsection{$(3+1)$-decomposition}

Now let us rewrite the complex chiral action in the canonical form.
First note that the chiral gravitational Lagrangian density
(\ref{ChiralLagrangianDensity:Gravity}) is written from
Eq.(\ref{Star:Commutativity}) and the self-duality of the chiral
combinations as
\begin{equation}
\chiral\L_\r{G}=\pm{2i\over\kappa^2}[\chiral\Sigma_{0I}\wedge\chiral\F_{0I}
-d(\chiral\Sigma_{0I}\wedge\chiral\A_{0I})].
\end{equation}
This equation shows that we only have to calculate the $(0I)$
components.

In order to decompose the curvature form, let us introduce the
vector-type notation as in \S2.2:
\begin{eqnarray}
&&\chiral\A_\mu:=(\chiral\A_{I\mu}):=({1\over\kappa^2}\chiral\A_{0I\mu}),\\
&&\chiral\B_\mu:=(\chiral\B_{I\mu}):=(\epsilon_{IJK}\chiral\A_{JK\mu}).
\end{eqnarray}
In contrast to the real connection, $\chiral\A_\mu$ and $\chiral\B_\mu$
are not independent but are related owing to the self-duality as
\begin{equation}
\chiral\B_\mu = \mp 2i\kappa^2 \chiral\A_\mu,
\end{equation}
which will play a crucial role later.  In terms of these quantities
the relevant curvature form is expressed as
\begin{equation}
\chiral\F_{0I}=\kappa^2[\partial_t\chiral \A_j-\D_j\chiral\A_t]_Idt\wedge dx^j
\pm{i\over4}\chiral\F_{Ijk}dx^j\wedge dx^k,
\end{equation}
where
\begin{eqnarray}
&&\D_j\chiral \A_t:=\partial_j\chiral \A_t -
\chiral\B_j\times\chiral\A_t,\\
&&\chiral\F_{jk}:=\partial_j\chiral\B_k-\partial_k\chiral\B_j-\chiral\B_j\times\chiral\B_k.
\end{eqnarray}

Next to rewrite $\chiral\Sigma_{0I}$ let us rotate the tetrad $e_a$ by
some appropriate proper Lorentz transformation so that $e_0$ is
orthogonal to the constant-time hypersurfaces.  We denote this new
tetrad by $\hat e_a$ and define $\tilde e^j$, $q^{jk}$, $N^j$ and
$\utilde N$ from this tetrad as in \S2.2.  There we had to restrict
the tetrad variable to this special one in order to arrive at the
canonical formalism. In the present case we do not have to do such a
gauge fixing.  This is because $\chiral\Sigma_{0I}$ for the general
tetrad is related to $\chiral\hat\Sigma_{0I}$ for the special tetrad
by $\chiral\Sigma_{0I}=\chiral\hat\Sigma_{0J}\chiral O_{JI}$ with some
matrix $O\in SO(3,\CF)$ as was shown in \S2.3.1.  With the help of
this relation, if we use the chirally rotated quantity $\E^j$ defined
by
\begin{equation}
\chiral\E^j:=(\chiral\E^{Ij}):=(\tilde e^{Jj}\chiral O_{JI})
\end{equation}
in stead of $\tilde e^j$, $\chiral\Sigma_{0I}$ is simply written from
Eqs.(\ref{Sigma:decomposition1}) and (\ref{Sigma:decomposition2}) as
\begin{equation}
\chiral\Sigma_{0I}=\pm{i\over4}\epsilon_{jkl}\chiral\E^{Il}dx^j\wedge dx^k
\mp {i\over2} \epsilon_{jkl}[N^k\chiral\E^l\mp{i\over2}\utilde N
\chiral\E^k\times\chiral\E^l]^I dt\wedge dx^j.
\label{ChiralSigmaByE}\end{equation}
{}From these equations $\chiral\L_\r{G}$ is put into the following
canonical form:
\begin{eqnarray}
&2\chiral\L_\r{G}&=2\chiral\dot\E^j\cdot\chiral\A_j
-2\chiral\A_t\cdot\D_j\chiral\E^j
\nonumber\\
&&\quad \pm{i\over\kappa^2}N^j\chiral\E^k\cdot\chiral\F_{jk}
+{1\over2\kappa^2}\utilde
N(\chiral\E^j\times\chiral\E^k)\cdot\chiral\F_{jk} \nonumber\\
&&\quad +\partial_j[(N^k\chiral\E^j-N^j\chiral\E^k\mp i\utilde
N\chiral\E^k\times\chiral\E^j)\cdot\chiral A_k].
\end{eqnarray}

Next we rewrite the Lagrangian densities of the matter fields.  First
for the scalar field, introducing the momentum $\Pi$ conjugate to
$\Phi$ by Eq.(\ref{Pi:def}), $\L_\r{S}$ is written as
\begin{eqnarray}
&\L_\r{S}&=\Pi\cdot\dot\Phi + e\Pi\cdot\bm{A}^P_t\bm{T}_P\Phi -
N^j\Pi\cdot\bm{D}_j\Phi \nonumber\\ &&\quad -\utilde
N[{1\over2}\Pi^2+{1\over2}qq^{jk}(\bm{D}_j\Phi)\cdot(\bm{D}_k\Phi)+qV(\Phi)].
\end{eqnarray}
Here, since $\chiral O \in SO(3,\CF)$,  $q^{jk}$ and $q$ are expressed in
terms of $\chiral \E^j$ as
\begin{eqnarray}
&&qq^{jk}=\tilde e^j\cdot\tilde e^k = \chiral\E^j\cdot\chiral\E^k,\\
&&q={1\over3!}\epsilon_{jkl}(\tilde e^j\times\tilde e^k)\cdot\tilde
e^l
={1\over3!}\epsilon_{jkl}(\chiral\E^j\times\chiral\E^k)\cdot\chiral\E^l.
\end{eqnarray}

Next for the gauge field only $\bm{A}_j$ has the conjugate momentum
$\bm{E}^j$ defined by Eq.(\ref{E:def}) and the Lagrangian density is
written as
\begin{eqnarray}
&\L_\r{YM}&=\bm{E}^j\cdot\dot{\bm{A}}_j -
\partial_j(\bm{A}_t\cdot\bm{E}^j)+\bm{A}_t\cdot\bm{D}_j\bm{E}^j+N^j\bm{F}_{jk}\cdot\bm{E}^k
\nonumber\\ && - \utilde
N\left({1\over2}q_{jk}\bm{E}^j\cdot\bm{E}^k+{1\over4}qq^{jk}q^{lm}\bm{F}_{jl}\cdot\bm{F}_{km}\right),
\end{eqnarray}
where
\begin{equation}
\bm{D}_j\bm{E}^{Pj} = \partial\bm{E}^{Pj} + e f^P{}_{QR}\bm{A}^Q_j\bm{E}^{Rj}.
\end{equation}

Finally in order to rewrite the Lagrangian density for the spinor
field, we recall the argument on the relation of the chiral
representation of $SO_+(3,1)$ to $SL(2,\CF)$ and to $SO(3,\CF)$ in
\S2.3.1.  From the equations there $V\in SL(2,\CF)$ and ${}^+\!O_{IJ}$
are related for the left-chiral representation by
\begin{equation}
(V^\dagger)^{-1}\sigma_I V^\dagger = {}^+\!O_{IJ}\sigma_J.
\end{equation}
With the aid of this relation and
Eq.(\ref{RightChiralRepresentation:spinor}) we can express $\sigma^a
e_a$ in Eq.(\ref{ChiralAction:LeftSpinor}) in terms of $N$, $N^j$,
$\chiral\E^j$ and $V$. Putting this expression into the chiral
Lagrangian density for the spinor field yields
\begin{eqnarray}
&{}^+\!\L_\r{F}
&=\eta(\partial_t+\kappa^2\;{}^+\!\A_{It}\sigma_I+e\bm{A}_t\cdot\bm{T})\xi
-
N^j\eta(\partial_j+\kappa^2\;{}^+\!\A_{Ij}\sigma_I+e\bm{A}_j\cdot\bm{T})\xi\nonumber\\
&&-\utilde N[\chiral
\E^{jI}\eta\sigma_I(\partial_j+\kappa^2{}^+\!\A_{Ij}\sigma_I+\bm{A}_j)\xi+q\xi_c^\dagger
M\xi],
\end{eqnarray}
where
\begin{equation}
\eta:=i\sqrt{q}\xi^\dagger VV^\dagger.
\end{equation}

With these equations altogether, we finally obtain the following
canonical Lagrangian for the total system:
\begin{eqnarray}
&{}^\pm\!L &=\int_\Sigma d^3x\big[2{}\dot\E^j\cdot{}\A_j+\Pi\cdot\dot\Phi
+\bm{E}^j\cdot\dot{\bm{A}}_j+\eta\dot\xi \nonumber\\
&& -(\bm{A}_t\cdot\bm{C}_{\sbm{A}} + \A_t\cdot\C_{\A}
+N^j\C_{\r{M}j}+\utilde N \C_\r{H})\big] \nonumber\\
&&+\int_{\partial\Sigma}dS_j (N^k\E^j-N^j\E^k\mp i\utilde N
\E^k\times\E^j)\cdot\A_k,
\label{ChiralLagrangian:CanonicalForm}\end{eqnarray}
where
\begin{eqnarray}
&\bm{C}_{\sbm{A}}&:=-e\Pi\cdot\bm{T}\Phi
-\bm{D}_j\bm{E}^j+e\eta\bm{T}\xi,\\ &\C_{\A}&:=2\D_j{}\E^j \mp
\kappa^2\eta\bg{\sigma}\xi,\\
&\C_{\r{M}j}&:=\mp{i\over\kappa^2}{}\E^k\cdot{}\F_{jk}+\Pi\cdot\bm{D}_j\Phi-\bm{F}_{jk}\cdot\bm{E}^k+\eta\bm{D}_j\xi,\\
&\C_\r{H}&:=-{1\over2\kappa^2}({}\E^j\times{}\E^k)\cdot{}\F_{jk}
+{1\over2}(\Pi^2+{}\E^j\cdot{}\E^k\bm{D}_j\Phi\cdot\bm{D}_k\Phi)
+qV(\Phi) \nonumber\\ &&
\quad+{1\over4q}(\E^j\cdot{}\E^k)(\E^l\cdot{}\E^m)
[\epsilon_{pjl}\epsilon_{pkm}\bm{E}^p\cdot\bm{E}^q+\bm{F}_{jk}\cdot\bm{F}_{lm}]\nonumber\\
&& \quad \pm\E^j\cdot(\eta\bg{\sigma}\bm{D}_j\xi) + q\Tp\xi
M(\Phi)\xi.
\end{eqnarray}
Here in the expression for ${}^-\!L$, $\xi$ should be replaced by the
right chiral field $\xi_c$, $\eta$ by
$\eta_c:=-i\sqrt{q}\xi^\dagger_c(VV^\dagger)^{-1}$, and the covariant
derivative $\bm{D}_j\xi$ by
\begin{equation}
\bm{D}_j\xi_c =(\partial_j-\kappa^2\;{}^-\!\A_j\cdot\bg{\sigma} +
e\bm{A}_j\cdot\bm{T})\xi_c.
\end{equation}

In these equations we have omitted the suffix $\pm$ to distinguish the
left and the right chiral variables.  We will adopt this
simplification throughout the paper from now on unless it is necessary
to distinguish the left and the right chiral variables.  The $\pm$ or
$\mp$ signs in the equations are always to be understood that the
upper sign corresponds to the left chiral variables and the lower sign
to the right chiral ones.

In contrast to the real tetrad approach, $\B_j$ which corresponds to
$Q_j$ is not independent from $\A_j$, and its elimination does not
lead to any new constraint.  Hence by treating the complex variables
$\E^j$ and $\A_j$ as the fundamental canonical variables for the
gravitational field and setting the Poisson brackets among them as
\begin{eqnarray}
&&\{ \E^{Ij}(x), \A_{Jk}(y) \}={1\over2}\delta^I_J \delta^j_k
\delta^3(x-y),
\label{PoissonBracket:Chiral1}\\
&&\{ \E^{Ij}(x), \E^{Jk}(y) \}=0, \qquad
\{ \A_{Ij}(x), \A_{Jk}(y) \}=0, \label{PoissonBracket:Chiral2}
\end{eqnarray}
the time evolution of a functional of the canonical variables is given
by the canonical equation of motion
\begin{equation}
\dot F = \{ F, H \}; \qquad F=F(\E, \A, \Phi, \Pi, \bm{A}, \bm{E}, \xi, \eta)
\end{equation}
with the complex Hamiltonian
\begin{equation}
H := C_{\sbm{A}}(\bm{A}{}_t) +\C_\A(\A_t) + \C_\r{M}(\bm{N}) +
\C_\r{H}(\utilde N).
\end{equation}

This Hamiltonian is a differential polynomial of $\E^j$ and $\A_j$
which is of degree three in $\E^j$ and two in $\A_j$ if the material
gauge fields do not exist.  Including the material gauge fields breaks
this polynomiality.  If one makes it polynomial by rescaling $\utilde
N$ to include $1/q$, the degree of the Hamiltonian in $\E^j$ increases
by three.  Hence as far as the degree of the polynomialized
Hamiltonian is concerned, the gain of introducing the complex chiral
variable is small: it decreases the degree only by three compared to
the metric approach.  However, if we inspect the expression in detail,
we find that its structure is much simplified: $qq^{jk}$ which is a
complicated combination of $q_{jk}$ in the metric approach is replaced
by a simple expression $\E^j\cdot\E^k$, and the cumbersome term
$\omega^I_{Jj}$ in the real triad approach does not appear.

\subsubsection{Gauge invariance}

In the real triad approach we had to fix the local $SO_+(3,1)$ gauge
freedom of the tetrad partially in order to put the theory in the
canonical form. As a result the gauge symmetry of the theory was
reduced from $SO_+(3,1)$ to $SO(3)$.  In the chiral canonical theory
this reduction of symmetry does not occur because the original local
$SO_+(3,1)$ transformation is faithfully represented as the local
$SO(3,\CF)$ transformation on the canonical variables.  In fact it is
easily checked that the chiral canonical Lagrangian
(\ref{ChiralLagrangian:CanonicalForm}) is invariant under the local
$SO(3,\CF)$ transformation by noting that $\D_j$ is the connection
with respect to this gauge symmetry and $\F_{jk}$ is its curvature
form.

As in the metric approach and the real triad approach, the constraint
functionals $C_{\sbm{A}}(\bg{\Lambda})$, $\C_{\A}(\lambda)$ and
$\C_\r{D}(L)$ defined by
\begin{equation}
\C_\r{D}(L):=\C_\r{M}(L) + C_{\sbm{A}}(L^j\bm{A}\!{}_j)+\C_{\A}(L^j\A_j),
\end{equation}
are the generators of the infinitesimal canonical transformations
corresponding to the material gauge, the local $SO(3,\CF)$ and the spatial
coordinate transformations, respectively.  In particular the local
$SO(3,\CF)$ and the spatial coordinate transformations of the complex
canonical variables $\E^j$ and $\A_j$ are given by
\begin{eqnarray}
&& \delta_{\lambda}\E(\phi) = \{ \C_{\A}(\lambda), \E(\phi) \} = \mp
2i\kappa^2 \E(\lambda\times\phi),\\
&& \delta_{\lambda}\A(\alpha) = \{ \C_{\A}(\lambda), \A(\alpha) \}
= - \D\lambda(\alpha),\\
&& \delta_L\E(\phi) = \{ \C_\r{D}(L), \E(\phi)\} = \E(\Lie_L\phi),\\
&& \delta_L\A(\alpha) = \{ \C_\r{D}(L), \A(\alpha)\} = \A(\Lie_L\alpha),
\end{eqnarray}
where
\begin{equation}
\E(\phi):= \int_\Sigma d^3x \E^j\cdot \phi_j, \qquad
\A(\alpha):= \int_\Sigma d^3x \A_j \cdot \alpha^j.
\end{equation}
Due to this group theoretical property these constraint functionals form
an algebra isomorphic to the Lie algebra of the corresponding groups
with respect to the Poisson brackets as in the real canonical theories:
\begin{eqnarray}
&& \{ C_{\sbm{A}}(\bg{\Lambda}\!{}_1),C_{\sbm{A}}(\bg{\Lambda}\!{}_2)\} =
e C_{\sbm{A}}([\bg{\Lambda}\!{}_1,\bg{\Lambda}\!{}_2]);\quad
[\bm{\Lambda}\!{}_1,\bm{\Lambda}\!{}_2]^\alpha=f^\alpha{}_{\beta\gamma}\bm{\Lambda}\!{}_1^\beta
\bm{\Lambda}\!{}_2^\gamma,\\
&&\{\C_{\A}(\lambda_1),\C_{\A}(\lambda_2)\} = \mp 2\kappa^2 i
\C_{\A}(\lambda_1\times\lambda_2),\\
&& \{\C_{\A}(\lambda),C_{\sbm{A}}(\bg{\Lambda})\} = 0,\\
&& \{ \C_\r{D}(L),C_{\sbm{A}}(\bg{\Lambda})\}=C_\r{A}(\Lie_L \bg{\Lambda}),\\
&& \{ \C_\r{D}(L), \C_{\A}(\lambda)\} = \C_{\A}(\Lie_L\lambda),\\
&& \{ \C_\r{D}(L_1), \C_\r{D}(L_2)\} = \C_\r{D}([L_1,L_2]).
\end{eqnarray}
For the same reason the Poisson brackets of these constraints with the
Hamiltonian constraint are given by
\begin{eqnarray}
&& \{ C_{\sbm{A}}(\bg{\Lambda}), \C_\r{H}(\utilde T)\} = \{ C_{\A}(\lambda),
\C_\r{H}(\utilde T)\} =0,\\
&& \{ C_\r{D}(L),\C_\r{H}(\utilde T)\} = \C_\r{H}(\Lie_L \utilde T),
\end{eqnarray}
where $\utilde T$ implies that it behaves as a scalar density of
weight $-1$.  Finally the Poisson bracket of $\C_\r{H}$ is given by
\begin{eqnarray}
&\{\C_\r{H}(\utilde T{}_1),\C_\r{H}(\utilde T{}_2)\} &=
\C_\r{M}(\E\cdot\E(\utilde T{}_1\partial \utilde T{}_2 -\utilde
T{}_2\partial \utilde T{}_1)) \nonumber\\
&&= \int_\Sigma d^3x
\E^j\cdot\E^k(\utilde T{}_1\partial_j \utilde T{}_2 -\utilde
T{}_2\partial_j \utilde T{}_1)\C_{\r{M}j}.
\end{eqnarray}

\subsubsection{Reality condition}

In the complex canonical theory the dynamical degrees of freedom for
the gravitational fields are doubled compared with the real triad
formalism.  Hence in order for the theory to be equivalent to the
original Einstein theory, some additional constraints should be
imposed.  These constraints are obtain from the requirement that the
spatial metric calculated from $\E^j$ is real and its consistency with
the time evolution equations.

First from the relation $qq^{jk}=\E^j\cdot\E^k$ the reality condition
for the spatial metric is written as
\begin{equation}
\overline{(\E^j\cdot\E^k)} = (\E^j\cdot\E^k).
\label{ClassicalRealityCondition1}\end{equation}
Since the time evolution of $\E^j\cdot\E^k$ is given by
\begin{eqnarray}
&(\E^j\cdot\E^k)\dot{} &= -2\partial_l
N^{(j}(\E^{k)}\cdot\E^l)+2\partial_lN^l(\E^j\cdot\E^k) +
N^l\partial_l(\E^j\cdot\E^k) \nonumber\\ && \quad - \kappa^2(N^j\E^k +
N^k\E^j)\cdot\C_{\A} + 2i\utilde N \psi^{jk},
\end{eqnarray}
where
\begin{equation}
\psi^{jk} := (\E^{(j}\times\D_l\E^{k)})\cdot\E^l \mp
{i\over2}\kappa^2(\E^j\cdot\E^k)\eta\xi,
\label{PhiConstraint:def}\end{equation}
the consistency of the condition (\ref{ClassicalRealityCondition1})
with the time evolution equations yields a further constraint
\begin{equation}
\overline{\psi^{jk}}=-\psi^{jk}.
\label{ClassicalRealityCondition2}\end{equation}
It can be shown that the time derivative of this constraint yields no
new constraint.

Though the second condition contains the spinor fields, their
contribution is trivial.  In order to see this, recall the relation
(\ref{ConnectionForm:SpinorContribution}).  This relation implies that
the connection form consists of a geometrical part and a spinor
current part.  As was shown in \S2.2, the terms arising from the
latter spinor part cancels in the total action.  Thus it is expected
that if we decompose the connection form as
\begin{equation}
A_{ab} = A_{ab}' \mp {\kappa^2\over4}\epsilon_{abcd}S^c\theta^d,
\end{equation}
the spinor contribution in Eq.(\ref{PhiConstraint:def}) will be
separated.  This expectation is true.  Actually by noting that this
decomposition is written in terms of the chiral connection as
\begin{eqnarray}
&\B_j & = \B'_j \pm {\kappa^2\over4} O^{-1}(\hat S_0\hat\theta_j \pm i
\hat\bm{S}\times\hat\theta_j)\nonumber\\ && = \B'_j \mp
i{\kappa^2\over4q}\epsilon_{jkl}[(\eta\xi)\E^k\times\E^l \pm
i(\eta\bg{\sigma}\xi)\times(\E^k\times\E^l)],
\end{eqnarray}
we can show that $\psi^{jk}$ is written as
\begin{equation}
\psi^{jk} = (\E^{(j}\times\D'_l\E^{k)})\cdot\E^l,
\end{equation}
where $\D'_j$ is the chiral covariant derivative with $\B_j$ replaced
by $\B'_j$.  Further in the same way we can show that the constraint
function $\C_{\A}$ is written as
\begin{equation}
\D'_j\E^j = 0.
\label{D'E}\end{equation}

With this separation of the spinorial contribution we can find the
geometrical meaning of the reality condition.  First note that the
condition (\ref{ClassicalRealityCondition1}) implies that with some
appropriate $SO(3,\CF)$ rotation $\E^j$ can be made real, which is
equivalent to take the spatial gauge for the tetrad $\E^j = \tilde
e^j$.  In this gauge if we decompose $\A'_\mu$ and $\B'_\mu$ as
\begin{equation}
\A'_\mu = P_\mu \pm {i\over2\kappa^2}Q_\mu, \quad
\B'_\mu = Q_\mu \mp 2\kappa^2i P_\mu,
\end{equation}
the real part of $\psi^{jk}$ is proportional to $\phi^{jk}$ defined in
Eq.(\ref{phi:def}):
\begin{equation}
\r{Re}\psi^{jk}=q^{3/2}\phi^{jk}.
\end{equation}
Further from Eq.(\ref{D'E}) the constraint function $\C_{\A}$ is
decomposed as
\begin{equation}
\C_{\A} = C_\r{B} \mp 2\kappa^2i C_\r{R}.
\label{CAbyCBandCR}\end{equation}
Hence from the argument in \S2.2 it follows that the constraint
(\ref{ClassicalRealityCondition2}) supplemented with the constraint
$\C_{\A}=0$ is equivalent in the spatial gauge to the condition that
$Q_j$ is given by the Riemannian three-dimensional connection
determined from $e^j$.

Thus the reality condition reduces the dynamical degree of freedom of
the gravitational fields to the correct one at the level of the
equation of motion.  However, if one tries to impose the reality
condition within the framework of the canonical theory, one meets a
difficulty. To see this, neglecting the spinor contribution for the
moment for simplicity, let us decompose the chiral variables into
their real parts and the imaginary parts as
\begin{equation}
\E^j = \E^j_1 \mp i\E^j_2, \quad
\A_j = P_j   \pm {i\over2\kappa^2}Q_j.
\end{equation}
Putting this decomposition into the canonical Lagrangian, its kinetic
part for the gravitational field is written as
\begin{equation}
-2\E^j\cdot\dot\A_j = -2(\E^j_1\cdot\dot P^j +
{1\over2\kappa^2}\E^j_2\cdot\dot Q_j) \pm 2i(\E^j_2\cdot\dot P_j -
{1\over2\kappa^2}\E^j_1\cdot\dot Q_j).
\end{equation}
Hence if the real and the imaginary parts of the variables were
regarded as independent dynamical freedoms, one would get two sets of
canonical systems with different canonical structures after
decomposing the Lagrangian into its real part and imaginary part. This
implies that one cannot treat the real and imaginary parts of the
complex variables as independent canonical degrees of freedom in the
complex canonical theory, and that the reality condition is not a
constraint in the dynamical complex phase space but exact relations
among the complex variables to be satisfied on the image of embedding
of a dynamical real phase space into the formal complex phase space.

For example, in the spatial gauge $\E^j=\tilde e^j$, from the argument
above, the expression
\begin{equation}
\A_j = P_j \pm {i\over2\kappa^2}Q_j(e)
\label{ClassicalRealityCondition:SpatialGauge}\end{equation}
with $Q_j(e)$ given by Eq.(\ref{QbyOmega}) and the real $P_j$ yields a
natural embedding, consistent with the reality condition, of the
canonical variables in the real triad approach into the complex phase
space. The real canonical theory induced by this embedding from the
complex canonical theory is equivalent to the one given in \S2.2.
Actually Ashtekar constructed the complex canonical theory by this
embedding when he first proposed the
theory\CITE{Ashtekar86a,Ashtekar87}.  Here we give a rough sketch of
the proof for the pure gravity case since some equations appearing in
the course of the proof will be utilized later.

First note that from the relation
\begin{equation}
\F^I_{jk}=F^I_{jk} +4\kappa^4(P_j\times P_k)^I \mp 2\kappa^2i(D_jP_k-D_kP_j)^I,
\end{equation}
the constraint functions $\C_\r{M}$ and $\C_\r{H}$ are
 decomposed into real parts and imaginary parts as
\begin{eqnarray}
&&\C_{\r{M}j}=C_{\r{M}j} \pm 2\kappa^2iP_j\cdot C_\r{R}
\mp{i\over\kappa^2}\tilde e^k\cdot F_{jk},\\
&&\C_\r{H} = C_\r{H}
\pm i\partial_j(\tilde e^j\cdot C_\r{R}) \mp 2i D_j(\tilde
e^j\times \tilde e^k)\cdot P_k.
\end{eqnarray}
For $Q_j$ given by Eq.(\ref{QbyOmega}), the last term on the
right-hand side of the first equation $\tilde e^j\cdot F_{jk}$
vanishes due to the first Bianchi identity and the last term on the
right-hand side of the second equation is written from
Eq.(\ref{DeByGamma}) as
\begin{equation}
D_j(\tilde e^j\times \tilde e^k)\cdot P_k = {1\over2}(\partial_j\sqrt{q})
e^j\cdot
C_\r{R}.
\end{equation}
Hence together with Eq.(\ref{CAbyCBandCR}) the constraint functions
are written as linear combinations of those in the real tetrad
approach:%
\begin{eqnarray}
&&\C_{\A} = \mp 2\kappa^2i C_\r{R},\\
&&\C_{\r{M}j}=C_{\r{M}j} \pm 2\kappa^2i P_j\cdot C_\r{R},\\
&&\C_\r{H} = C_\r{H} \pm i\partial_j(\tilde
e^j\cdot C_\r{R}) \mp i(\partial_j\sqrt{q})e^j\cdot
C_\r{R}.
\end{eqnarray}

Putting these expressions into the chiral canonical Lagrangian
(\ref{ChiralLagrangian:CanonicalForm}), and setting the value of the
arbitrary function $P_t$(the real part of $\A_t$) to be the one given
in Eq.(\ref{Pt:SpatialGauge}), the chiral canonical Lagrangian reduces
to
\begin{equation}
{}^\pm\!L = L \pm \int d^3x {i\over\kappa^2}\dot {\tilde e}{}^j\cdot
Q_j.
\end{equation}
Due to the identity equation
\begin{equation}
2\dot {\tilde e}{}^j\cdot Q_j(e) =
(\sqrt{q}\epsilon_{IJK}\theta_{Ij}\partial_k e^{Jj}e^{Kk})\dot{} +
\partial_k(\sqrt{q}\epsilon_{IJK}\dot\theta_{Ij}e^{Jj}e^{Kk}),
\label{IdentityForQ}\end{equation}
the imaginary part of the chiral Lagrangian reduces to a total time
derivative.  This proves the equivalence.

Finally note that there exist an infinite number of embeddings
consistent with the reality condition.  They are all obtained from the
above embedding by some appropriate local $SO(3,\CF)$ gauge
transformations and are mutually equivalent.  What is important here
is that the dynamics is always described by the same complex theory
irrespective of the embedding.  Actually the detail of the embedding
becomes relevant only when the correspondence between the theory and
observations is concerned.  In particular, when we are concerned only
with the quantities invariant under the local $SO(3,\CF)$ gauge
transformations, we do not even have to fix the embedding.  This point
becomes important in quantizing the theory.

\section{Quantum Theory of Totally Constrained Systems}

As we have seen in the previous section, the Hamiltonian in the canonical
theory of general relativity is written as a linear combination of
the first-class constraint functions, irrespective of the approaches
to construct the canonical theory.  We call such a canonical system
{\it a totally constrained system}. Classically this special structure
of the theory does not give rise to any serious problem, and can be
treated like ordinary canonical systems with constraints.  In
contrast, however, when one tries to quantize such a system, some
serious problems arise.

The first problem is the formulation of dynamics, i.e., the time
evolution in the quantum theory.  For example, let us consider a
totally constrained system with a Hamiltonian
\begin{equation}
H = \sum_\alpha \lambda^\alpha C_\alpha,
\end{equation}
where $C_\alpha$ are constraint functions and $\lambda^\alpha$ are
arbitrary function of time.  If one quantizes this system by the Dirac
method, classical constraints $C_\alpha=0$ are translated to a set of
constraints on physical states:
\begin{equation}
\hat C_\alpha |\Psi> =0,
\end{equation}
where $\hat C_\alpha$ are the operators corresponding to $C_\alpha$.
However, since the Hamiltonian operator $\hat H$ is written as a
linear combination of $\hat C_\alpha$, these quantum constraints yield
\begin{equation}
\hat H |\Psi> =0,
\end{equation}
which implies that the Schr\"odinger equation becomes trivial for the
physical states.  Hence one loses dynamics in this formal quantization
method.  This difficulty arises due to the time-reparametrization
invariance of the theory which implies that the formal time variable
of the system has no observational meaning\CITE{Kuchar81}.  Therefore
it is closely connected with the following set of problems: Is it
possible to find natural time variables expressed in terms of the
canonical variables?  Are the time variables to be included among
observables? Are the time variables represented as operators? and so
on.  These problems constitute the issue of time in quantum gravity.

The second problem is concerned with observables.  In ordinary gauge
field theories observables are restricted to the gauge-invariant
quantities.  The same restriction formally applies to totally
constrained systems since it can be generally shown that each of the
first-class constraints corresponds to a generator of some generalized
gauge transformation\CITE{HalliwellHartle91}.  However, in totally
constrained systems, some subtle problems arise.

First, as was stated above, totally constrained systems have the
time-reparametrization invariance, which is a generalized gauge
invariance generated by the constraint function $C^t$ whose Poisson
bracket with the time variable $t$ does not vanish.  Since
time-reparametrization invariant quantities are essentially constant
of motion as we will see later, one would lose the chance to extract
dynamics in the Dirac quantization method if one restricts observables
to those invariant under the time-reparametrization.

Second as we have seen in \S2, a part of the constraint functions are
the generators of the spatial coordinate transformation or the spatial
diffeomorphism. Though this is a kind of gauge transformation in a
generalized sense, invariance under this transformation has a meaning
which is quite different from the usual gauge invariance.  In particular the
observational meaning of quantities invariant under the spatial
coordinate transformation is not a priori clear since all the observational
information of physical quantities are expressed in a coordinate
dependent fashion.

These problems have a long history and a lot of work has been done
on them ( see \cite{Kuchar81} on the early history and the difficulty
in finding the internal time variables, and \cite{AshtekarStachel91} on
the recent arguments).  Nevertheless, no resolution which satisfies most
of people has not been found yet.  If technical difficulties are neglected,
the main controversial point is whether operators should be assigned to the
time variables or not.  The formulation of canonical quantum gravity, in
particular the treatment of observables and states, changes significantly
depending on which standpoint is taken.

Though this is a review article, we adopt the standpoint that the time
variables should be included into the observables represented by
operators since the other standpoint seems to be unsatisfactory to the
author.  In the subsequent part of this section we explain the reason
for that by analyzing the above problems in detail both from the
classical and the quantum points of view.  Then from this standpoint
we summarize the formal structure of the canonical quantization
program of gravity.  Further, after critical comments on the approaches
proposed so far we propose one possible new approach on the treatment
of the quantum hamiltonian constraint.

Finally we should remark the reader that, although there exist some
researchers who take standpoints similar to ours( for example, see
Kucha\v{r}'s comment in the discussion part of Rovelli's article in
\cite{AshtekarStachel91}), our argument may not be convincing enough for
people taking different standpoints( including C. Rovelli).

\subsection{Gauge-invariant Quantities and Dynamics}

Since most of the difficulties associated with quantization of a
totally constrained system are caused by the gauge invariance of the
system, we expect that a part of them are eliminated if we can
separate the gauge invariant freedoms and extract the true physical
freedoms at the classical level.  However, in general, this
separation procedure provokes various problems which do not arise in
the ordinary gauge field theories.  In this subsection we look at
these problems and discuss how they should be handled in the classical
framework.

\subsubsection{Gauge-fixing method vs. gauge invariant formalism}

There are in general two approaches to eliminate the gauge freedom.
The first is the gauge-fixing method.  In this method one imposes a set of
additional constraints, the gauge fixing condition, by fixing the
values of some canonical quantities as $\psi^\alpha = f^\alpha(t)$.
If $\psi^\alpha$ is selected so that $\{C_\alpha, \psi^\beta\}$ is a
regular matrix, the consistency of the gauge condition with the
canonical time evolution equation,
\begin{equation}
\dot f^\alpha = \lambda^\beta \{C_\alpha, \psi^\beta\},
\end{equation}
completely determines the arbitrary functions $\lambda^\alpha$.  The
only feature specific to the totally constrained system is that
$f^\alpha$ should be explicitly time-dependent since all the
$\lambda^\alpha$ should not vanish simultaneously.

The gauge fixing constraints with the original constraints are of the
second class.  Thus in general they should be solved explicitly before
quantization.  In realistic theories such as general relativity this
is practically impossible.  Of course if one modifies the canonical
structure by the Dirac method or reformulates the theory by
introducing ghost fields so that it is BRS-invariant, one does not
need to solve the constraints.  However, even in these methods, one
needs the inverse of the matrix $\{C_\alpha, \psi^\beta\}$.  Since
this inversion is to solve non-linear partial differential equations,
it is also practically quite difficult.

Besides this technical difficulty there exits a more fundamental
problem specific to general relativity.  In general relativity
practical coordinate-fixing conditions are always local and can not be
applied to the whole spacetime.  Furthermore whether a given
coordinate condition is appropriate or not depends on a solution of
the Einstein equation.  This feature makes it difficult to formulate
the quantum theory.

 The second approach is to find all the gauge-invariant quantities in
the classical canonical theory and construct quantum theory based only
on them.  Since the constraint functions are the generators of the
gauge transformations, the condition for a function $F$ on the
canonical phase space $\Gamma$ to be gauge invariant is expressed as
\begin{equation}
\{C_\alpha, F\} = 0  \qquad \forall \alpha.
\end{equation}
Here for the generators which exactly form a Lie algebra the equations
should hold strongly, but those for the others such as the Hamiltonian
constraint are required to hold only modulo the constraints. Since the
Hamiltonian is written as a linear combination of the constraint
functions, this condition implies that $F$ is a constant of motion.
Conversely any canonical quantity which is conserved for any choice of
the arbitrary functions $\lambda^\alpha$ is gauge invariant.

In general finding all the constants of motion is equivalent to
solving the canonical equation of motion.  This also holds for the
present case.  To see this, let $\Gamma_*$ be the subspace of the
original phase space $\Gamma$ determined by the constraints, $\O$ be
the set of all the function on $\Gamma$, $\O_\r{cons}$ be its subset
consisting of the constants of motion, and $X_j$ be a maximal subset
of functions in $\O_\r{cons}$ which are independent on $\Gamma_*$.
Since $\Gamma_*$ is invariant under the canonical transformations
generated by the constraint functions, $\Gamma_*$ is decomposed to a
set of orbits each of which has the same dimension as the number of
independent constraint functions, and one orbit is picked up if the
values of all $X_j$ are specified. Hence, if we select a set of
functions $\psi^\alpha$ in $\O$ such that
$\Delta^\alpha_\beta:=\{\psi^\alpha, C_\beta\}$ is a regular matrix,
$X_j$ and $\psi^\alpha$ defines a local coordinate system of
$\Gamma_*$. Now let $S$ be a gauge orbit in $\Gamma_*$ and $\gamma$ be
an arbitrary curve parametrized by $t$ in $S$.  Then, since $X_j$ is
constant along $\gamma$, we can find functions $g^\alpha(\psi)$ such
that along $\gamma$ $\psi$ satisfies the differential equation
\begin{equation}
{d\psi^\alpha \over dt} = g^\alpha(\psi).
\end{equation}
Hence, if $\lambda^a$ is a function on $\Gamma$ determined by
\begin{equation}
g^\alpha(\psi) = \Delta^\alpha_\beta(\psi,X)\lambda^\beta(\psi,X),
\end{equation}
any function $F$ in $\O$ satisfies the equation
\begin{equation}
{dF \over dt} = {\partial F\over \partial
\psi^\alpha}g^\alpha(\psi)=\{F,\lambda^\alpha C_\alpha\}
\end{equation}
along $\gamma$.  This implies that the curve $\gamma$ is a solution of
the canonical equations of motion.

\subsubsection{Physical meaning of the general covariance}

The argument above shows that the gauge-invariant quantities carries
all the dynamical information in the classical totally constrained
systems.  In particular the dynamics is determined only by the
canonical structure and the constraint functions.  If we assign
operators only to the gauge-invariant quantities in quantization,
however, we meet a difficulty of losing dynamics as stated at the
beginning of this section.  In order to find out what is wrong with
this approach, let us analyze the meaning of the three types of gauge
invariances which appear in general relativity: the usual internal
gauge invariance associated with the gauge fields and the tetrad
field, the spatial coordinate transformation invariance, and the time
coordinate transformation invariance.

In the usual gauge field theories gauge-dependent quantities are
introduced to express the relations among directly measurable
quantities in terms of equations local in the space-time coordinates.
Thus they are genuinely theoretical entities and the restriction of
observables to gauge-invariant quantities or relations is the
fundamental postulate of the formulation. The spontaneous symmetry
breaking does not change the situation: it is essentially breaking of
a global symmetry at the phenomenological level and can be described
only in terms of the gauge-invariant quantities.  The freedom of the
tetrad rotation in general relativity is of the same nature as the
internal gauge transformation.

In contrast to the gauge field theories some of the quantities
represented by space-time-coordinate dependent fields are measurable
in general relativity.  For example, let us consider the phase $\phi$
of electromagnetic fields which is a scalar quantity in general
relativity, and an apparatus which measures the value of $\phi$ at the
space-time location of the apparatus.  Clearly the apparatus yields a
definite value of $\phi$ by measurement.  Of course the collection of
the measurements at various spacetime points is just a set of real or
complex numbers and is not described by a field.

 However, if one performs similar measurements of other independent
scalar quantities $\phi^1, \ldots, \phi^n$($n\ge4$), each measurement
yields a definite set of values $(\phi, \phi^1, \ldots, \phi^n)$, and
the collection of the results at various space-time points will form a
definite four-dimensional submanifold of the $(n+1)$-dimensional
linear space. Let us select four quantities $x^0,\ldots,x^3$ from
$\phi^1,\ldots,\phi^n$.  If the projection of an open subset of this
submanifold to a four-dimensional space $(x^0,\ldots,x^3)$ is
one-to-one, then the observational data corresponding to this subset
is represented by a function $\phi(x^0,\ldots,x^3)$. If another choice
of four scalar fields $y^0,\ldots,y^3$ has the same property, one will
obtain another expression for the data, $\phi(y^0,\ldots,y^3)$.
Clearly these functions behave as a scalar field if $x^0,\ldots,x^3$
and $y^0,\ldots,y^3$ are regarded as local coordinate systems of
space-time.

This observation shows that a scalar field $\phi(x^0,\ldots,x^3)$ in
general relativity is a symbol to represent the set of relations of a
measurable quantity $\phi$ to four other independent scalar quantities
describing the whole observational data of a spacetime phenomenon, and
that the space-time coordinate transformations are just the change of
the choice of scalar quantities to which $\phi$ is related.  Thus if
one denotes the set of all possible configurations of the scalar field
by $U$, the coordinate transformations induce transformations of $U$
and to each orbit of the transformation group corresponds one
space-time phenomenon as a whole.  Since the gauge-invariant functions
on $U$ take constant values on each orbit, they yield a set of label
to classify the orbits or the space-time phenomena.

There appear of course much more complicated quantities such as
tensors or tensor densities in general relativity.  However, since
they all are concepts theoretically derived from scalar fields or
introduced to describe the structure of the manifold formed by the set
of values of scalar quantities, the above argument can be essentially
extended to the whole theory.  Thus as in the gauge field theories it
is true that only the values of the gauge-invariant quantities or
functionals are determined by observations, but their meaning is
different: the gauge-invariant quantities are not directly measurable
ones in general but are abstract objects to describe the relations of
the measurable quantities in each space-time phenomenon.

\subsubsection{Dynamics in the invariant formalism}

With the arguments so far in mind, let us analyze the problem of time
evolution.  This problem is intimately connected with the fundamental
nature of the canonical approach as a framework to describe the
physical law.  In the canonical theories one decomposes the set of
measured relations among physical quantities describing each spacetime
phenomenon as a whole to a one-dimensional sequence of subsets, and
formulate the physical law as the relations among the subsets
independent of the phenomena.  Usually this decomposition is done in
terms of a special measurable quantity $t$ called the time variable,
and by slicing the four-dimensional data set to subsets on each of
which the value of $t$ is constant.  Hence the quantities which are
invariant under the space-time coordinate transformations and describe
a space-time phenomenon as a whole are not sufficient to describe the
law. One needs quantities to describe the relations in each data set
with constant time.

Taking account of the arguments above on what are determined by
observations, the most natural quantities to describe the information
at each time slice are the functions on the phase space $\Gamma$ which
are invariant under the internal gauge transformations and the spatial
coordinate transformations.  In fact we can write the canonical
gravity theory only in terms of such invariant functions.  Let
$\O_\r{inv}$ be the set of all such functions on $\Gamma$.
$F\in\O_\r{inv}$ is characterized by the conditions
\begin{eqnarray}
&&\{C_\r{G}(\Lambda), F\}=0 \quad \forall
\Lambda,\label{ClassicalGIQ:def}\\ &&\{C_\r{D}(L), F\}=0 \quad \forall
L^j, \label{ClassicalDIQ:def}
\end{eqnarray}
where $C_\r{G}$ represents the constraint functions corresponding to
the material gauge transformations and the tetrad rotations.
$\O_\r{inv}$ can be regarded as functions on the orbit space
$\Gamma_\r{inv}$ of the gauge transformation group acting on $\Gamma$.
Each point of $\Gamma_\r{inv}$ represent a state of the system on a
time slice. Due to the first-class nature of $C_\r{G}$ and $C_\r{D}$,
$\{F_1, F_2\}$ is again invariant if $F_1, F_2 \in \O_\r{inv}$.  Hence
the Poisson brackets in $\O_\r{inv}$ is naturally determined from that
in $\O$.

The constraint functions $C_\r{G}(\Lambda)$ and $C_\r{D}(L)$ do not
belong to $\O_\r{inv}$.  However, if we replace the smoothing
functions $\Lambda$ and $L$ by functionals $\hat\Lambda$ and $\hat L$
which are expressed only in terms of the canonical quantities and have
the same transformation behavior as $\Lambda$ and $L$ respectively,
they can be made  invariant:
\begin{eqnarray}
&&\{ C_\r{G}(\Lambda'), C_\r{G}(\hat\Lambda)\}=
\{ C_\r{D}(L'), C_\r{G}(\hat\Lambda)\}=0,\\
&&\{ C_\r{G}(\Lambda'), C_\r{D}(\hat L)\}=
\{ C_\r{D}(L'), C_\r{D}(\hat L)\}=0.
\end{eqnarray}
Hence the gauge and diffeomorphism constraints are translated to the
constraints in $\Gamma_\r{inv}$, which determines a subspace
$\Gamma^*_\r{inv}$.  Each point of $\Gamma^*_\r{inv}$ determines a
physical state( for the parametrization of $\Gamma^*_\r{inv}$ see
\cite{NewmanRovelli92}).  Points in $\Gamma^*_\r{inv}$ are
distinguished by the functions in $\O_\r{inv}$ that does not vanish
identically on $\Gamma^*_\r{inv}$,
\begin{equation}
\O^*_\r{inv}:=\O_\r{inv}/\O_\r{inv}[C_\r{G}, C_\r{D}].
\end{equation}

Like $C_\r{G}$ and $C_\r{D}$ the Hamiltonian constraint function
$C_\r{H}(T)$ can be made  invariant by replacing the smoothing
function $T$ by an  invariant functional $\hat T$:
\begin{equation}
\{ C_\r{G}(\Lambda), C_\r{H}(\hat T)\}=
\{ C_\r{D}(L), C_\r{H}(\hat T)\}=0.
\end{equation}
Hence the canonical equation of motion in $\Gamma$ can be rewritten as
the canonical equation for $F$ in $\O_\r{inv}=\O(\Gamma_\r{inv})$,
\begin{equation}
{d F\over dt} = \{F, C_\r{H}(\hat N)
\},\label{ClassicalEOM:InvariantPhaseSpace}
\end{equation}
with the constraints
\begin{equation}
\C_\r{G}(\hat\Lambda)=0, \quad C_\r{D}(\hat  L)=0, \quad C_\r{H}(\hat T)=0.
\end{equation}
If the states are restricted to $\Gamma^*_\r{inv}$, the first two of
these constraints become redundant.

So far we have implicitly assumed that some time-slicing is given, but
we did not have to specify the slicing.  As a result of this,
$\Gamma_\r{inv}$ contains all the states obtained by all possible time
slicings of the whole space-time phenomena.  Actually, since the
condition to specify a time slice, $t=$const, is by itself a gauge and
spatial diffeomorphism invariant relation of the measured quantities,
the time-slicing condition can be expressed in terms of functions in
$\O_\r{inv}$.  To be precise, corresponding to the freedom of the
lapse function, we need formally $\infty^3$ numbers of the
invariant functions to completely specify a time slicing within
the framework of the canonical gravity.  This point is reflected in
Eq.(\ref{ClassicalEOM:InvariantPhaseSpace}) as the freedom of the
invariant canonical function $\hat N$. Solutions to this equation
obtained by changing the choice of $\hat N$ with a fixed initial
condition correspond to different time slicing of a single space-time
solution.

Of course $\O_\r{cons}$ is a subset of $\O_\r{inv}$ and the space-time
solutions are completely classified by the functions in $\O_\r{cons}$.
But as we have stated, we cannot restrict the variables to
$\O_\r{cons}$ to treat dynamics in the canonical approach.

\subsection{Quantization}

On the basis of the analysis on classical totally constrained systems
in the previous subsection, we describe the framework of the quantum
gravity program and discuss the difficulties associated with it.

\subsubsection{Program}

As explained in the previous subsection, a state on each time slice is
specified by the values of the gauge and the spatial-diffeomorphism
invariant functionals $\O_\r{inv}$ which is closed with respect to the
Poisson brackets in the phase space $\Gamma$, and the constraints are
represented by a special set of functionals in $\O_\r{inv}$,
$C_\r{G}(\hat \Lambda)$, $C_\r{D}(\hat L)$, and $C_\r{H}(\hat T)$.
Thus ideally the process to construct the canonical quantum gravity
theory is summarized as follows:
\begin{itemize}
\item[1)] Find all the invariant functionals  $\O_\r{inv}$ among the
gauge-dependent functionals on the phase space $\Gamma$ consisting of the triad
fields and their conjugate momentums, or its chiral version $\Gamma_{\CF}$.
\item[2)] Select a set of fundamental invariant variables from which
$\O_\r{inv}$ is generated.
\item[3)] Construct an operator algebra $\hat\O_\r{inv}$ by defining the
commutation relations among the operators corresponding to the classical
fundamental variables based on their Poisson bracket structure.
\item[4)] Assign a set of operators $\hat C^\alpha_\r{G}$, $\hat C^\beta_\r{D}$
and $\hat C^\gamma_\r{H}$ in $\hat\O_\r{inv}$ to the classical constraint
functionals $C_\r{G}(\hat \Lambda)$, $C_\r{D}(\hat L)$ and $C_\r{H}(\hat T)$.
\item[5)] Construct a representation of $\hat\O_\r{inv}$ on a linear space
$\V_{inv}$.
\item[6)] Define the involutive anti-linear operation $\hat F \rightarrow \hat
F^\star$ in $\hat\O_\r{inv}$ which corresponds to the complex conjugation in
$\O_\r{inv}$.
\item[7)] Construct a Hilbert space $\H$ from a subset of $\V_\r{inv}$ by
defining an inner product such that $\hat F^\star$ becomes the hermitian
conjugation of $\hat F$.
\item[8)] Define the dynamics by the constraint operators and find the
interpretation.
\end{itemize}

There are however lots of problems in this simple-minded program.
First no one has succeeded even in clearing the step 1) so far. Thus
one would not be able to discuss any aspect of quantum gravity for a
rather long while in the future if one exactly follows this program.
Fortunately, however, there are several ways to study some partial
aspects of the quantum gravity theory in advance of its full
construction.

One is to start from a subset of functions in $\O$ which are invariant
under a subgroup of the full gauge and diffeomorphism group.  Since
the steps 2) to 5) are easily passed if the step 1) are cleared, one
can construct a formal quantum theory by replacing $\O_\r{inv}$ by the
subset. Then it becomes possible to seek for the path to reach the
final goal staying within the quantum framework, or to investigate a
part of the problems which can be discussed without the full knowledge
of the final theory.

Thus various approaches are possible depending on at which stage of
the classical theory one proceeds to the quantum framework, as is
shown in Table\ref{QuantizationProgram}.  There $\O_\r{G}$ in the
classical framework denotes the set of functionals which are gauge
invariant, and $\hat\O_\r{G}$ in the quantum framework denotes the
subalgebra of $\hat\O$ formed by the operators which commute with
$\hat C_\r{G}$:
\begin{equation}
\hat\O_\r{G}:=\{ \hat F\in\hat\O \mid [\hat C_\r{G}(\Lambda),\hat F]=0 \quad
\forall \Lambda\}.
\end{equation}
Similarly $\hat\O_\r{inv}$ is defined as
\begin{equation}
\hat\O_\r{inv}:=\{ \hat F\in \hat\O_\r{G} \mid [\hat C_\r{D}(L),\hat F]=0 \quad
\forall  L \}.
\end{equation}
$*$ on the functional algebra means that it is restricted to the
subspace of $\Gamma$ determined by the constraints relevant at each
stage.  This restriction corresponds in the quantum framework to
restricting the state vectors to the subspaces of $\V$, $\V_\r{G}$ or
$\V_\r{inv}$ defined by
\begin{eqnarray}
&& \V_\r{G} := \{ |\Phi> \in \V \mid \hat C_\r{G}(\hat
\Lambda)|\Phi>=0 \quad \forall \hat\Lambda\},\\
&& \V_\r{inv} := \{ |\Phi> \in \V_\r{G} \mid \hat C_\r{D}(\hat L)|\Phi>=0
\quad \forall\hat L \}.
\end{eqnarray}

\begin{table}
\begin{tabular}{lllclll}
& \multicolumn{2}{c}{Classical Stages} & & &
\multicolumn{2}{c}{Quantum Stages} \\ &&&&&& \\ (I$_C$) & $\O,
\Gamma$;&$C_\r{G},C_\r{D},C_\r{H}$ & $\Longrightarrow$ &(I$_Q$) &
$\hat\O, \V$;&$\hat C_\r{G},\hat C_\r{D},\hat C_\r{H}$ \\ &
$\downarrow$ &&&& $\downarrow$ &\\ (II$_C$) & $\O_\r{G},
\Gamma_\r{G}$;&$C_\r{G}(\hat\Lambda),C_\r{D},C_\r{H}$ &
$\Longrightarrow$ &(II$_Q$) & $\hat\O_\r{G}, \V$;&$\hat
C_\r{G}(\hat\Lambda),\hat C_\r{D},\hat C_\r{H}$ \\ & $\downarrow$ &&&&
$\downarrow$ &\\ (II$^*_C$) & $\O^*_\r{G},
\Gamma^*_\r{G}$;&$C_\r{D},C_\r{H}$ & $\Longrightarrow$ &(II$^*_Q$) &
$\hat\O_\r{G}, \V_\r{G}$;&$\hat C_\r{D},\hat C_\r{H}$ \\ &
$\downarrow$ &&&& $\downarrow$ &\\ (III$_C$) & $\O_\r{inv},
\Gamma_\r{inv}$;&$C_\r{D}(\hat L),C_\r{H}(\hat T)$ & $\Longrightarrow$
&(III$_Q$) & $\hat\O_\r{inv}, \V_\r{G}$;&$\hat C_\r{D}(\hat L),\hat
C_\r{H}(\hat T)$ \\ & $\downarrow$ &&&& $\downarrow$ &\\ (III$^*_C$) &
$\O^*_\r{inv}, \Gamma^*_\r{inv}$;&$C_\r{H}(\hat T)$ &
$\Longrightarrow$ &(III$^*_Q$) & $\hat\O_\r{inv}, \V_\r{inv}$;&$\hat
C_\r{H}(\hat T)$ \\ & $\downarrow$ &&&& $\downarrow$ &\\ (IV$^*_C$) &
$\O^*_\r{cons}, \Gamma^*_\r{inv}$ && &(IV$^*_Q$) & $\hat\O_\r{inv},
(\H^*,\H)$;&$\hat C_\r{H}(\hat T)$ \\
\end{tabular}
\caption{\label{QuantizationProgram}Flow of Quantization Program}
\end{table}

For example, in the classical canonical theory obtained in the metric
approach in \S2.1, the fundamental variables $q_{jk}$ and $p^{jk}$ for
the pure gravity case do not depend on the tetrad, hence are already
gauge-invariant.  Therefore one can move to the quantum framework at
the stage II$_C$ to get the Wheeler-DeWitt theory\CITE{DeWitt67a}.
Actually a large fraction of the work done on the canonical quantum
gravity so far has followed this approach.  At present, however, there
exists no practical method to proceed to the next stage III$_Q$ in
this approach.

Another example is the loop space quantization of the chiral canonical
theory, which starts from the stage I$_C$ , proceeds to the stage
II$^*_C$ by constructing the gauge-invariant functionals in terms of
the Wilson loop integrals, and then moves to II$^*_Q$.  The future
prospect of this approach is brighter than the Wheeler-DeWitt approach
as we will see in detail in \S4.

The second way to bypass the step 1) is to restrict the considerations
to minisuperspace models.  Since they are finite-dimensional, it is
easy to go to the stage III$_Q$ and to study how to get to the final
step corresponding to 8) in the above program.  Lots of interesting
results are obtained in this approach centered around the work by
Hawking, Hartle, and
Vilenkin\CITE{HartleHawking83,Hawking84,Vilenkin82,Vilenkin83,Vilenkin84}.
Unfortunately, however, study in this approach is largely hampered at
present because it cannot discuss any realistic problem for its
oversimplicity.

The third way is to jump from the stage I$_C$ to the final goal
corresponding to the step 8) by constructing the quantum theory in
terms of the path integral\CITE{Hawking79,HalliwellHartle91}. A lot of
work in this approach has been done stimulated by the no-boundary
proposal for the wavefunction of the Universe based on the Euclidean
path-integral by Hartle and Hawking\CITE{HartleHawking83}.  The
present status of this approach is not good either.  The most serious
obstacle is the bad behavior of the Euclidean path integral specific
to the general relativity theory.  Detailed investigations of its
behavior for mini-superspace
models\CITE{HalliwellLouko89a,HalliwellLouko89b,HalliwellHartle90,GarayHalliwellMarugan91}
have revealed that the integration should be done along complex paths
which are along the Lorentzian path to make the path integral
well-defined. Further it is also shown that there remain large
ambiguities in the choice of the integration contour even if one
requires that the path integral converges.  Thus it is difficult to
give a sound basis to the Hartle-Hawking proposal.

The fourth way is to study the quantization of the lower dimensional
versions of the general relativity as a preliminary step. For example,
in the $(2+1)$-dimensional space-times, the pure gravity has only a
finite degrees of freedom which are all topological because the
space-time solutions are all isometric to Minkowski space-time
locally. Hence the exact quantum theory can be formulated if one can
find an appropriate parametrization of the topological
freedom\CITE{Witten88c}. Some interesting work has be done on the
possibility of quantum topology changes following this
approach\CITE{FujiwaraEtAl91a,FujiwaraEtAl91b,FujiwaraEtAl92a}.
However, for the cases in which the spatial slices are compact
surfaces with genus higher than 2, the good parametrization of the
freedom, which corresponds to constructing $\Gamma_\r{inv}$, is not
found.
Hence
the relation of the low-dimensional theories to the four-dimensional
quantum gravity is not clear at present.

\subsubsection{Hamiltonian constraint and dynamics}

In the previous subsection we have not touched upon the treatment of
the hamiltonian constraint in the quantum theory.  Since all the
dynamical information is contained in the constraint functions and the
constraints other than the hamiltonian constraint are purely
group-theoretical, this problem constitutes the central part of
constructing the physical quantum gravity theory.

Usually the hamiltonian constraint is treated like the other
constraints, and in the quantum framework it is assumed to be
represented as the quantum constraint on the physical state vectors,
\begin{equation}
\hat C_\r{H}|\Psi>=0.
\label{StrongQuantumHamiltonianConstraint}\end{equation}
However, this treatment leads to a theory with no dynamics as stated
at the beginning of this section.

Various approaches have been proposed to resolve this difficulty so
far. First is the WKB approach proposed in the context of quantum
cosmology\CITE{Banks85,HalliwellHawking85}.  In this approach one
first classifies the canonical variables into two classes, macroscopic
variables $(Q_A, P^A)$ and quantum variables $(q_I,p^I)$, and assumes
that $C_H$ has the structure
\begin{equation}
C_H = C_H^1(Q,P) + C_H^2(q,p,Q)
\end{equation}
with
\begin{equation}
C_H^1(Q,P)={1\over2}\G_{AB}(Q)P^A P^B + \V_A(Q)P^A + \U(Q).
\end{equation}
According to this classification one decomposes the state space as
$\V_\r{inv}=\V_1\otimes\V_2$, and expands the state vector $|\Psi>$ in
terms of the eigenstates of $\hat Q$ in $\V_1$ as
\begin{equation}
|\Psi> = \int dQ \,|Q>\!\otimes |\Phi(Q)>.
\label{Psi:decomposition}\end{equation}
Here the criterion for $(Q, P)$ to be regarded as macroscopic is that
$|\Phi(Q)> \in \V_2$ can be decomposed into a rapidly changing phase
$e^{iS(Q)}$ expressed in terms of a solution $S(Q)$ to the
Hamilton-Jacobi equation
\begin{equation}
C_H^1(Q,\partial_Q S)=0,
\end{equation}
and a vector $|\Phi'(Q)>$ changing slowly with $Q$, as
\begin{equation}
|\Phi(Q)> = e^{iS(Q)}|\Phi'(Q)>.
\end{equation}
The solution to the Hamilton-Jacobi equation yields an ensemble of
solutions to the classical equation of motion determined by $C_H^1$,
which foliate the $Q$-configuration space into a family of
trajectories.  Along each trajectory one can introduce a time variable
by
\begin{equation}
\dot Q = \{ Q, C_H^1(\hat N) \}.
\end{equation}
Then, owing to the Hamilton-Jacobi equation and the assumption on
$|\Phi'(Q)>$, one can show that $|\Phi'(Q)>$ approximately satisfies
the Schr\"odinger equation along each trajectory:
\begin{equation}
i\partial_t |\Phi'(Q)> \simeq \hat C_H^2(\hat N)|\Phi'(Q)>.
\end{equation}

Apart from the problem of the unitarity, this approach has some
intrinsic ambiguities.  First it is not clear how to separate the
macroscopic variables in the realistic situations in which the vector
$|\Psi>$ or the wavefunction has a quite complicated structure.
Second it does not give the information which trajectory should be
taken.  These two ambiguities are intimately connected because there
exists no variable which behaves always classically.  Though a couple
of proposals based on the idea of decoherence of history have been
proposed so far to resolve these ambiguities\CITE{Hartle91}, they are
far from satisfactory.

The second approach to extract dynamics from
Eq.(\ref{StrongQuantumHamiltonianConstraint}) is based on the
deparametrization of the theory\CITE{Kuchar81}.  In this approach one
tries to find a canonical transformation of the original variables
$Q_A$ and $P^A$ to $t_\mu$, $\pi^\mu$, $q_I$ and $p^I$, where
$t_\mu$'s are the time variables which specify a time slicing and
$\pi^\mu$'s are their conjugates(the index $\mu$ includes the freedom
of the spatial coordinates).  The requirement on the canonical
transformation is that $C_H$ can been written as
\begin{equation}
C_H(N) = N_\mu[\pi^\mu + h^\mu(q,p,t)].
\end{equation}
If this requirement is satisfied, the vector $|\Phi(t)>$ obtained by
the decomposition of $|\Psi>$ as in Eq.(\ref{Psi:decomposition}) with
$Q$ replaced by $t^\mu$ exactly satisfies the functional Schr\"odinger
equation
\begin{equation}
i{\delta \over \delta t_\mu}|\Phi(t)>=\hat h^\mu(t)|\Phi(t)>.
\end{equation}

Since the time variables are introduced in the classical level, this
approach is essentially of the gauge-fixing type.  Hence it is
expected to share the same difficulties with the gauge-fixing method.
In fact it is shown\CITE{Kuchar81} that, though one can find good time
variables in the class of variables whose conjugates are written as
linear combinations of $P^A$ if $C_H$ has the same structure as
$C_H^1$ in the WKB approach and there exists a conformal Killing
vector to the super metric $\G_{AB}$ which keeps $\V_A$ and $\U$
invariant, it cannot be extended to the generic case.  Of course this
does not exclude the possibility of good time variables existing in
the generic case.  For example, it is shown by
Ashtekar\CITE{AshtekarStachel91} within the framework of the complex
canonical theory that in the weak field approximation the variables
conjugate to the term corresponding to the superpotential $\U$ can be
taken as good time variables in the above sense.  However, it is not
clear whether the same method applies to the generic case.

In these two approaches the vector $|\Phi'(Q)>$ or $|\Phi(t)>$ obtained
by the decomposition of $|\Psi>$ is to be regarded as the quantum
state vector.  Hence the hamiltonian constraint is not imposed on the
state vectors in the genuine sense. In contrast to these, in the third
approach called the frozen approach, the hamiltonian constraint is
imposed strictly on state vectors.  Since the operators consistent
with this quantum constraint in general commute with $\hat C_\r{H}$
weakly, i.e., satisfy the condition
\begin{equation}
[\hat C_\r{H}(\hat T), \hat F]=\hat C_\r{H}(\hat T'),
\end{equation}
they correspond to the constants of motion $\O_\r{cons}$ in the
classical theory.  Hence in this approach the observables
are limited to the constants of motion $\hat\O_\r{cons}$.  In particular
the time variables cannot be included in the observables because
the classical time variables $t_\mu$ must satisfy the condition
\begin{equation}
\delta_T t_\mu= \{C_\r{H}(\hat T), t_\mu\}\not=0 \qquad \exists \mu\quad
\forall \hat T,
\end{equation}
which implies that the corresponding operators $\hat t^\mu$ does not commute
with  $\hat C_\r{H}(\hat T)$.

Thus the only way to introduce the time evolution concept into the
theory in this approach is to regard the time as the parameter of a
special unitary transformation $U(t)$ in the space of the solutions to
Eq.(\ref{StrongQuantumHamiltonianConstraint}), $\V_\r{cons}$.  Of
course there exists no principle to select $U(t)$ within the frozen
formalism.  However, if one can find good time variables in the
deparametrization approach explained above, it is possible to
construct a natural one-parameter family of unitary transformations,
$U(t)$.

To see this, first note that the set of functions on the
classical phase space $(q_I, p^I)$ in
the deparametrization approach is isomorphic $\O_\r{cons}$.  This
isomorphism induces a one-to-one correspondence between $|\Psi>\in
\V_\r{cons}$ and the initial state $|\Phi>\in \V_1$ of $|\Phi(t)>$.
Hence through this correspondence the time evolution operator $U(t)$
in $\V_1$, which is formally expressed as
\begin{equation}
U(t) = T \exp[-i \int dt_\mu \hat h^\mu(t)],
\end{equation}
is transferred to the operator on $\V_\r{cons}$.  For example for the
minisuperspace models of class A Bianchi type I, II, VI$_0$ and VII$_0$,
Ashtekar and others\CITE{AshtekarTateUggla92} have succeeded in
defining an inner product in $\V_\r{cons}$ and constructing a natural
unitary operator $U(t)$.

In canonical quantum gravity theory we cannot assume in general that
there is something like spacetime.  Nevertheless, it is natural assume
that there exist a set of observable quantities which can be regarded
as time variables in a given set of phenomena, since otherwise we lose
the possibility of discussing dynamics.  Under this assumption the
whole observational data of phenomena are divided into subsets for
each of which an appropriate set of time variables take specific
values, and the relations among each subset is represented by a
quantum state.  We call this decomposition into subsets a time slicing
in analogy with the classical theory.

Thus we must give a special role to the time variables in the canonical
quantum gravity. However, this does not mean that quantities chosen
as the time variables in a given situation have any special intrinsic nature
compared with the other quantities.  In fact almost any local quantity
can be used as a time variable in a certain situation.
This consideration suggests that all the observational quantities
 should be treated at equal footing in the formulation and that which
quantities should be regarded as the time variables is determined by
the observational data.

In the above three approaches on the treatment of the time variables
this point
is not realized.  The time variables are treated differently from the
other variables at the basic level of formulation.  In particular
they are regarded as classical variables.  This feature seems to be
quite unsatisfactory to the author. For example the above approaches
cannot deal with the situations in which different sets of time variables
are measured at the state setting and at its observation.  Further, if there
does not exist a universal set of good time variables, which seems to be
the case in the generic situation, the formalism itself breaks down.

If we include the time variables into observables represented by
operators, we cannot impose the hamiltonian constraint on the state
vectors since the operators corresponding to the time variables do not
commute with $\hat C_H$.  Then how should we implement the hamiltonian
constraint into the quantum theory?  Here we give one possible
formulation which is not mathematically well-defined yet but formally
answers to this question.

First note that in the quantum framework the dynamics is
eventually represented by the probability for each possible result of
measurements.  In the conventional quantum mechanics this probability
is expressed in terms of some time-dependent state vector $|\Psi(t)>$
as
\begin{equation}
\Pr(\hat A_\alpha \rightarrow a_\alpha) = |<\Psi(t)|a_\alpha>|^2.
\end{equation}
In this equation, though $|\Psi(t)>$ and $|a_\alpha>$ both belong to
the state space $\H$, they can be regarded as playing different
roles:$|a_\alpha>$ is an object to represent the observational data,
while $|\Psi(t)>$ is an object to assign the probability amplitude to
each possible observational data represented by $|a_\alpha>$.  If we
look at this equation in this fashion, the most natural way to assign
the probability to the state vectors which carry information on the
time $t$ is to replace the bra $<\Psi(t)|$ by a linear functional
$\Psi[\Phi]$ on the state space $\H$ and express the probability as
\begin{equation}
\Pr(\hat A_\alpha \rightarrow a_\alpha) = |\Psi[|a_\alpha>]|^2.
\end{equation}
Since this expression yields the probabilistic prediction on the
results of measurements on arbitrary time slice and since $\Psi$ does
not carry an information on the time slicing any longer, $\Psi$
represents the quantum counterpart of the complete data on a
space-time phenomenon as a whole in the classical theory.  Hence it
cannot be a state vector unlike in the quantum mechanics.  In
particular it cannot be a continuous functional on $\H$ since from
Riesz's theorem each continuous linear functional on the Hilbert space
$\H$ is in one-to-one correspondence to a vector in $\H$.  We do not
even require it to be defined on the whole $\H$.  As a result, $\Psi$
is not normalizable in general, and $\Pr$ in the above expression
should be understood to represent only the relative probability.

In spite of these properties we formally write the functional
$\Psi[\Phi]$ from now on as
\begin{equation}
\Psi[\Phi] = <\Psi|\Phi>,
\end{equation}
for the notational simplicity, and call $<\Psi|$ the {\it probability
amplitude functional}.

Since the dynamical information is carried by $<\Psi|$ in this
formulation, it is the most natural to translate the classical
hamiltonian constraint to the constraint on $<\Psi|$ as
\begin{equation}
<\Psi|\hat C_\r{H}(\hat T)=0 \qquad \forall \hat T\in \hat\O.
\label{WeakQuantumHamiltonianConstraint}\end{equation}
The exact meaning of this formal expression is
\begin{equation}
<\Psi|\hat C_\r{H}(\hat T)|\Phi>=0 \qquad \forall |\Phi>\in\H, \forall
\hat T\in\hat\O.
\end{equation}
We call this condition {\it the weak quantum hamiltonian constraint}.

In the limit the Planck constant $\hbar$ vanishes the weak quantum
hamiltonian constraint reduces to the classical one.  To see this, let
us represent the constraint in terms of the coherent representation.
Let $|Q, P>$ be the basis of the coherent representation of $\H$, and
assume that the generalized version of the completeness condition on
$|Q,P>$,
\begin{equation}
<\Psi|\Phi>=\int DQ DP <\Psi|Q,P> <Q,P|\Phi>,
\end{equation}
holds.  Further let $\hat C^\mu_\r{H}$ be the generating operators for
$\hat C_\r{H}(\hat T)$ such that
\begin{equation}
\hat C_\r{H}(\hat T) = \hat C^\mu_\r{H}\hat T_\mu \qquad \exists \hat T_\mu \in
\O_\r{inv}.
\end{equation}
Then the weak quantum hamiltonian constraint is represented as
\begin{eqnarray}
& <\Psi|\hat C_\r{H}^\mu|Q,P> & = \int DQ'DP' <\Psi|Q'P'><Q'P'|\hat
C_\r{H}^\mu|Q,P> \nonumber\\ &&\simeq <\Psi|Q,P> C_\r{H}^\mu(Q,P),
\end{eqnarray}
where $\simeq$ means that it becomes the exact equality in the limit
$\hbar \rightarrow 0$.  From this equation it follows that for
$<\Psi|QP>\not=0$, $C_\r{H}^\mu(Q,P)\simeq0$, which implies that the
hamiltonian constraint holds in the classical limit.  The origin of
this result is the fact that the weak hamiltonian constraint in the
configuration-space representation $\Psi(Q)=\overline{<\Psi|Q>}$ is nothing
but the generalized Wheeler-DeWitt equation
\begin{equation}
\bar C_\r{H}^\mu(Q,-i{\partial\over \partial Q})\Psi(Q)=0.
\end{equation}

Further, since the hermitian operators in $\hat \O_\r{cons}$ commute
weakly with $\hat C_\r{H}^\mu$, there exist a class of probability
amplitude functionals each of which satisfies the equation
\begin{equation}
<\Psi|\hat F|\Phi> \simeq f <\Psi|\Phi> \qquad \forall |\Phi> \in \H,
\label{WavePacketConditionOnPsi}\end{equation}
 for any hermitian operator $\hat F$ in $\hat \O_\r{cons}$ where $f$
is a real number depend on the operator $\hat F$.  In the coherent
representation this equation is written as
\begin{equation}
<\Psi|QP> (F(Q,P)-f) \simeq 0.
\end{equation}
This implies that $<\Psi|QP>$ does not vanish only around a subspace
of $\Gamma$ corresponding to the classical solution specified by the
conditions on the constants of motion, $F(Q,P)=f$, provided the
condition (\ref{WavePacketConditionOnPsi}) is satisfied.

Thus at least in the sense of the word used in the quantum mechanics
the above formulation reduces to the classical general relativity
theory in the limit $\hbar\rightarrow0$.  Further, though its
appearance is largely different from the conventional quantum
mechanics, it can be shown that the above formalism is an extension of
the latter in the sense that the quantum mechanics can be rewritten in
the same form, as will be illustrated in the next subsection.

In order to complete the above formalism, we must give the
procedure to determine $<\Psi|$ from the observational data.  This
problem is intimately connected with the choice of the time operators
$\hat t_\mu$.  To see this, let us assume that the observational data
determine the eigenvalues of a maximal commuting set $\hat q_I$
and $\hat t_\mu$ of $\hat \O_\r{inv}$.  Then if one can find a
decomposition $\hat C_\r{H}^\mu = \hat h^\mu + \hat \pi^\mu$ and a set
of operators $\hat n_\mu$ such that $\hat \pi:=\hat \pi^\mu\hat n_\mu$ and
$\hat h:=\hat h^\mu\hat n_\mu$ are represented as
\begin{eqnarray}
&&<\Psi|\hat h|q, t(\tau)> =
h(q,i{\partial\over\partial q},
t)\overline{\Psi(q,t(\tau))},\\
&&<\Psi|\hat \pi|q,t(\tau)> =
i{\partial\over\partial \tau}\overline{\Psi(q,t(\tau))}
\end{eqnarray}
for some one-parameter family $t_\mu(\tau)$ where
\begin{equation}
\Psi(q,t):=\overline{<\Psi|q, t>},
\end{equation}
then the weak hamiltonian constraint is expressed as
\begin{equation}
i\partial_\tau \Psi(q,t(\tau))=
h(q,-i{\partial\over\partial q},t)\Psi(q,t(\tau)).
\end{equation}
Thus $\Psi(q,t(\tau))$ satisfies the Schr\"odinger equation with the
time variable $\tau$ and is determined if the value at $\tau=0$ is
given.  In particular if the measurements give data
$(q^0_I,t_\mu(0))$, the weak hamiltonian constraint determines
$<\Psi|$ at least on the subset $|q,t(\tau)>$.

Here the reader will notice that this argument to derive the Schr\"odinger
equation is quite similar to that
in the deparametrization approach.  Actually the above condition on
the structure of $\hat C_H$ is the quantum analogue of the condition
for the good time variables to exist in the deparametrization approach.
Further, if we expand $<\Psi|$ formally as
\begin{equation}
<\Psi| = \int dt \, <t|\otimes \!<\Phi(t)|,
\end{equation}
the ket $|\Phi(t)>$  dual to the bra $<\Phi(t)|$ given by
\begin{equation}
|\Phi(t)> = \int dq \Psi(q,t)|q>
\end{equation}
satisfies the same functional Schr\"odinger equation as in the
deparametrization approach.

Our formalism, however, is not completely equivalent to the
deparametrization approach.  It is because it does not depend on a
special choice of the time variables and hence is more flexible.  In
particular it is potentially applicable to cases in which no universal
set of good time variables exists.  Of course in such cases a large
ambiguity comes into the formulation in determining $<\Psi|$ by
measurements. To cope with such a situation, we must impose some
additional constraints on $<\Psi|$ to remove the ambiguity in $<\Psi|$
which remains even if the initial data are given, or limit the
predictions to those insensitive to it. At present it is not clear
which prescription is better.  More detailed studies of variety of
examples should be done to make complete the formulation.

\subsection{Examples}

In the previous subsection we proposed a new implementation of the
hamiltonian constraint into the quantum framework and discussed its
potential problems associated with the time variable and dynamics.  In
this subsection we will see how the formalism work in some simple
examples and analyze the problems in more detail.

\subsubsection{Quantum mechanics}

As is well-known, the equation of motion of non-relativistic particles
can be always written in a time-reparametrization invariant
form\CITE{Kuchar81}.  For example, the variational equation for a
non-relativistic particle
\begin{equation}
\delta S= \delta \int dt\left({1\over2}m\Frac({dx}/{dt})^2 - V(x)\right)=0
\end{equation}
is equivalent to
\begin{equation}
\delta S' = \delta\int d\tau (p\dot x + \pi \dot t - N C_\r{H}),
\label{TRIAction:NRP}\end{equation}
where the dot denotes the derivative with respect to the parameter
time $\tau$ and
\begin{equation}
C_\r{H} = \pi + h(x,p);\qquad h={p^2 \over 2m} + V(x).
\end{equation}
The system described by the action $S'$ is a totally constrained
system with a single constraint function $C_\r{H}$.

Since there exists no symmetry corresponding to the gauge or the
spatial diffeomorphism, $\Gamma_\r{inv}$ and $\O_\r{inv}$ are given by
$\Gamma=\{(x,p,t,\pi)\}$ and $\O=\O(\Gamma)$ in the present case.
Hence $\hat \O=\hat\O_\r{inv}$ is constructed from the fundamental
operators $\hat x$, $\hat p$, $\hat t$ and $\hat\pi$ satisfying the
commutation relations
\begin{equation}
[\hat x, \hat p]=i,\quad [\hat t, \hat \pi]= i, \quad \r{others}=0.
\end{equation}

{}From the structure of $C_\r{H}$, in the standard coordinate
representation for which $\hat x$ and $\hat t$ are diagonal, the weak
quantum hamiltonian constraint is written as
\begin{equation}
\overline{<\Psi|\hat C_\r{H}|x,t>} = -i\partial_t\Psi(x,t)+\hat h \Psi(x,t)=0,
\end{equation}
where
\begin{equation}
\Psi(x,t)=\overline{<\Psi|x,t>}.
\end{equation}
Hence in the present case the weak hamiltonian constraint is exactly
identical to the Schr\"odinger equation if $\Psi(x,t)$ is interpreted
as the wavefunction. This interpretation is consistent with the role
of $<\Psi|$ as the probability amplitude functional in our formalism.
{}From the Schr\"odinger equation, the norm of $\Psi(x,t)$ calculated by
formally regarding it as the state vector is divergent:
\begin{equation}
<\Psi|\Psi>:=\int dxdt|\Psi(x,t)|^2= \int dt |\Psi(x,0)|^2 = +\infty.
\end{equation}
Thus $<\Psi|$ is not continuous functional on the state space $\H$ as
discussed in the previous subsection in a general setting.

In the present case the dynamics is treated as follows.  First let us
assume that the measurements of two commuting observables $\hat
A(x,p)$ and $\hat t$ gave the results $a_0$ and $t_0$.  This result of
measurements is represented by the eigen state vector $|a_0,t_0>\in
\H$,
\begin{equation}
\hat A|a_0,t_0>=a_0|a_0,t_0>, \qquad \hat t|a_0,t_0>=t_0|a_0,t_0>.
\end{equation}
Then the probability amplitude functional $<\Psi|$ is determined as a
solution to the weak hamiltonian constraint satisfying the condition
\begin{equation}
<\Psi|a,t_0>=0 \quad \forall a\not=a_0.
\end{equation}
Since this condition is equivalent to fixing $\Psi(x,t_0)$ as
$\Psi(x,t_0)=\Phi_{a_0}(x)$ with the eigenfunction $\Phi_{a_0}(x)$ of
the operator $\hat A$, the equivalence of the weak hamiltonian
constraint with the Schr\"odinger equation guarantees that $<\Psi|$ is
uniquely determined by this condition apart from the total
normalization.  This unique probability vector yields the
probabilistic prediction on the results of measurements on other
commuting operators including the time variable.

Thus the conventional quantum mechanics can be rewritten with our
formalism.  Of course the conventional one will be more convenient to
solve practical standard problems.  However, our formalism is much
more flexible in treating dynamics than the conventional one as noted
in the previous subsection. For example, the time variable used in
prediction can be different from the one used to fix the probability
amplitude functional.  This feature seems to become important in
quantum gravity since there exists in general no global time variable
in general relativity.

\subsubsection{Relativistic free particle}

The classical action of a relativistic free particle
\begin{equation}
S=-m\int ds = -m \int d\tau [-\eta_{\mu\nu}\dot x^\mu \dot
x^\nu]^{1/2}
\end{equation}
is invariant under the reparametrization of $\tau$, and is equivalent
to the action with a structure similar to the previous example:
\begin{equation}
S=\int d\tau \left( {1\over 2N}\dot x^\mu \dot x^\nu - {N\over 2}m^2
\right).
\end{equation}
Hence in the canonical form it becomes a totally constrained system:
\begin{eqnarray}
&&L=\dot x^\mu p_\mu - N C_\r{H},\\ &&C_\r{H}={1\over2}(p^\mu p_\mu +
m^2).
\end{eqnarray}

This classical system can be easily quantized like the previous
example.  The operator algebra is generated from the fundamental
operators $\hat x^\mu$ and $\hat p_\mu$ with the commutation relations
\begin{equation}
[\hat x^\mu, \hat p_\nu]=i\delta^\mu_\nu, \quad [\hat x^\mu, \hat
x^\nu]=[\hat p_\mu, \hat p_\nu]=0,
\end{equation}
and the Hilbert space $\H$ as its irreducible representation space is
uniquely determined modulo the unitary equivalence.

In contrast to the example of non-relativistic particles, the
constraint function of this system is quadratic in all of the momentum
variables.  Hence the difficulty associated with the time variable and
dynamics discussed in \S3.2 is expected to occur.  Actually, since in
the coordinate representation the weak quantum hamiltonian constraint
becomes the Klein-Gordon equation for the wavefunction $\Psi(x)$:
\begin{equation}
(\partial^2 -m^2)\Psi(x)=0,
\label{WQHC:RFP}\end{equation}
the probability amplitude functional $<\Psi|$ is not uniquely
determined by its value on a spacelike time slice.

In the present case, however, this difficulty is easily eliminated by
imposing a natural additional constraint on the probability amplitude
functionals.  To see this, first note that the solutions of the
hamiltonian constraint in the classical theory consist of two
disconnected submanifolds $p^0=\pm
\omega(\bm{p})$($\omega(\bm{p})=(\bm{p}^2+m^2)^{1/2}$).  Since the two
solutions passing through $(x^\mu,\omega(\bm{p}),\bm{p})$ and
$(x^\mu,-\omega(\bm{p}),-\bm{p})$ in these submanifolds correspond to
the same trajectory in $x$-space, two points connected by the
transformation $p^\mu \rightarrow -p^\mu$ represent the same classical
state.

This degeneracy of the classical phase space should be taken into
account in quantum theory.  Since it is not consistent with the
commutation relation to require the observables to be invariant under
the transformation, the only possible way is to impose the constraint
$p^0>0$ (or $p^0<0$).  In the probability vector formalism this
constraint is expressed as
\begin{equation}
<\Psi|\P_+ = <\Psi|,
\label{PositiveEnergyCondition}\end{equation}
where $\P_+$ is the projection operator defined by
\begin{equation}
\P_+ := \int d\bm{p} \int_0^\infty dp^0 |p><p|.
\end{equation}
Under this constraint the general solution to Eq.(\ref{WQHC:RFP}) is
written as
\begin{equation}
\Psi(x) = \int d\bm{p} \Psi_+(\bm{p})e^{-i\omega t + i\bm{p}\cdot \bm{x}},
\end{equation}
where $t=x^0$. Since this solution obeys the Schr\"odinger equation
\begin{equation}
i\partial_t \Psi(x) = \omega(\hat\bm{p})\Psi(x),
\end{equation}
the probability vector is uniquely fixed by its value on the states
with a fixed value of $t$.  Thus $t=x^0$ becomes a good time variable
under the constraint (\ref{PositiveEnergyCondition}).

As explained in \S3.2 the treatment in this and the previous example
is essentially equivalent to the deparametrization approach.  Hence
the same treatment is possible for the cases in which the latter
approach works.  For such examples see \cite{Kuchar81} and
\cite{AshtekarTateUggla92}.

\subsubsection{Minisuperspace model}

As the final example, we consider the system of a spatially
homogeneous real scalar field $\phi$ coupled to  a closed Robertson-Walker
geometry.  We parametrize the Robertson-Walker metric as
\begin{equation}
ds^2= {2\kappa^2\over\Omega}[-N^2dt^2 + \sigma \chi^I\chi^I],
\end{equation}
where $\chi^I$ is the basis of the invariant forms on the Euclidean
sphere normalized as
\begin{equation}
d\chi^I = {1\over2}\epsilon_{IJK}\chi^J\wedge\chi^K,
\end{equation}
and $\Omega$ is the coordinate volume defined by
\begin{equation}
\Omega=\int |\chi|d^3x = 16\pi^2.
\end{equation}
Then the Lagrangian of the system is written as
\begin{eqnarray}
&&L=p\dot\sigma +\pi\dot\phi - NC_\r{H},\\
&&C_\r{H}=\sqrt{\sigma}\left[-{1\over6}p^2-{3\over2}+{1\over2\sigma^2}\pi^2+\sigma
V(\phi)\right]. \label{CH:RW+phi}
\end{eqnarray}

In contrast to the previous examples quantization of this system is
not so straightforward.  First it is not obvious which quantities
should be taken as the fundamental variables: we can use, for example,
$a=\sqrt{\sigma}$ and $2ap$, or $\alpha={1\over2}\ln\sigma$ and
$2p\sigma$ in place of $\sigma$ and $p$. Classically they are all
equivalent, but after quantization they may lead to different
theories.  For example, since $a$ should be positive, its conjugate
momentum cannot be hermitian, while the momentum conjugate to $\alpha$
can be represented as a hermitian operator.  A similar problem occurs
for the choice $\sigma$ and $p$ depending on whether we restrict the
range of $\sigma$ to $\sigma>0$ or not.

Second we can always multiply $C_\r{H}$ by any non-degenerate quantity
by rescaling $N$.  For example, in the chiral canonical theory, $N$ is
rescaled to $\utilde N=N/\sigma^{3/2}$, hence the expression with
$\sqrt{\sigma}$ replaced by $\sigma^2$ in Eq.(\ref{CH:RW+phi}) should
be used as $C_\r{H}$.

Third there is a large ambiguity in the operator ordering for $\hat
C_\r{H}$.  In quantum mechanics the ambiguity in operator ordering is
often reduced by the requirement of hermiticity.  However, this does
not apply to the present case since $\hat C_\r{H}$ need not be
observable.

These ambiguities may be partly eliminated from some consistency
conditions in the general quantum gravity theory, as we will see in
\S4.  However, in the present case, they remain as the freedom in
quantization.  In the rest of this subsection we adopt $\sigma$ and
$p$ as the fundamental variables and rescale $N$ so that
$\sqrt{\sigma}$ in Eq.(\ref{CH:RW+phi}) is eliminated for simplicity.

Let us first consider the case in which the scalar field contribution
is neglected except for its constant potential energy acting as the
cosmological constant.  In this case the weak quantum hamiltonian
constraint in the $\sigma$ representation,
$\Psi(\sigma)=\overline{<\Psi|\hat C_\r{H}|\sigma>}$, is written under
the natural ordering as
\begin{equation}
\left({d^2\over d\sigma^2} - 9 + \lambda \sigma\right)\Psi(\sigma)=0.
\end{equation}
Since this equation is of second-order, the same problem as in the
relativistic free particle system occurs if $\sigma$ is taken as the
time variable.  In contrast to that case, however, there exists no
natural constraint on $\Psi(\sigma)$ to eliminate this problem in the
present case since there exists no degeneracy in the phase space. This
does not imply no good time variable exists in this system.  In fact
since the hamiltonian constraint is written in the $p$-representation
as
\begin{equation}
i{d\over dp}\Psi(p) = -{1\over\lambda}\left(p^2 +
{3\over2}\lambda\right)\Psi(p),
\end{equation}
$-p$ can be taken as a good time variable.  Classically this
corresponds to taking the expansion rate of the universe as time.
Obviously the quantum theory becomes trivial for this choice.

Unfortunately this success is specific to this simple case.  The
situation changes significantly if we take account of the scalar field
dynamics.  For example let us consider the operator ordering of $\hat
C_\r{H}$ given by
\begin{equation}
\hat C_\r{H}=-{1\over6}\hat \sigma^{-1/2}\hat  p\hat\sigma \hat
p\hat\sigma^{-1/2} -{3\over2} + {\hat \pi^2\over 2\hat \sigma^2} + \hat\sigma
V(\hat\phi).
\end{equation}
Then for the potential of the form
\begin{equation}
V(\phi)=\sqrt{3}\left( a\cosh(\phi/\sqrt{3}) +
b\sinh(\phi/\sqrt{3})\right)
\end{equation}
the hamiltonian constraint is exactly
soluble\CITE{GarayHalliwellMarugan91}.  In fact in terms of the
operators
\begin{eqnarray}
&&\hat x= \sqrt{3}\hat\sigma\cosh(\hat\phi/\sqrt{3}), \\ &&\hat y=
\sqrt{3}\hat\sigma\sinh(\hat\phi/\sqrt{3}),\\ &&\hat
p_x={1\over\sqrt{3}}\left(\hat p
-{i\over2\hat\sigma}\right)\cosh{\hat\phi\over\sqrt{3}} - {\hat
\pi\over\hat\sigma}\sinh{\hat\phi\over\sqrt{3}},\\ &&\hat
p_y=-{1\over\sqrt{3}}\left(\hat p
-{i\over2\hat\sigma}\right)\sinh{\hat\phi\over\sqrt{3}} + {\hat
\pi\over\hat\sigma}\cosh{\hat\phi\over\sqrt{3}},
\end{eqnarray}
which satisfy the standard commutation relations
\begin{equation}
[\hat x, \hat p_x]=i, \qquad [\hat y, \hat p_y]=i,\qquad \r{others}=0,
\end{equation}
$\hat C_\r{H}$ is written as
\begin{equation}
\hat C_\r{H}={1\over2}(-\hat p_x^2 + \hat p_y^2)-{3\over2}+a \hat x + b\hat y,
\end{equation}
which is a sum of two simple decoupled systems.

The last equation suggests that, if $a\not=0$, $\hat p_x$ can be a
good time variable as in the pure gravity case above.  However, it is
not the case.  The trouble is caused by the fact that $\hat p_x$ is
symmetric but not hermitian.  Due to this property the
$(p_x,y)$-representation and $(\sigma,p)$-representation are not
unitary equivalent.

This analysis of the simple example suggests that there exists no
quantity which can be used as a good time variable universally.  One
should probably impose some constraint on the wavefunctions as a
fundamental postulate such as the Hartle-Hawking ansatz, or restrict
the observables.  In the former case the choice of good time variables
depends on the constraint, while in the latter case what can be
predicted depends on the time variable chosen.  At present we can say
nothing about which approach is better.

\section{Quantization of the Complex Canonical Theory}

As noted in \S3.2.1, in the ADM-WD theory based on the canonical
theory in the metric approach, there is no gauge symmetry associated
with the tetrad rotation in the phase space $\Gamma_\r{ADM}=\{(q_{jk},
p^{jk},\ldots)\}$.  Hence as far as the gravitational fields are
concerned one can start from the stage II$^*_C$ in Table
\ref{QuantizationProgram}, and easily proceeds to the stage II$^*_Q$
in the quantum framework at least formally.  However, it is quite
difficult to go to the next stage either in the classical framework or
in the quantum framework.

Mathematically one can define an abstract space, called {\it a
superspace}, as the set of diffeomorphism equivalent classes of
metrics and fields on some three-dimensional manifold.  If the
superspace is constructed explicitly, then the cotangent bundle of the
superspace will play the role of $\Gamma_\r{inv}$ in the classical
framework.  This has been widely accepted as the ideal program to
build the canonical quantum gravity since its proposal by Wheeler.
However, little progress has so far been made on the quantum aspect of
this program.

In the quantum framework the construction of the superspace is
replaced by solving the spatial diffeomorphism constraint
\begin{equation}
\hat C_\r{D}(L)|\Phi>=0,
\label{QuantumDiffeomorphismConstraint}\end{equation}
and the construction of its cotangent bundle by finding the
diffeomorphism invariant operators $\hat \O_\r{inv}\subset\hat\O$.
For the pure gravity case, in the representation in which $\hat
q_{jk}$ is diagonal, Eq.(\ref{QuantumDiffeomorphismConstraint}) is
written as
\begin{equation}
i\delta_L \Phi[q] = -i \int d^3x \Lie_L q_{jk}(x) {\delta\over\delta
q_{jk}(x)}\Phi[q] =0.
\end{equation}
Thus the former problem is to find all the functionals on the space of
metrics which are invariant under the spatial diffeomorphism
transformations.  Since this is almost equivalent to constructing the
superspace and find a projection to it from the space of metrics, it
is intractable like the corresponding classical problem. Thus, a
generic class of solutions to
Eq.(\ref{QuantumDiffeomorphismConstraint}) has not been constructed
explicitly so far.

On the other hand the latter problem appears to be easily solved once
the former problem is solved since $\hat\O_\r{inv}$ modulo the
operators which vanish on $\V_\r{inv}$ is in one-to-one correspondence
with the operators in the subspace $\V_\r{inv}$.  However, this is not
sufficient since one needs to know this correspondence explicitly to
find the physical interpretation of the invariant operators.

In contrast to the ADM-WD theory the chiral canonical theory is
formulated on the complex phase space $\Gamma_{\CF}=\{(\E^j,\A_j,\ldots)\}$
which is not gauge invariant.  Hence one may suspect that it would be
much more difficult to construct the quantum gravity theory starting
from it than the ADM theory.  However, it is not the case.  One reason
is that quite different constructions of the gauge-invariant
quantities become possible in the chiral canonical theory.  Another
reason is the much simpler structure of equations.  Owing to these
features one can go at least to the same stage as in the ADM-WD
approach and sometimes can go further.

In this section we will see these points in greater detail by
constructing the quantum theory of the chiral canonical theory
following the scheme presented in the previous section.  We limit the
consideration to the case of pure gravity with the cosmological
constant since the cases in which material fields are coupled have not
been studied well yet.  Further we should remind the reader that most
considerations in this section are formal and much is still to be done
to make them mathematically well-defined.

\subsection{Operator Algebra}

First we follow the path which goes to the quantum framework at the
stage I in Table \ref{QuantizationProgram}.  In this path we take
$\E^j$ and $\A_j$ as the fundamental variables and assign to them the
operators $\hat\E^j$ and $\hat \A_j$ with the commutation relations
\begin{eqnarray}
&& [\hat\E^{Ij}(x),
\hat\A_{Jk}(y)]={i\over2}\delta^I_J\delta^j_k\delta^3(x-y),
\nonumber\\
&& [\hat\E^{Ij}(x), \hat\E^{Jk}(y)]=[\hat\A_{Ij}(x),
\hat\A_{Jk}(y)]=0.
\label{CommutationRelation:ChiralVariables}\end{eqnarray}
The full operator algebra $\hat \O$ is easily constructed from them.
Thus apart from the regularization problem the remaining algebraic
problem is to embed the constraint functions into this operator
algebra by fixing the operator ordering.

\subsubsection{Constraint operators}

First there is no ordering ambiguity in the gauge constraint operator
$\hat \C_\r{G}:=\hat \C_{\A}$ since
\begin{equation}
[\hat\A_{[J|j|}(x), \hat \E^j_{K]}(y)]=0.
\end{equation}
Further it is easily checked that $\hat C_\r{G}$ yields the generator
of the $SO(3,\CF)$ gauge transformation in $\hat \O$:
\begin{eqnarray}
&&-i[\hat\C_\r{G}(\lambda),
\hat\E(\phi)]=\mp2i\kappa^2\hat\E(\lambda\times\phi)=\delta_\lambda
\hat\E(\phi),\\ &&-i[\hat\C_\r{G}(\lambda),
\hat\A(\alpha)]=-\hat\D\lambda(\alpha)=\delta_\lambda \hat\A(\alpha).
\end{eqnarray}

In contrast there are some ordering ambiguities in $\hat \C_\r{D}$ due
to the commutation relations
\begin{eqnarray}
&&[\hat\E^{kI}(x), \hat\F_{Ijk}(y)]=\mp
3\kappa^2\partial_j\delta^3(x-y),\\ &&[\hat\C^I_\r{G}(x),
\hat\A_{Ij}(y)]=-3\kappa^2\partial_j\delta^3(x-y).
\end{eqnarray}
However, this type of ambiguities can be easily eliminated by defining
the products of operators at the same spatial point by
\begin{equation}
\hat X(x)\hat Y(x) := \lim{1\over2}\left(
\hat X(\alpha)\hat Y(\beta)+\hat X(\beta)\hat Y(\alpha) \right),
\end{equation}
where $\hat X(\alpha)$ denotes $\int d^3x \alpha(x)\hat X(x)$, and
$\lim$ implies the limit $\alpha(y)\rightarrow \delta^3(y-x)$ and
$\beta(y)\rightarrow \delta^3(y-x)$.  Since this definition is quite
natural, we adopt it throughout this paper and call it {\it the
symmetric regularization}.  Under the symmetric regularization there
remains no ordering ambiguity in $\hat \C_\r{D}$ as far as one does
not decompose the combinations $\hat \F_{jk}$ and $\hat\C_\r{G}$, and
it generates the infinitesimal coordinate transformation in $\hat\O$:
\begin{eqnarray}
&&-i[\hat\C_\r{D}(L),
\hat\E(\phi)]=\hat\E(\Lie_L\phi)=\delta_L\hat\E(\phi),\\
&&-i[\hat\C_\r{D}(L),
\hat\A(\alpha)]=\hat\A(\Lie_L\alpha)=\delta_L\hat\A(\alpha).
\end{eqnarray}

Since $\hat \C_\r{G}$ and $\hat \C_\r{D}$ are the generators of the
transformations in $\hat \O$, their commutation relations are
isomorphic to the corresponding Lie algebra as in the classical theory
and given by
\begin{eqnarray}
&&[\hat\C_\r{G}(\lambda_1), \hat\C_\r{G}(\lambda_2)]=\pm
2\kappa^2\hat\C_\r{G}(\lambda_1\times\lambda_2),\\ &&[\hat\C_\r{D}(L),
\hat\C_\r{G}(\lambda)]=i\hat\C_\r{G}(\Lie_L\lambda),\\
&&[\hat\C_\r{D}(L_1), \hat\C_\r{D}(L_2)]=i\hat\C_\r{D}([L_1,L_2]).
\end{eqnarray}
For the same reason the commutation relations of them with the
Hamiltonian constraint operator $\hat \C_\r{H}$ have the same
structure as in the classical theory regardless of the ordering in
$\hat \C_\r{H}$:
\begin{eqnarray}
&&[\hat\C_\r{G}(\lambda), \hat \C_\r{H}(\utilde T)]=0,\\
&&[\hat\C_\r{D}(L), \hat \C_\r{H}(\utilde
T)]=i\hat\C_\r{H}(\Lie_L\utilde T).
\end{eqnarray}

Unlike $\hat \C_\r{D}$ the ordering ambiguities in $\hat\C_\r{H}$
cannot be removed by the symmetric regularization since
\begin{equation}
[(\hat\E^j\times\hat\E^k)^I(x),
\hat\F_{Ijk}(y)]=8i\kappa^4(\hat\E^j(x)\cdot\hat\A_j(y) + \hat
\A_j(y)\cdot\hat\E^j(x))\delta^3(x-y).
\end{equation}
In this paper we only consider two orderings: the one in which all the
$\hat\E$'s are located at the left to $\hat\A$ and the reversed one
$\hat\E\cdots\hat\A\cdots$.  For these orderings the commutation
relations among $\hat\C_\r{H}$ are given by
\begin{equation}
[\hat\C_\r{H}(\utilde T{}_1), \hat\C_\r{H}(\utilde
T{}_2)]=i\hat\C_\r{M}(\hat L) = i(\hat\C_\r{D}(\hat L) -
\hat\C_\r{G}(\hat L\cdot\hat A)),
\end{equation}
where $\hat L$ is given by
\begin{equation}
\hat L^j=(\hat\E^j\cdot\hat\E^k)(\utilde T{}_2\partial_k \utilde T{}_1 -
\utilde T{}_1\partial_k \utilde T{}_2),
\end{equation}
and the same ordering as in $\hat\C_\r{H}$ is supposed to be taken in
$\hat\C_\r{M}(\hat L)$.

\subsubsection{Reality condition}

As explained in \S2.3.5, in order for the complex canonical theory to
be equivalent to the general relativity theory, the reality condition
must be satisfied.  In the classical theory this condition had to to
be imposed not as a constraint on states but as a condition to reduce
the freedom to the physical one and cut out a physical phase space
from the formal complex phase space.  Accordingly in the quantum
theory it must be implemented as operator equations.  Since the
complex conjugation is replaced by the star operation in the quantum
theory, from Eqs.(\ref{ClassicalRealityCondition1}) and
(\ref{ClassicalRealityCondition2}), they are expressed at the formal
algebraic level as the polynomial relations
\begin{eqnarray}
&&(\hat\E^j\cdot\hat\E^k)^\star =
\hat\E^j\cdot\hat\E^k,\label{QuantumRealityCondition1}\\
&&(\hat\psi^{jk})^\star =-\hat\psi^{jk}.
\label{QuantumRealityCondition2}
\end{eqnarray}

Though these conditions are not expressed in terms of the invariant
operators, the conditions themselves are invariant under the gauge and
the spatial diffeomorphism transformations.  Hence they are expected
to be consistently translated to conditions on the star operation on
the invariant operators.  Then by requiring the star operation to be
realized as the hermitian conjugation, they yield the conditions on
the inner product to be constructed in $\hat\V_\r{inv}$.

Thus the reality condition can be consistently built into the operator
formalism formally.  However, it is expected to be highly non-trivial
task to construct the inner product consistent with it.  To see this,
let us consider the case in which the spatial gauge is imposed before
quantization.  In this gauge $\E^j=\tilde e^j$ is restricted to be
real, and the second reality condition is explicitly solved to be
expressed by Eq.(\ref{ClassicalRealityCondition:SpatialGauge}) with
the real quantity $P_j$ in the classical theory.  Hence after
quantization the reality conditions are expressed as
\begin{eqnarray}
&&(\hat\E^j)^\star =\hat \E^j,\\ &&(\hat\A_j)^\star = \hat \A_j \mp
{i\over\kappa^2}Q_j(\hat\E).
\end{eqnarray}
Since the gauge freedom of the tetrad is restricted to the local
$SO(3)$ rotation in this gauge, these conditions are gauge and spatial
diffeomorphism invariant again.

Here note that from the identity equation (\ref{IdentityForQ})
$Q_j(e)$ is expressed as
\begin{equation}
Q_j(e) = \kappa^2{\delta W(\tilde e)\over \delta \tilde e^j},
\end{equation}
where $W$ is the functional
\begin{equation}
W(\E):= {1\over2\kappa^2}\int d^3x\,(\E^j\times \E_k)\cdot\partial_j
\E^k.
\end{equation}
Here $\E_j=(\E_{Ij})$ is the inverse of $\E^j=(\E^{Ij})$.  $W$ is a
homogeneous functional of degree one, spatial diffeomorphism
invariant, and transforms under the infinitesimal triad rotation
\begin{equation}
\delta\tilde e^j=2\kappa^2 \lambda \times \tilde e^j\quad(\lambda:\r{real}),
\end{equation}
as
\begin{equation}
\delta W(\tilde e) = 2 \int d^3x\, \tilde e^j\cdot \partial_j\lambda.
\label{GeneratingFunctional}\end{equation}
{}From these properties it follows that the quantity $Z$ defined by
\begin{equation}
Z(\A,\E):= 2\int d^3x\, \E^j\cdot \A_j \mp i W(\E)
\label{Z:def}\end{equation}
is a functional invariant under the gauge and spatial diffeomorphism
transformations as well as the star operation. This quantity will be
used in the formal construction of the inner product in the next section.

\subsection{Connection Representation}

In order to construct a definite quantum theory, we must construct a
representation of the operator algebra on a linear vector space where
some inner product is eventually introduced.  In contrast to the
quantum mechanics of systems with finite degrees of freedom the
explicit construction of representation is quite important in systems
with infinite degrees of freedom since different representations in
general lead to different quantum theories.

In the ADM or the real canonical theories the most natural
representation at the stage I$_Q$ is the one in which the operator
corresponding to the metric or the triad is diagonal.  In contrast in
the complex canonical theory another representation in which the
connection operator is diagonal is possible and may be much better
than the triad representation at least as far as we start from the
stage I$_Q$.  Since this representation corresponds to a kind of the
Bargmann representation or the coherent state representation of the
real canonical theory, we first recapitulate some basic facts of the
Bargmann representation in a somewhat generalized form.

\subsubsection{Bargmann representation}

For a quantum mechanical system with the fundamental canonical
variables $\hat q$ and $\hat p$ let us consider the operator
$\hat\alpha$ defined by
\begin{equation}
\hat \alpha := \hat p + i\partial_q w(\hat q) =e^{w(\hat q)}\hat p e^{-w(\hat
q)}.
\label{alpha:Bargmann}\end{equation}
Though in the standard Bargmann theory $w(q)$ is given by a positive
definite quadratic functions, we do not impose such restriction here
with the application to the gravity theory in mind.

Since $\hat \alpha$ is not hermitian, the eigenvalues of its hermitian
conjugate $\hat \alpha^\dagger$ are complex.  Let us label the
corresponding eigenvectors as
\begin{equation}
\hat\alpha^\dagger|\alpha> = \bar\alpha|\alpha>.
\end{equation}
We normalize them in terms of their $q$ representation as
\begin{equation}
<q|\alpha>= e^{i\bar\alpha q - w(q)}.
\end{equation}
These eigenvectors are not mutually orthogonal:
\begin{equation}
<\alpha'|\alpha> = \int dq\, e^{i(\bar\alpha-\alpha')q-2w(q)}.
\end{equation}
However, they are overcomplete and satisfy the simple completeness
relation
\begin{equation}
\bm{1} = \int{d\alpha d\bar\alpha\over
4\pi^2}\mu(2\r{Im}\alpha)|\alpha><\alpha|,
\label{completeness:Bargmann}\end{equation}
where $d\alpha d\bar\alpha=2d\alpha_1d\alpha_2$ for
$\alpha=\alpha_1+i\alpha_2$ and $\mu$ is given by
\begin{equation}
\mu(x):=\int dq\,e^{-ixq + 2w(iq)}.
\label{measure:Bargmann}\end{equation}

In terms of these eigenvectors let us assign a function $\Phi(\alpha)$
on the complex space to each state vector $|\Phi>$ by
\begin{equation}
\Phi(\alpha):=<\alpha|\Phi>=\int dq e^{-i\alpha q - w(q)}\tilde\Phi(q),
\label{QtoA:Bargmann}\end{equation}
where $\tilde\Phi(q)$ is a wavefunction in the $q$-representation.
Then in the representation defined by this assignment the operators
are represented as
\begin{eqnarray}
&\hat\alpha|\Phi> & \rightarrow
\quad\alpha\Phi(\alpha),\label{BargmannRep1}\\ &\hat q|\Phi> &
\rightarrow \quad i{\partial\over\partial \alpha}\Phi(\alpha).
\label{BargmannRep2}\end{eqnarray}
If $w(q)$ is chosen so that the integrand in Eq.(\ref{QtoA:Bargmann})
is square integrable for any value of $\alpha$ and any square
integrable function $\tilde \Phi(q)$, $\Phi(\alpha)$ yields a
representation of the state vector by a holomorphic function.  This is
the most important features of the Bargmann representation.

{}From the completeness relation (\ref{completeness:Bargmann}) the inner
product is expressed as
\begin{equation}
<\Phi_1|\Phi_2> =\int{d\alpha d\bar\alpha\over
4\pi^2}\mu(2\r{Im}\alpha)\overline{\Phi_1(\alpha)}\Phi_2(\alpha).
\label{InnerProduct:Bargmann}\end{equation}
The completeness relation also yields the inversion formula of
Eq.(\ref{QtoA:Bargmann}):
\begin{equation}
\tilde\Phi(q) = e^{-w(q)}\int{d\alpha d\bar\alpha\over
4\pi^2}\mu(2\r{Im}\alpha)e^{i\bar\alpha q}\Phi(\alpha).
\label{AtoQ:Bargmann1}\end{equation}
In the case $\Phi(\alpha)$ is holomorphic, it is completely determined
from its values on the real axis.  As a result the inversion formula
can be also expressed in terms of a contour integral
\begin{equation}
\tilde \Phi(q) = e^{w(q)} \int_C d\alpha\, e^{i\alpha q}\Phi(\alpha),
\end{equation}
where the integration contour $C$ is the real axis.

Some comments are in order here.  First, though we have considered a
one-dimensional system so far, this restriction is not essential and
the extension of the formulation to multi-dimensional cases are
straightforward.  Second the transformation (\ref{QtoA:Bargmann}) in
general yields holomorphic functions for a wider class of functional
space $V$ than $L_2(\RF^n)$.  In this extension the holomorphic
functions corresponding to the normalizable states are characterized
by the condition that the inner product
Eq.(\ref{InnerProduct:Bargmann}) is finite.  Third we can regard
Eqs.(\ref{BargmannRep1}) and (\ref{BargmannRep2}) as defining a
representation of the operator algebra on the space of holomorphic
functions.  In this case, we can show that if we require that the
hermitian conjugate operator to $\hat\alpha$ is given by
$\hat\alpha^\dagger=\hat \alpha - 2i w(\hat q)$ and the inner product
is expressed in the form (\ref{InnerProduct:Bargmann}), $\mu$ is
uniquely determined to be given by Eq.(\ref{measure:Bargmann}).  We
call this representation {\it the holomorphic representation}.

\subsubsection{Holomorphic connection representation}

The chiral connection $\A_j$ is a complex quantity, and has an
expression similar to the variable $\alpha$ in
Eq(\ref{alpha:Bargmann}) in the spatial gauge as explained in \S4.1.2.
This suggests that it is natural to consider the holomorphic
representation of the chiral canonical theory\CITE{JacobsonSmolin88b}.

Let us define a holomorphic functional $\Phi[\A]$ as a functional on
the space $\C$ of chiral connection $\A_j$ such that
$\Phi[\A+\sum_{a=1}^A z_a \phi^a]$ is a holomorphic function of
$(z_a)\in \CF^A$ for any set of smooth field $\phi^a_{Ij}(x)$ and any
number $A$.  Then the holomorphic representation of the chiral
operators $\hat \A_j$ and $\hat \E^j$ on the space $\C^\omega$ of
holomorphic functional $\Phi[\A]$ is defined by
\begin{eqnarray}
&\hat\A(\alpha)|\Phi> &\rightarrow \quad \A(\alpha)\Phi[\A],\\
&\hat\E(\phi)|\Phi> &\rightarrow \quad {i\over2}\int
d^3x\,\phi_{Ij}(x){\delta\over\delta\A_{Ij}(x)}\Phi[\A]:=\left.
{d\over dz}\Phi[\A + z\phi]\right|_{z=0}.
\end{eqnarray}
It is easily checked that the commutation relations
(\ref{CommutationRelation:ChiralVariables}) are corrected represented.
We call this {\it the holomorphic connection representation}.

By extending the correspondence between $\Phi(\alpha)$ and $\tilde
\Phi(q)$ in the Bargmann representation, we can formally construct a
transformation between the holomorphic functional $\Phi[\A]$ and the
wavefunctional $\tilde\Phi[\tilde e]$ in the triad representation of
the real canonical formalism defined by
\begin{eqnarray}
&\hat \E(\phi)|\Phi> & \rightarrow \quad \tilde e(\phi) \tilde
\Phi[\tilde e],\\&\hat P(\chi)|\Phi> & \rightarrow \quad
{1\over2i}\int d^3x\,\chi^{Ij}(x){\delta\over\delta \tilde e^{Ij}(x)}
\tilde \Phi[\tilde e],
\end{eqnarray}
as
\begin{equation}
\Phi[\A] = \int [D\tilde e]e^{-2i\A(\tilde e) \mp W(\tilde e)}\tilde
\Phi[\tilde e]=\int [D\tilde e] e^{-iZ(\A,\tilde e)}\tilde \Phi[\tilde e].
\label{EtoA:ConnectionRep}\end{equation}
Further we can define the formal inner product in $\C^\omega$ by
\begin{equation}
<\Phi_1|\Phi_2>:= \int [D\A
D\bar\A]\Delta[\A]\mu[4\r{Im}\A]\overline{\Phi_1[\A]}\Phi_2[\A],
\label{InnerProduct:ConnectionRep}\end{equation}
where, by noting that $W[\tilde e]$ is a homogeneous functional of
$\tilde e^j$ of degree one, $\mu[\A]$ is expressed as
\begin{equation}
\mu[X]:=\int [D\tilde e]e^{-i\tilde e(X) \pm 2iW(\tilde e)}.
\label{measure:ConnectionRep}\end{equation}
This inner product respects the reality condition formally at least if
$\Delta[\A]$ is taken to be unity.

As we will see soon, $|\Phi>\in\V_\r{inv}$ is represented by a gauge
and spatial diffeomorphism invariant functional.  If the measure
$[D\tilde e]$ is invariant under the gauge and the
spatial-diffeomorphism, this invariance of the wavefunction is
respected by the transform (\ref{EtoA:ConnectionRep}) since $Z(\A,\E)$
is an invariant functional as shown in \S4.1.2.  For the same reason
$\mu[4\r{Im}\A]$ is also an invariant functional.  This invariance,
however, makes the definition of the inner product ill-defined since
the functional integration along the noncompact gauge-orbits will
diverge if the measure $[D\A D\bar\A]\Delta[\A]$ is invariant.  Thus
in order to make the inner product (\ref{InnerProduct:ConnectionRep})
well-defined for the invariant states, we must take a gauge-variant
functional or distribution as $\Delta[\A]$.

For real $SO(3)$ gauge fields it is shown by Ashtekar and
Isham\CITE{AshtekarIsham92} using the Gel'fand spectral theory that
there exists a natural measure which leads to a well-defined inner
product for the gauge-invariant wavefunctions.  However, nothing
definite is known in the cases in which the symmetry is extended to
the gauge theories with non-compact groups or to the spatial
diffeomorphism.  Thus at present we can say nothing about whether we
can find an appropriate measure $\Delta[\A]$.

\subsubsection{Invariant states}

In the holomorphic connection representation the action of the gauge
constraint operator is expressed as
\begin{eqnarray}
&\hat\C_\r{G}(\lambda)\Phi[\A] &= -i\int d^3x\,\D_j\lambda\cdot{\delta
\over \delta \A_j}\Phi[\A] \nonumber\\
&& =i\left.{d\over
dz}\Phi[\A-z\D_j\lambda]\right|_{z=0} \nonumber\\
&& = i\delta_\lambda
\Phi[\A].
\end{eqnarray}
Hence the gauge constraint is equivalent to the condition that
$\Phi[\A]$ is gauge invariant.  Similarly, since the action of the
diffeomorphism constraint operator is expressed as
\begin{eqnarray}
&\hat\C_\r{D}(L)\Phi[\A] &= -i\int d^3x\,\Lie_L\A_j\cdot{\delta \over
\delta \A_j}\Phi[\A] \nonumber\\
&& =i\left.{d\over dz}\Phi[\A-z\Lie_L\A]\right|_{z=0} \nonumber\\
&& = i\delta_L
\Phi[\A],
\end{eqnarray}
the diffeomorphism constraint implies that $\Phi[\A]$ is
spatial-diffeomorphism invariant.

Thus $\V_\r{inv}$ is characterized as the linear space of gauge and
spatial-diffeo\-morphism invariant holomorphic functionals in the
holomorphic connection representation.  Though this characterization
is mathematically clear enough, it is too implicit to get a clear
image of the structure of the state space and find the physical
interpretation of states, like the superspace construction of
$\V_\r{inv}$ in the ADM-WD approach.  We need a more explicit
parametrization of the states. One of the most important features of
the complex canonical theory is that we can construct such a
parametrization with the help of the Wilson-loop type
integral\CITE{JacobsonSmolin88b}.

Let $\gamma$ be a closed curve
$\gamma(s)=(x^j(s))(\gamma(0)=\gamma(1))$ and define the matrix
representing the parallel transport with respect to the $SL(2,\CF)$
connection $\A_j$ along $\gamma$ from $\gamma(s_0)$ to $\gamma(s)$ by
\begin{eqnarray}
&U_\gamma(s_0,s) &:= \r{P}_\gamma
\exp\left[-\kappa^2\int_{\gamma(s_0)}^{\gamma(s)}dx^j\A_j\cdot\bg{\sigma}\right]
\nonumber\\ &&= \r{P}_\gamma
\exp\left[\mp{i\over2}\int_{\gamma(s_0)}^{\gamma(s)}dx^j\B_j\cdot\bg{\sigma}\right],
\end{eqnarray}
where $\bg{\sigma}=(\sigma_I)$ is the Pauli matrix and $\r{P}_\gamma$
is the path-ordered product whose precise definition is given by the
differential equation for $U_\gamma(s_0,s)$,
\begin{eqnarray}
&&{d\over ds}U_\gamma(s_0,s)=-\kappa^2 U_\gamma(s_0,s)\dot
x^j\A_j\cdot\bg{\sigma},\label{DifEq:U}\\ && U_\gamma(s_0,s_0)=1.
\end{eqnarray}

Since $\A_j\cdot\bg{\sigma}$ transforms by the gauge transformation
$V(x)\in SL(2,\CF)$ as
\begin{equation}
\A_j\cdot\bg{\sigma} \rightarrow V\A_j\cdot\bg{\sigma}V^{-1}
-\kappa^{-2}\partial_jV V^{-1},
\end{equation}
it is shown from the above differential equation that
$U_\gamma(s_0,s)$ transforms as
\begin{equation}
U_\gamma(s_0,s) \rightarrow
V(\gamma(s_0))U_\gamma(s_0,s)V(\gamma(s))^{-1}.
\end{equation}
Hence the quantity defined by
\begin{equation}
T_\gamma[\A]:=\Tr U_\gamma(0,1)
\label{T:def}\end{equation}
is gauge invariant.

An important property of these gauge invariant functionals is that if
the values of $T_\gamma[(i/2\kappa^2)Q]$ for all the closed loops are
given, the $SO(3)$ gauge-equivalent class of the real connection $Q$
is uniquely specified.  An elegant proof is given by Ashtekar and
Isham\CITE{AshtekarIsham92} using a general theory of embedding the
holonomy group into a compact group.  Here we give a more direct
proof.

Let $\Omega(x_0)$ be the space of loops which pass through $x_0$, and
$U_\gamma$ be the $SU(2)$ matrix obtained from $U_\gamma(0,1)$ with
$\A_j$ replaced by $(i/2\kappa^2)Q_j$.  Then $U_\gamma$ and its trace
$T_\gamma$ are expressed in terms of some vector
$\bg{\theta}_\gamma=(\theta_\gamma^I)$ as
\begin{eqnarray}
&&U_\gamma = e^{{i\over2}\bg{\theta}_\gamma\cdot\bg{\sigma}}
=\cos{\theta_\gamma\over2} +
i\hat\theta_\gamma\cdot\bg{\sigma}\sin{\theta_\gamma\over2},\\
&&T_\gamma = 2\cos{\theta_\gamma\over 2},
\end{eqnarray}
where $\theta_\gamma=|\bg{\theta}_\gamma|$ and
$\hat\theta_\gamma=\bg{\theta}_\gamma/\theta_\gamma$.  Since two
connections are equivalent if $U_\gamma$ coincides for all closed
loops in $\Omega(x_0)$ apart from the inner automorphism by a constant
$SU(2)$ matrix, the statement is proved if we can show that
$\bg{\theta}_\gamma$ is completely determined by the values of
$T_\gamma$ apart from its rotational freedom independent of $\gamma$.

First for the case $T_\gamma=\pm2$ for any loop $\gamma$, $U_\gamma$
is determined to be $U_\gamma=T_\gamma/2$, which implies the
connection is trivial apart from the possible $\bm{Z}_2$
representations of the fundamental group $\pi_1$ of the space. Next
for the case $T_\gamma$ takes values different from $\pm2$ for some
loops, let us take one such loop $\alpha$ and fix it.  By an
appropriate $SU(2)$ transformation we can put $\hat\theta_\alpha$ to
$(0,0,1)$.  Then since $T_{\gamma\alpha}$ is expressed as
\begin{equation}
T_{\gamma\alpha}=\Tr U_\gamma U_\alpha =
2\cos{\theta_\gamma\over2}\cos{\theta_\alpha\over2} - 2
\hat\theta_\gamma\cdot\hat\theta_\alpha
\sin{\theta_\gamma\over2}\sin{\theta_\alpha\over2},
\end{equation}
we can determine $\theta_\gamma^3$ by the values of $T_\gamma$,
$T_\alpha$ and $T_{\gamma\alpha}$.  If $\hat\theta_\gamma^3=\pm1$ for
any loop $\gamma$, the connection is reducible to $U(1)$-connection
and $U_\gamma$ is completely determined from $T_\gamma$.  On the other
hand if there exists a loop $\beta$ for which
$\hat\theta_\beta^3\not=\pm1$, by an appropriate $SU(2)$ rotation
which leaves $\hat\theta_\alpha$ invariant, we can put $\hat\theta_\beta$ to be
$(0,\hat\theta^2_\beta,\hat\theta^3_\beta)$.  Since
$\hat\theta^3_\beta$ is known, the value of
$\hat\theta^2_\beta$($\not=0$) is also fixed.  Hence by the same
argument as on $\hat\theta_\alpha$, we can determine $\theta_\gamma^2$
by the values of $T_\gamma$, $T_\beta$ and $T_{\gamma\beta}$.  Since
the angle $\theta_\gamma$ is determined only by $T_\gamma$, this
completes the proof.

This property of the gauge-invariant functionals $T_\gamma[\A]$
implies that any holomorphic gauge-invariant functional $\Phi[\A]$ can
be expressed as some functional on the set of $T_\gamma[\A]$, since
$\Phi[\A]$ is determined by $\Phi[i\r{Im}\A]$.  This in turn implies
that a holomorphic functional $\Phi[\A]$ which is invariant under the
$SO(3)$ gauge transformations is automatically $SO(3,\CF)$ gauge
invariant.

Another important property of $T_\gamma$ is that they are not
independent.  First from its definition the relation
\begin{equation}
T_{\gamma\alpha\alpha^{-1}}=T_\gamma
\label{T:Relation1}\end{equation}
holds for any loops $\gamma$ and $\alpha$.  Second from the relation
for $U\in SL(2,\CF)$,
\begin{equation}
\sigma_2 U\sigma_2 = \Tp U^{-1},
\end{equation}
it follows that
\begin{equation}
\sigma_2 U_\gamma(0,1)\sigma_2 = \Tp U_{\gamma^{-1}}(0,1).
\end{equation}
Hence $T_\gamma$ is independent of the orientation of the loop:
\begin{equation}
T_\gamma = T_{\gamma^{-1}}.
\label{T:Relation2}\end{equation}
Third from the equation
\begin{equation}
T_{\alpha\beta}= {1\over2}T_\alpha T_\beta -
2\hat\theta_\alpha\cdot\hat\theta_\beta
\sin{\theta_\alpha\over2}\sin{\theta_\beta\over2},
\end{equation}
noting that $\hat\theta_{\beta^{-1}}=-\hat\theta_\beta$, we obtain the
relations
\begin{eqnarray}
&& T_{\alpha\beta}=T_{\beta\alpha},\label{T:Relation3}\\ &&
T_{\alpha\beta}+T_{\alpha\beta^{-1}}=T_\alpha T_\beta.
\label{T:Relation4}
\end{eqnarray}

The final relation shows that any product of a finite number of
$T_\gamma$'s is expressed as a linear combination of $T_\gamma$'s.
Thus together with the gauge-orbit separating property of $T_\gamma$
proved above, it is expected that a wide class of holomorphic
gauge-invariant functionals $\Phi[\A]$ are expressed formally as
\begin{equation}
\Phi[\A] = \int_{\gamma\in\Omega}\mu(\gamma) \Phi[\gamma]T_\gamma[\A],
\label{PhiByT}\end{equation}
where $\mu(\gamma)$ is some fixed measure in the loop space $\Omega$
of loops with no fixed point and $\Phi[\gamma]$ is a functional on
$\Omega$.

For the class of functionals of this form, the requirement of
diffeomorphism invariance is translated to a condition on the measure
$\mu(\gamma)\Phi[\gamma]$.  To see this, first note that from the
differential equation (\ref{DifEq:U}) the change of $U_\gamma(s_0,s)$
by an infinitesimal coordinate transformation $x^j\rightarrow x^j+L^j$
is represented as
\begin{equation}
U_\gamma(s_0,s;\A-\Lie_L\A) = U_{\gamma-L}(s_0,s;\A).
\end{equation}
{}From this equation it follows that the action of a spatial
diffeomorphism $g$ on $T_\gamma$ is expressed as
\begin{equation}
g T_\gamma[\A]:=T_\gamma[g^{-1}\A]=T_{g\gamma}[\A].
\end{equation}
Hence $\Phi[\A]$ is diffeomorphism invariant if the measure
$\mu(\gamma)\Phi[\gamma]$ is diffeomorphism invariant:
\begin{equation}
\mu(g\gamma)\Phi[g\gamma]=\mu(\gamma)\Phi[\gamma].
\end{equation}
Since any diffeomorphism invariant functional on the loop space
depends on loops only through the knot invariants of them, this
implies that each diffeomorphism functional in the class represented
by Eq.(\ref{PhiByT}) is expressed formally in terms of a functional
$\Phi[L]=\Phi(L_1,L_2,\ldots)$ on the knot invariants and a fixed
measure $\mu(L)$ of the space of knot invariants as
\begin{equation}
\Phi[\A]=\int_\Omega \mu(L(\gamma))\Phi[L(\gamma)]T_\gamma[\A].
\end{equation}

Thus if we restrict the states to those expressed as in
Eq.(\ref{PhiByT}), the construction of the explicit parametrization of
the invariant states is reduced to the following two problems:
\begin{itemize}
\item[i)] Determination of all the knot invariants,
\item[ii)] Construction of the measure $\mu[L]$ on the space of knot
invariants.\end{itemize}

However, there exist some hidden difficulties in this parametrization.
First it is not clear whether the states of the form (\ref{PhiByT})
are dense in the physical state space $\H$.  Second, as
$T_\gamma[\A]$'s are not independent, $\Phi[\A]$ is not in one-to-one
correspondence with $\Phi[\gamma]$, but is parametrized by some
equivalent class of loop space functionals.  We cannot determine this
equivalent class explicitly at present since we do not know the full
linear relations among $T_\gamma$'s.

\subsubsection{Invariant operators}

In order to give the correspondence between the theory and
observations, one must construct the invariant operators
$\hat\O_\r{inv}$ from the gauge-variant quantities and determine their
operation on the invariant state space $\V_\r{inv}$.

As for the gauge-invariant operators one possible generating set can
be constructed with the help of the Wilson-loop type integrals.  To
see this, we first show that the set of gauge-invariant functionals on
the complex phase space, $T_\gamma[\A]$ defined by Eq.(\ref{T:def})
and
$T^{(n)}_\gamma[\A,\E]=T^{j_1,\cdots,j_n}_\gamma(s_1,\ldots,s_n)[\A,\E]$
defined by
\begin{eqnarray}
&T^{j_1,\cdots,j_n}_\gamma(s_1,\ldots,s_n)[\A,\E]&:=\Tr\left[U_\gamma(0,s_1)\E^{j_1}(\gamma(s_1))\cdot\bg{\sigma}U_\gamma(s_1,s_2)\cdots\right.
\nonumber\\ &&\qquad
\left.\cdots\E^{j_n}(\gamma(s_n))\cdot\bg{\sigma}U_\gamma(s_n,1)\right]
\end{eqnarray}
generate the holomorphic subset of $\O_\r{G}(\Gamma_{\CF})$.  Due to
the holomorphic requirement we can restrict the consideration to the
subspace of $\Gamma_{\CF}=\{(\E,\A)\}$ such that $\A$ is a pure
imaginary $SO(3)$ connection, and the gauge group to $SO(3)$.  Under
this restriction $T_\gamma[\A]$ separates the gauge equivalent class
of $\A$.  Hence what we have to prove is that we can determine $\E$
from the values of $T^{(1)}_\gamma$, $T^{(2)}_\gamma$, $\ldots$, for a
fixed $\A$ modulo the $SO(3)$ gauge transformations which leave $\A$
invariant.

We only need to consider the case in which the base manifold $M$ is
connected.  Let us fix a point $x_0\in M$ and assign to each point
$x\in M$ a path $\gamma_x$ connecting $x_0$ and $x$.  Then $\E^j(x)$
is completely determined by the values of $\tilde \E^j(x)$ defined by
\begin{equation}
\tilde
\E^j(x)\cdot\bg{\sigma}:=U_{\gamma^{-1}_x}\E^j(x)\cdot\bg{\sigma}U_{\gamma_x},
\end{equation}
which transforms under the gauge transformation $V(x)\in SU(2)$ as
\begin{equation}
\tilde \E^j(x)\cdot\bg{\sigma} \rightarrow V(x_0) \tilde
\E^j(x)\cdot\bg{\sigma} V(x_0)^{-1}.
\end{equation}
In terms of $U_\alpha$ used in the previous subsection
$T^{(1)}_{\gamma_x\alpha\gamma_x^{-1}}$ with $\E$ inserted at $x$ is
expressed as
\begin{equation}
\Tr\tilde \E^j(x)\cdot\bg{\sigma} U_\alpha = \tilde \E^j\cdot \hat\theta_\alpha
\sin{\theta_\alpha\over2}.
\end{equation}
Hence if the vector space spanned by
$\{\hat\theta_\alpha|\alpha\in\Omega_{x_0}\}$ is three-dimensional,
$T^{(1)}_{\gamma_x\alpha\gamma_x^{-1}}$ determines $\E^j$ completely.

On the other hand if the dimension of the vector space is smaller than
3, the connection $\A$ is reducible to a $U(1)$ or trivial one.  First
for the case $U(1)$ reductive, $U_\alpha$ can be put in a form
$U_\alpha=\cos{\theta_\alpha\over2}+i\sigma_3
\sin{\theta_\alpha\over2}$.  Hence $\tilde \E^{3j}$ is determined by
$T^{(1)}_{\gamma_x\alpha\gamma_x^{-1}}$.  Further in this gauge the
residual gauge transformation $V$ is written as
$V=\cos{\theta\over2}+i\sigma_3 \sin{\theta\over2}$ with a constant
$\theta$.  The independent combinations of $\tilde\E^j(x)$ invariant
under this constant rotation are given by $\tilde\E^{3j}$ and
$\sum_{I=1,2}\tilde \E^{Ij}(x)\tilde\E^{Ik}(x')$.  The latter can be
replaced by
$\Tr[\tilde\E^j(x)\cdot\bg{\sigma}\tilde\E^k(x')\cdot\bg{\sigma}]$
which is of the type $T^{(2)}_\gamma$.  Hence the gauge equivalent
class of $\E^j$ is completely determined by $T^{(1)}_\gamma$ and
$T^{(2)}_\gamma$ in this case.  Finally for the case the connection is
trivial, the residual gauge freedom is represented by an arbitrary
constant matrix of $SU(2)$ under the gauge $\A=0$.  The independent
combinations of $\tilde \E^j$ invariant under this residual gauge
freedom is given by $\tilde\E^j\cdot\tilde\E^k$.  Hence the gauge
equivalent class is determined by $T^{(2)}_\gamma$. $T^{(1)}_\gamma$
vanishes in this case. This completes the proof of the statement.

Thus we can take $T_\gamma[\A]$ and $T^{(1)}_\gamma[\A,\E]$ as
fundamental gauge-invariant variables in the classical theory since
the reductive connections form a subset of measure zero(for a more
detailed treatment on the completeness of the loop variables, see
\cite{GoldbergLewandowskiStornaiolo92}).  This suggests that
$T_\gamma[\hat\A]$ and $T^{(1)}_\gamma[\hat\A,\hat\E]$ can be taken as
the fundamental operators in $\hat\O_\r{G}$.

In contrast to the construction of the invariant state, there is a
serious problem in constructing a generating set of diffeomorphism
invariant operators from these loop-integral-type operators.  Of
course as for the operators depending only on $\hat\A$, a formal
construction is possible with the help of the loop space integration
introduced in the previous subsection as
\begin{equation}
\hat T_\Phi= \int _\Omega \mu(L(\gamma))\Phi[L(\gamma)]T_\gamma[\hat\A].
\end{equation}
However, this formal prescription does not work in constructing
invariant operators depending on $\hat\E^j$ from
$T^{(n)}_\gamma[\hat\A,\hat\E]$ since we need covariant tensor
densities to make them scalar.  The only covariant tensor density
which does not include the inverse of $\hat\E^j$ is $\epsilon_{jkl}$.
However, this is not sufficient to produce all the invariant
operators(cf. \cite{Smolin92}).

For example let us consider the gauge and diffeomorphism invariant
quantity $Z$ introduced in \S4.1.2.  This quantity has a clear
physical meaning. In fact from Eq.(\ref{PbyK}) $Z$ is expressed in
terms of the ADM variables as
\begin{equation}
Z(\A,\tilde e) = 2 \int d^3x\,\tilde e^j\cdot P_j =
{1\over\kappa^2}\int d^3x\,\sqrt{q}K.
\end{equation}
Hence noting that $\sqrt{q}K$ is written from Eq.(\ref{K:def}) as
\begin{equation}
\sqrt{q}K = {1\over N}\left(-\dot{\sqrt{q}} + \partial_j(\sqrt{q}N^j)\right),
\end{equation}
we see that $Z$ represents the expansion rate of the total spatial
volume in the spatially compact cases.

Since the definition of $Z$ can be rewritten as
\begin{eqnarray}
&Z & = \pm {i\over2\kappa^2}\int
d^3x\,q^{-1}q_{jk}(\E^j\times\D_l\E^k)\cdot\E^l \nonumber\\ && = \pm
{i\over2\kappa^2}\int d^3x\,(\E\cdot\E)^{-1}_{jk}\psi^{jk},
\label{ZbyEandpsi}\end{eqnarray}
where $(\E\cdot\E)^{-1}_{jk}$ is the inverse matrix of
$\E^j\cdot\E^k$,
$Z$ is guaranteed to be real by the
generic reality conditions in the classical framework.  This expression,
however, causes a problem in the quantum framework since it contains the
inverse of $\E^j$.  In particular it seems to be quite hard to give
$Z$ a regular expression in terms of the loop variables. Nevertheless,
as it is a natural physical quantity, it should not be excluded from
$\O_\r{inv}$.  This simple example suggests that we should allow
quantities with some singularities to be included in $\hat O_\r{inv}$,
though all the expressions are simple polynomials at the stage where
the spatial diffeomorphism invariance is not respected.

Beside this there occurs another subtle problem if we take the loop
variables as the fundamental operators. To see this, let us consider
the problem to express in terms of the loop variables the local
gauge-invariant quantities written in terms of $\E^j$ and $\A_j$.  Due
to the local nature of the gauge transformation, most of them are
written as products of fields at the same spatial point.  This implies
that the local gauge-invariant quantities can be constructed only as a
limit.  For example $qq^{jk}$ is expressed as
\begin{equation}
q(x)q^{jk}(x)=\E^j(x)\cdot\E^k(x)={1\over2}\lim_{\gamma\rightarrow
x}T^{jk}_\gamma,
\end{equation}
where $\gamma\rightarrow x$ implies to shrink the loop to a point $x$.

The expressions for the constraint functions become much more
intricate because they contain the curvature tensor.  To derive them,
consider a one-parameter-family of loops
$\gamma_\tau(s)=(x^j(s,\tau))$ with a common base point
$\gamma_\tau(0)$.  Then by differentiating Eq.(\ref{DifEq:U}) by
$\tau$, we obtain the following differential equation for
$U=U_{\gamma_\tau}(s_0,s)$:
\begin{equation}
\partial_s\left[(\partial_\tau U + \kappa^2U\bg{\sigma}\cdot\A_j\partial_\tau
x^j)U^{-1}\right]={i\over2}U\bg{\sigma}\cdot\F_{jk}U^{-1}\partial_s
x^j\partial_\tau x^k.
\end{equation}
If the curves $\gamma_\tau$ differ only within an interval
$s_0<s<s_1$, integration of this equation yields
\begin{equation}
\partial_\tau U(s_0,s_1)={i\over2}\int_{s_0}^{s_1}ds\,
U(s_0,s)\bg{\sigma}\cdot\F_{jk}U(s,s_1)  \partial_s x^j\partial_\tau x^k.
\end{equation}
{}From this equation we find that the variation of $T^{(n)}_\gamma$ by
an infinitesimal deformation to the $k$-coordinate direction of the
curve $\gamma$ at a point $\gamma(s)$, when divided by the area swept
by the deformation, yields $T^{(n+1)}_\gamma$ obtained by inserting
$\pm(i/2)\dot\gamma^j\F_{jk}\cdot\bg{\sigma}$ to $T^{(n)}_\gamma$ at
$\gamma(s)$.  Let us denote this area derivative by
$\dot\gamma^j(s)\Delta_{jk}(s)T^{(n)}_\gamma$.  Then the momentum and
the hamiltonian constraint functions are expressed in terms of the
loop variables as
\begin{eqnarray}
&&\C_{\r{M}j}(x) = -{1\over\kappa^2}\lim_{\gamma_j\rightarrow
x}\Delta_{jk}(s)T^k_\gamma(s),\label{MomentumConstraintByLoop}\\
&&\C_{\r{H}}(x) = \pm{1\over\kappa^2}\sum_j\lim_{\gamma_j\rightarrow
x}\Delta_{jk}(s)T^{jk}_\gamma(s,s),
\label{HamiltonianConstraintByLoop}
\end{eqnarray}
where the loop $\gamma_j$ is chosen so that $\dot
\gamma_j^k(s)=\delta_j^k$.

\subsection{Loop Space Representation}

In the previous subsection we have seen that the classical loop
variables $T_\gamma[\A]$ and $T^{(1)}_\gamma[\A,\E]$ are complete at
least in the space of the holomorphic functionals on the complex phase
space and that a family of gauge-invariant operators can be
constructed from them.  We have also seen that a wide class of
gauge-invariant states are represented by functionals on loop space at
least formally. These results strongly suggest that one may construct
a quantum theory at the stage II$_Q$ by directly quantizing the loop
variables and representing them on the space of functionals on the
loop space without referring to the connection or the tetrad.
Actually Rovelli and Smolin showed that this observation is correct
and proposed a new quantization program of the chiral canonical theory
called {\it the loop space representation}\CITE{RovelliSmolin88}.
Since lots of excellent reviews have been published on this
approach\CITE{Smolin89,Rovelli91a,Ashtekar91a,Smolin92}, we present here
only its basic features and potential problems.

\subsubsection{Algebra of loop variables}

In order to construct a quantum theory based only on the loop
variables, we must first show that they form a closed algebra under
the Poisson bracket in the classical framework.  It is easy to see
that this requirement is satisfied at least formally.  In fact the
Poisson brackets among $T_\gamma[\A]$ and $T^j_\gamma[\A,\E]$ can be
calculated from Eqs.(\ref{PoissonBracket:Chiral1}) and
(\ref{PoissonBracket:Chiral2}) as
\begin{eqnarray}
&&\{T_\alpha, T_\beta \}=0, \label{PoissonBracket:Loop1} \\
&&\{T_\alpha,T^j_\beta(s)\}=\kappa^2\Delta^j[\beta,\alpha](s)(T_{\alpha\beta}-T_{\alpha\beta^{-1}}),\label{PoissonBracket:Loop2}
\\
&&\{T^j_\alpha(s),T^k_\beta(t)\}=
\kappa^2\Delta^k[\beta,\alpha](t)(T^j_{\alpha\#_t\beta}-T^j_{\alpha\#_t\beta^{-1}})(u(s))
\nonumber \\ &&\qquad
-\kappa^2\Delta^j[\alpha,\beta](s)(T^k_{\beta\#_s\alpha}-T^k_{\beta\#_s\alpha^{-1}})(u(t)),\label{PoissonBracket:Loop3}
\end{eqnarray}
where $\alpha\#_t\beta$ denotes a loop formed from $\alpha$ and
$\beta$ by cutting both and reconnecting at $\beta(t)$ respecting the
orientations, $u(s)$ is the value of the parameter $u$ at the point
$\alpha(s)$ of the curve $\alpha\#_t\beta$ normalized to the range
$0\le u\le1$, and $\Delta^j$ is defined by
\begin{equation}
\Delta^j[\beta,\alpha](s):={1\over2}\int_0^1 dt
\dot\alpha^j\delta^3(\beta(s)-\alpha(t)).
\end{equation}

Since these equations do not contain $\A_j$ or $\E^j$ explicitly, they
guarantee that $T_\gamma$ and $T^{(1)}_\gamma$ form a closed algebra
with respect to the Poisson bracket.  However, they have one
uncomfortable feature: they are not mathematically well-defined since
the coefficient $\Delta^j$ is singular. This implies that we must
introduce some regularization in order to make them well-defined.  A
similar situation occurs in the field theories when local fields are
taken as the fundamental variables as in the connection
representation.  There the expressions can be made well-defined by
smoothing the local variables by smooth test functions.  In the
present case this method cannot be applied because the loop variables
has no explicit dependence on the spatial coordinate.

One solution to this difficulty is to replace the loop variable
$T^{(1)}_\gamma$ by the strip variable which depends only on a
one-parameter family of loops
$\Sigma=\{\gamma_\tau(s)=(x^j(s,\tau))|0\le s,\tau \le1\}$ and defined
by
\begin{equation}
T^{(1)}_\Sigma := \int_0^1ds\int_0^1d\tau\,\partial_s x^j\partial_\tau
x^k\epsilon_{jkl} T^l_{\gamma_\tau}(s).
\end{equation}
Since the two-dimensional smoothing completely eliminates the
$\delta$-function singularity, the Poisson brackets of the strip
variables are given by regular expressions:
\begin{eqnarray}
&&\{T_\alpha, T^{(1)}_\Sigma\} = {1\over2}\kappa^2
\sum_{\Sigma\cap\alpha}\r{sign}(\alpha:\Sigma)(T_{\Sigma\#\alpha}-T_{\Sigma\#\alpha^{-1}}),
\label{PoissonBracket:Strip1}\\
&&\{T^{(1)}_{\Sigma_1}, T^{(1)}_{\Sigma_2}\} =
\kappa^2\sum_{\Sigma_1\cap\Sigma_2}\r{sign}(\Sigma_1:\Sigma_2:\Sigma_1\#\Sigma_2)\left(T^{(1)}_{\Sigma_1\#\Sigma_2}
-T^{(1)}_{\Sigma_1\#\Sigma_2^{-1}}\right).\nonumber\\
&&
\label{PoissonBracket:Strip2}
\end{eqnarray}
In the first equation $\r{sign}(\alpha:\Sigma)$ denotes the
orientation of the three vectors $\{\partial_s \alpha,
\partial_t\gamma, \partial_\tau\gamma\}$ at each intersection point of
$\alpha$ and $\Sigma=\{\gamma_\tau(t)\}$, $\Sigma\#\alpha$ represents
a curve $\gamma\#\alpha$ formed from $\gamma\in\Sigma$ passing through
the intersection point, and the summation is taken over all the
intersection points.  In the second equation $\Sigma_1\#\Sigma_2$
represents a strip formed from one-parameter family of the pairs of
curves crossing at each connected segment in $\Sigma_1\cap\Sigma_2$,
and $\r{sign}(\Sigma_1:\Sigma_2:\Sigma_1\#\Sigma_2)$ is a sign
determined as follows: first fix the $\tau$-direction of the
intersection segment to form $\Sigma_1\#\Sigma_2$; then the relative
direction of this and the corresponding original $\tau$-directions of
$\Sigma_1$ and $\Sigma_2$ at the segment determines signs,
$\r{sign}(\Sigma_1:\Sigma_1\#\Sigma_2)$ and
$\r{sign}(\Sigma_2:\Sigma_1\#\Sigma_2)$; finally multiply these two
signs and the sign corresponding to the orientation of the triplet of
the segment vector and the tangent vectors to curves in $\Sigma_1$ and
$\Sigma_2$.

We can construct smoothened strip variables from the higher-order
variables $T^{(n)}$ in a similar way.  In particular the momentum and
the hamiltonian constraint functions can be expressed as some limits
of these quantities starting from Eqs.(\ref{MomentumConstraintByLoop})
and (\ref{HamiltonianConstraintByLoop}).  However, we do not give
their explicit expressions here because they are complicated and we do
not need them later.

\subsubsection{Representation on the multi-loop space}

Since the Poisson brackets among $T_\gamma$ and $T^j_\gamma(s)$(or
$T^{(1)}_\Sigma$) are written by their linear combination, we can
determine the commutation relations among the operators $\hat
T[\gamma]$ and $\hat T^j[\gamma](s)$($\hat T^{(1)}[\Sigma])$
corresponding them without ambiguity and construct the operator
algebra $\hat \O_\r{G}$.  Thus the remaining task in constructing the
quantum theory at the stage II$_Q$ is to find an appropriate
representation of them on the loop space functionals.

For that purpose let us recall Eq.(\ref{PhiByT}) which connects the
holomorphic functionals to loop space functionals.  Of course this
expression cannot be used to find the required representation since it
does not fix $\Phi[\gamma]$ due to overcompleteness of
$T_\gamma[\A]$'s. However, this situation reminds us of the relation
of the wavefunctions in the $q$-representation, $\tilde\Phi(q)$, and
in the Bargmann representation, $\Phi(\alpha)$.  If we compare the
equations in \S4.2.1 with the equations in the present case, we find
that $\tilde \Phi(q)$, $\Phi(\alpha)$ and $<q|\alpha>$ correspond to
$\Phi[\A]$, $\Phi[\gamma]$ and $T_\gamma[\A]$, respectively.  These
correspondences and Eq.(\ref{QtoA:Bargmann}) suggests the transform
\begin{eqnarray}
&\Phi[\gamma] &= \int [D\A
D\bar\A]\Delta[\A]\mu[4\r{Im}\A]\overline{T_\gamma[\A]}\Phi[\A]
\nonumber\\ && = <0|\hat T[\gamma]^\dagger|\Phi>. \label{LoopByA}
\end{eqnarray}
This formal expression is called {\it the Rovelli-Smolin
Transform}(cf. \cite{AshtekarIsham92}).

As shown in \S4.2.3, the products of $\overline{T_\gamma}$'s are
expressed as a linear combination of them.  Since the proof given
there can be applied even if $\E^j\cdot\bg{\sigma}$ is inserted into
the loop integral, this property holds also for
$\overline{T^{(n)}_\gamma}$.  Hence this transform can be used to find
a representation of the loop variables on the functional space on
$\Omega$. There exists, however, one subtle problem in this method: an
arbitrary functional $\Phi[\gamma]$ cannot be taken as representing a
state since $T_\gamma$'s are not linearly independent.  This situation
is similar to the Bargmann representation: there the wavefunctions
were required to be holomorphic.  In the present case, however, the
constraints to be satisfied by the functionals are not fully known
since the relations among $T_\gamma$'s are not fully determined yet.

The method adopted by Rovelli and Smolin to deal with this problem was
to utilize a functional $\Phi$ on the multi-loop space defined as a
set of functionals
\begin{equation}
\Phi = \{\Phi_0, \Phi_1[\alpha], \Phi_2[\beta_1,\beta_2], \ldots\},
\end{equation}
where $\Phi_n[\gamma_1,\ldots,\gamma_n]$ is a functional on $\Omega^n$
invariant under the permutations of $\gamma_1$, $\ldots$, $\gamma_n$.
Let $\V_\r{ML}$ be a linear space consisting of functionals of this
type.  Then by assigning to $\Phi_n[\gamma_1,\ldots,\gamma_n]$ the
expression obtained from Eq.(\ref{LoopByA}) by replacing $\hat
T[\gamma]$ by $\hat T[\gamma_1]\ldots \hat T[\gamma_n]$ we find that
$\hat T[\alpha]$ and $\hat T^{(1)}[\Sigma]$ are naturally represented
on $\V_\r{ML}$ as
\begin{eqnarray}
&&\hat
T[\alpha]\Phi[\gamma_1,\cdots,\gamma_n]:=\Phi[\alpha,\gamma_1,\cdots,\gamma_n],\\
&&\hat T^{(1)}[\Sigma]\Phi[\gamma_1,\cdots,\gamma_n]:=-i{\kappa^2\over2}
\sum_{a=1}^{n}\sum_{\Sigma\cap\gamma_a}\r{sign}(\gamma_a:\Sigma)\nonumber\\
&& \qquad\qquad \times(\Phi[\gamma_1,\ldots,\Sigma\#\gamma_a,\cdots]
-\Phi[\gamma_1,\ldots,\Sigma\#\gamma_a^{-1},\cdots]).
\end{eqnarray}
It is easily checked that the commutation relations corresponding to
Eqs.(\ref{PoissonBracket:Loop1})-(\ref{PoissonBracket:Loop3}) are
satisfied.

As stated above, the loop variables are not independent but related at
least by Eqs.(\ref{T:Relation1}), (\ref{T:Relation2}),
(\ref{T:Relation3}) and (\ref{T:Relation4}).  The operation defined
above are consistent with them only if the functional $\Phi$ satisfies
the following equations:
\begin{eqnarray}
&&\Phi[\gamma,\cdots]=\Phi[\gamma^{-1},\cdots],\label{Phi[loops]:Relation1}\\
&&\Phi[\alpha\beta,\cdots]=\Phi[\beta\alpha,\cdots],\label{Phi[loops]:Relation2}\\
&&\Phi[\alpha,\beta,\cdots]=\Phi[\alpha\beta,\cdots]+\Phi[\alpha\beta^{-1},\cdots],\label{Phi[loops]:Relation3}\\
&&\Phi[\alpha\gamma\gamma^{-1},\cdots]=\Phi[\alpha,\cdots].\label{Phi[loops]:Relation4}
\end{eqnarray}
Of course these do not exhaust all the relations.  Further relations
are obtained from the consistency of the operations of $\hat T$ and
$\hat T^{(1)}$ on these relations.  Besides there may exist others
derived from relations among the loop variables presently unknown.
The true representation space $\V_\r{G}$ is given by the quotient
space of $\V_\r{ML}$ by the linear subspace $\R_\r{ML}$ spanned by
these full relations.

{}From the relation (\ref{Phi[loops]:Relation3}) it follows that each
multi-loop functional $\Phi$ is completely determined by its
single-loop component(and null-loop component) as noted above.
However, the full reduction to it is not possible until the complete
knowledge on $\R_\r{ML}$ are obtained.  Thus in the multi-loops space
approach the construction of a representation is divided to two steps:
the construction of the formal representation of the operators and the
determination of $\R_\r{ML}$.  In this approach problems independent
of the structure of $\R_\r{ML}$ can be studied before the complete
construction.  This is the advantage of the multi-loop approach over
the single-loop approach.

In order to complete the quantization program, one must construct
invariant operators $\hat\O_\r{inv}$ and an invariant state space
$\V_\r{inv}\subset\V_\r{G}$, and introduce an inner product into the
latter.  As is expected from the argument in \S4.2.4, the
diffeomorphism invariant states are represented by the loop-space
functionals which depend only on the link invariants.  Hence the
second problem is reduced to finding all the link invariants
$L=(L_1,L_2,\ldots)$ and one invariant measure $\mu(\gamma)$ on the
loop space.  On the other hand the full construction of the
diffeomorphism invariant operators cannot be reduced to such a
well-defined problem.  Of course, as for the operators corresponding
to the variables expressed only in terms of $\A_j$'s, the invariant
operators are constructed by taking the average of products of loop
variables $\hat T[\gamma]$ with diffeomorphism invariant measures on
the loop space.  However, as for those containing $\E^j$'s, we need
additional covariant tensors to make scalar quantities.  The only such
tensor we have at hand is $\epsilon_{jkl}$.  Though a few interesting
geometrical invariants have been constructed with the help of
$\epsilon_{jkl}$\CITE{Smolin92}, it is obvious that such quantities do
not exhaust the full set of invariant operators as was discussed in
\S4.1.2. Finally the construction of the inner product is extremely
difficult in this approach, since we do not know how to express the
reality condition in terms of the loop variables.  On this point the
approach based on the connection representation appears to be more
hopeful since the formal expression satisfying the reality condition
can be given at least as explained in \S4.2.

\subsection{Solutions to the Hamiltonian Constraint}

As explained in \S3, all the dynamical information of the theory is
contained in the quantum Hamiltonian constraint in the canonical
quantum gravity.  In the ADM-WD approach or the real tetrad approach
no exact solutions to it has been so far obtained except for the
minisuperspace models.  In contrast an infinite number of exact
solutions have been found in the chiral canonical approach.  This
point is one of the most fascinating features of the quantization
program based on the chiral canonical theory.  In this subsection we
briefly summarize the present status of the problem in this approach.

\subsubsection{Loop integral solutions}

As discussed in \S3.2, the Hamiltonian constraint cannot be applied to
the quantum states. In order to resolve this difficulty, we proposed
there the probability amplitude functional formalism in which the
constraint is imposed on the probability functionals on the state
space. Though this formalism is apparently different from the
conventional one, there exists no difference between them at least as
far as the problem of solving the constraint is concerned, as noticed
in \S3.2.2.

In order to apply this formalism to the quantum theory in the
holomorphic connection representation, however, we must eliminate some
ambiguities concerned with the explicit expression of the constraint.
To see this, let us assume that the probability amplitude
$<\Psi|\Phi>$ is expressed in terms of a holomorphic functional
$\Psi[\A]$ as
\begin{equation}
<\Psi|\Phi> =
\int_{\C}[\D\A\D\bar\A]\mu[4\r{Im}\A]\Delta[\A]\overline{\Psi[\A]}\Phi[\A].
\end{equation}
Since the functional $\Delta[\A]$ should have a gauge-fixing nature,
it is expected that the right-hand-side of this equation depends not
on the values of $\Psi[\A]$ on the whole connection space $\C$ but in
the neighborhood of a subspace $\C_0$ transversal to all the gauge and
diffeomorphism orbits.  Hence holomorphic functionals which coincide
with each other around $\C_0$ will give the same probability amplitude
functional.  This ambiguity is removed if we require that $\Psi[\A]$
is gauge and diffeomorphism invariant like $\Phi[\A]$.

Another ambiguity arises from the non-hermitian nature of the
Hamiltonian constraint operator: we can formulate the constraint
either as $<\Psi|\hat \C_\r{H}=0$ or as $<\Psi|\hat
\C_\r{H}^\dagger=0$.  For the second choice, from the above integral
expression, the weak quantum Hamiltonian constraint is written as
\begin{equation}
\hat \C_\r{H}\Psi[\A]=0,
\end{equation}
while for the first choice we get an expression with $\hat \C_\r{H}$
replaced by $\hat \C_\r{H}^\dagger$.  Clearly the former equation,
which is the one commonly used, is much more tractable than the latter
choice in the connection representation.

Finally there is an ambiguity associated with the operator ordering.
As shown in \S4.1, $[\hat \C_\r{H}, \hat \C_\r{H}]$ is written in the
form $\hat \C_{\r{D}k}\hat\E^j\cdot\hat\E^k$ for the ordering
$\hat\A\cdots\hat\E\cdots$.  Thus for this ordering the consistency of
the Hamiltonian constraint yields a new constraint.  This problem does
not occur for the reversed ordering.

In most of the work on the Hamiltonian constraint the ordering
$\hat\A\cdots\hat\E\cdots$ is adopted.  In this ordering it is quite
easy to find an infinite family of solutions by taking linear
combinations of the functionals
\begin{equation}
\Psi[\A;\gamma_1,\cdots,\gamma_n]=T_{\gamma_1}[\A]\cdots T_{\gamma_n}[\A].
\end{equation}
It is because this functional itself is a solution provided that the
loops $\gamma_1$, \ldots, $\gamma_n$ are smooth and do not intersect
with each other.  In fact, since this functional has a form
$\Psi[\A]=f[X^I_a]\;(X^I_a(x)=\A^I_j(x)V^j_a(x))$ where $V^j_a$ is a
tangent vector to the loop passing through $x$, the operation of $\hat
\C_\r{H}$ on it is written as
\begin{equation}
8\kappa^2\hat\C_\r{H}(x)\Psi[\A]=\sum_{a,b}V^j_a(x)V^k_b(x)\epsilon^{IJK}\F^I_{jk}(x)
{\delta^2 f\over \delta X^a_J(x) \delta X^b_K(x)},
\end{equation}
where the summation is taken over all the tangent vectors to curves
passing through $x$.  The right-hand side of this equation vanishes
for the non-intersecting case for which $a=b$.  To be precise, this
proof is too rough in that the functional derivatives of the loop
variables produce distributional singular terms.  For the exact proof
taking account of the regularization see \cite{JacobsonSmolin88b}.

In the cases the loops have intersections one must take linear
combinations of $\Phi[\A;\gamma_1,\cdots,\gamma_n]$ because the
right-hand side of the above equation does not vanish any longer.
Fortunately the condition for its cancellation can be shown to be
expressed by algebraic relations among the coefficients.  So far
solutions containing up to 5 intersecting loops have been explicitly
constructed and a general algorithm to find a general $n$-loop
solutions is found\CITE{Husain89,BruegmannPullin91}.

Though this family of solutions appear to be quite generic, they have a few
unpleasant features.  First they are obviously not diffeomorphism invariant.
It is often stated that this is a difficulty of them.  In our
formalism, however, this is not a difficulty because $\Psi[\A]$ as a
holomorphic representation of the probability amplitude functional need not be
diffeomorphism invariant.
Second it is shown that $<\Psi|\hat q(x)$ vanishes for all
the known solutions in this family\CITE{BruegmannPullin91}.
This is not an obvious
result since $\hat q(x)\Psi[\A]$ which is expressed in the relevant
cases as
\begin{equation}
\hat
q(x)\Psi[\A]=-{i\over48}\epsilon_{jkl}\sum_{a,b,c}V^j_a(x)V^k_b(x)V^l_c(x)\epsilon_{IJK}{\delta^3 f\over \delta X^I_a(x)\delta X^J_b(x) \delta X^K_c(x)}
\end{equation}
is not expected to vanish for the cases where more than two loops
intersect at a points.  One possible reason of this unexpected result
is the existence of an additional constraint for the ordering
$\hat\A\cdots\hat\E\cdots$ noticed above.

Though the latter feature apparently suggests that the solutions
represent spacetimes with degenerate spatial metrics, it is not really
clear whether it implies that the solutions are unphysical, since
$\hat q(x)$ is not a diffeomorphism-invariant operator.  Some authors
argue that the local operators such as $\hat q(x)$ are not good
operators because, when one tries to construct finite operators
corresponding to them from the loop variables by the point-splitting
regularization, the results depend on the background metric used in
the regularization\CITE{Smolin92}.  They instead propose some finite
geometrical operators which can be utilized to extract information on
spacetime structures for the above solutions under the assumption that
the classical spacetimes are defined only on scales which contain lots
of loops.

Finally we comment on the exact solutions in the loop space
representation found by Br\"ugmann, Gambini and
Pullin\CITE{BruegmannGambiniPullin92a,BruegmannGambiniPullin92b,BruegmannGambiniPullin92c}.
They first constructed an exact solution to all the quantum
constraints in terms of two knot invariants, for which $(\det
q)^{1/2}(x)\Psi[\gamma_1\gamma_2\gamma_3]$ does not vanish at a point
$x$ where the three loops intersect non-degenerately.  Later, by
taking the Rovelli-Smolin transform of the non-degenerate solution for
the pure gravity with non-vanishing cosmological constant $\Lambda$
given in the next subsection and by expanding it in terms of
$\Lambda$, they found that their first solution is a member of an
infinite series of exact solutions to the constraints with
$\Lambda=0$, all of which are intimately connected with the Jones
polynomial.  Though their results are quite interesting, it is not
clear at present whether they are genuine solutions because they are
expressed by the single-loop functional on which the constraints are
not fully known at present as stated in \S4.3.

\subsubsection{Non-degenerate solution in the connection representation}

In the holomorphic connection representation only one exact
non-degenerate solution to the quantum Hamiltonian constraint has been
found so far\CITE{Kodama90}.  It is a solution in the case of the pure
gravity with non-vanishing cosmological constant $\Lambda$ and given
by
\begin{eqnarray}
&&\Psi_\Lambda[\A] = e^{-iS_\Lambda};\\
&&S_\Lambda={2\kappa^2\over\Lambda}\int
d^3x\,\epsilon^{jkl}\left[\mp3i\A_j\cdot\partial_k\A_l
+2\kappa^2\A_j\cdot(\A_k\times\A_l)\right].
\end{eqnarray}
Since $S_\Lambda$ is the Chern-Simons functional, its functional
derivative with respect to $\A_j$ is proportional to the curvature:
\begin{equation}
{\delta S_\Lambda \over \delta \A} = {3\over
\Lambda}\epsilon^{jkl}\F_{kl}.
\label{dS0/dA}\end{equation}
Hence from the Bianchi identity $\epsilon^{jkl}\D_j\F_{kl}\equiv0$, it
satisfies the gauge constraint.  Further it is obviously
diffeomorphism invariant.  Finally since Eq.(\ref{dS0/dA}) is written
as
\begin{equation}
\hat \E^j \Psi_\Lambda[\A]={3\over
2\Lambda}\epsilon^{jkl}\F_{jk}\Psi_\Lambda[\A],
\end{equation}
the Hamiltonian constraint which is expressed in the present case as
\begin{equation}
\hat\C_\r{H}\Psi = -{1\over2\kappa^2}(\hat\E^j\times\hat\E^k)\cdot
\left(\hat\F_{jk}-{\Lambda\over3}\epsilon_{jkl}\hat\E^l\right)\Psi,
\end{equation}
is trivially satisfied for the ordering $\hat\E\cdots \hat\A\cdots$.

Since we do not have the full knowledge on the structure of the state
space and the invariant operators, we cannot explore the physical
interpretation of this solution exactly.  However, we can get some
insights by studying the behavior WKB orbits corresponding to the
wavefunction since the WKB structure is a gauge and spatial
diffeomorphism invariant property of the wavefunction.

As is seen from the above proof , $S_\Lambda[\A]$ is an exact solution to
the classical Hamilton-Jacobi equation.  Hence the WKB orbits
corresponding to $\Psi_\Lambda[\A]$ are given by the solutions to the
equations
\begin{eqnarray}
&& \F_{jk}={\Lambda\over3}\epsilon_{jkl}\E^l,\label{WKB:gen1}\\ &&
\dot \A_j=\{\A_j,H\}.
\end{eqnarray}
{}From Eq.(\ref{ChiralLagrangian:CanonicalForm}) the latter equation is
written as%
\begin{equation}
\F_{tj}=N^k\F_{kj} \pm i\utilde N \E^k\times\F_{jk}.
\end{equation}
Comparing the expression for $\chiral\F_{0I}$ calculated from these
equations taking account of Eq.(\ref{WKB:gen1}) with
Eq.(\ref{ChiralSigmaByE}), we find that the WKB equations are
equivalent to the equation
\begin{equation}
\chiral \F_{0I} = {\Lambda\over 6}\chiral\Sigma_{0I}.
\label{LorentzianWKBEq:chiral}\end{equation}

If we require that $\theta^a$ is real and $\chiral\F_{0I}$ is the
chiral combination obtained from the real curvature form $F_{ab}$,
this equation is further rewritten as
\begin{equation}
F_{ab}={\Lambda\over6}\Sigma_{ab}.
\label{LorentzianWKBEq}\end{equation}
By applying the second Bianchi identity (\ref{2ndBianchiIdentity}) to
this equation we obtain $\Theta^{[a}\wedge \theta^{b]}=0$ which is
equivalent to $\Theta^a=0$. Hence this equation guarantees that the
four-dimensional connection is Riemannian.  On the other hand for the
Riemannian connection the curvature form $\R_{ab}$ for the constant
curvature spacetime with sectional curvature $K$ is expressed as
\begin{equation}
\R_{ab}=K\theta_a\wedge\theta_b=K\Sigma_{ab}.
\end{equation}
Hence the solution to Eq.(\ref{LorentzianWKBEq}) represents the
constant curvature space time with the sectional curvature
$\Lambda\over6$.  Since all the constant curvature spacetimes with the
same sectional curvature are locally isometric, this implies that
there exists only one Lorentzian WKB orbit for the wavefunction
$\Psi_\Lambda[\A]$.  It is the de Sitter spacetime for $\Lambda>0$,
and the anti-de Sitter spacetime for $\Lambda<0$.  Thus the solution
is a quantum counter part of the classical (anti-)de Sitter spacetime,
and may be regarded as representing the ground state for the quantum
vacuum spacetime with non-vanishing cosmological constant.

We have so far assumed that $\theta^a$ is real.  However, this
assumption is to restrictive in the complex canonical theory. In fact
if we only require the reality condition on $\E^j\cdot\E^k$, a wider
possibility is allowed.  For example for the case $\E^j$, $N^j$ and
$\utilde N$ are real but $q=\det(\E^{Ij})$ is negative, $\theta^a$
becomes pure imaginary since $N$ is pure imaginary.  The spacetime
metric signature for this case is totally reversed and given by
$[+,-,-.-]$. Though this case is usually excluded, there exists no a
priori reason to regard it as unphysical because the causal structure
is normal.  Further this case occurs as a special sector of the single
wavefunction in the complex canonical theory.  Of course this does not
mean that the two sectors with the different signatures are
equivalent.

The reality condition also allows Euclidean WKB orbits.  Such orbits
correspond to the case $q<0$ and $N^j$ and $N$ are real or the case $q>0$
and $N^j$ are real but $N$ is pure imaginary.  For these cases, since
$\chiral\F_{0I}$ and $\chiral \Sigma_{0I}$ are pure imaginary and the
left and the right chiral variables become independent,
Eq.(\ref{LorentzianWKBEq:chiral}) does not lead to
Eq.(\ref{LorentzianWKBEq}).  Hence there exist an infinite number of
WKB
solutions\CITE{Kodama90,Samuel88,CapovillaJacobsonDell90,Uehara91}.

In order to see these points explicitly, let us examine the behavior of
the solution in the spatially homogeneous and isotropic sector.  In
this sector the chiral variables are expressed as
\begin{eqnarray}
&&\E^{Ij}={\kappa^2\over\Omega}\sigma |\chi|X^j_I,\\ &&\A_{Ij} =
{1\over6\kappa^2}A \chi^I_j,
\end{eqnarray}
where $X^j_I$ is the basis of the invariant vectors dual to $\chi^I_j$
introduced in \S3.3.3, and $\sigma$ and $A$ are variables independent
of the space coordinates.  The Poisson bracket between $\sigma$ and
$A$ is given by
\begin{equation}
\{\sigma, A\}=1,
\end{equation}
and the spacetime metric is expressed in terms of $\sigma$ and $N$ as
\begin{equation}
ds^2 = [-N^2dt^2 + {\kappa^2\over\Omega}\sigma \chi^I\chi^I].
\end{equation}

Since $S_\Lambda$ in this sector is simply given by
\begin{equation}
S_\Lambda = {\Omega\over18\kappa^2\Lambda}(2A^3\mp 9iA^2),
\end{equation}
the WKB equations are reduced to
\begin{eqnarray}
&&\sigma={\Omega\over3\kappa^2\Lambda}(A^2\mp 3iA),\\
&&\dot A =-(N\sigma^{-3/2}){\kappa\over\Omega^{1/2}}\Lambda\sigma^2.
\end{eqnarray}
In the present case $q$ is expressed as
$q=(\kappa^2/\Omega)^3\sigma^3|\chi|^2$.  Hence $q>0$ corresponds to
$\sigma>0$.  In the gauge $N=1$ the solution to these equations for
$\sigma>0$ and $\Lambda>0$ is given by the de Sitter solution $dS^4$
as is expected:
\begin{eqnarray}
&&A=-{3\over2}\sinh \xi,\\
&&\sigma={3\Omega\over4\kappa^2\Lambda}\cosh^2 \xi,\\ &&
\xi=\Frac(\Lambda/3)^{1/2}t.
\end{eqnarray}
The Euclidean solutions are obtained from this solution by the
analytic continuation of the time $t$ or $\xi$ to the imaginary
region.  For example, by replacing $\xi$ by $i\xi$, we obtain the four
dimensional Euclidean sphere $S^4$ for which $\sigma>0$ and $N$ is
pure imaginary.  On the other hand by the replacement $\xi \rightarrow
\xi + i\pi/2$ we get the four-dimensional hyperbolic spacetime $H^4$
for which $\sigma<0$.  In the present case these exhaust the Euclidean
WKB orbits for the wavefunction $\Psi_\Lambda$ with $\Lambda>0$.  Thus
in the WKB picture the wavefunction represents a sequence of two
Euclidean spacetimes $H^4$ and $S^4$, and one Lorentzian spacetime
$dS^4$.

This WKB picture for the wavefunction is slightly different from the
one proposed by Hawking, Vilenkin and others.  This difference comes
from the difference in the range of the variable $\sigma$.  To see
this, let us apply the general formula (\ref{EtoA:ConnectionRep}) to
the present case to find the corresponding wavefunction
$\tilde\Psi(\sigma)$ in the ADM-type representation:
\begin{equation}
\Psi(A)=C\int d\sigma e^{-iA\sigma \mp 3\sigma/2} \tilde\Psi(\sigma).
\label{AbyE:RW}\end{equation}
For the solution $\Psi_\Lambda(A)$ the inversion of this transform
yields
\begin{equation}
\tilde \Psi(\sigma) = C' \int_\gamma dz \exp\left[i\left({z^3\over3} + x
z\right)\right],
\end{equation}
where
\begin{equation}
x =
\Frac({9\Omega}/8\kappa^2\Lambda)^{2/3}\left(1-{4\kappa^2\Lambda\over3\Omega}\sigma\right).
\end{equation}
This is the integral expression for the Airy function as is expected.
The important point here is that Eq.(\ref{AbyE:RW}) yields the
original solution $\Psi_\Lambda$ only when $Ai(x)$ is taken as $\tilde
\Psi(\sigma)$ and the integration range in Eq.(\ref{AbyE:RW}) is taken
to be $-\infty < \sigma < \infty$\CITE{Kodama90}, while in the ADM
theory $\sigma$ is limited to the range $\sigma>0$.  Thus in the
ADM-type representation the present solution corresponds to the
Hartle-Hawking wavefunction extended to the classically forbidden
region $q<0$.  Though this is just an analytic continuation in the
spatially homogeneous and isotropic case, the wavefunction itself is
not such a simple mathematical extension of any wavefunction in the
ADM theory for the generic spacetime.  Actually for such generic case
in the ADM theory no exact solution is found and the Hartle-Hawking
proposal has no well-defined formulation. Thus the quantum gravity
theory based on the complex canonical theory yields a picture on the
quantum behavior of the universe different from that based on the ADM
theory if the universe can be described by a single wavefunction.

\begin{figure}
\vspace{10cm}
\includegraphics{BianchiIX.ps}
\caption{ADM-wavefunction $\tilde \Psi(E)$ in the Bianchi IX sector}
The behavior of $\tilde \Psi(E)$ on the $E_1=E_2$ section is shown.
The high peak at $E_1=E_2=E_3=0$ is truncated.
\end{figure}

The above WKB analysis of the wavefunction can be easily extended to
the generic Bianchi sector, and the characteristic behavior of the
solution and the structure of the WKB orbits are similar to the one in
the isotropic case except that much more abundant Euclidean spacetimes
appear.  For example, in the Bianchi IX sector, if we parametrize the
chiral variables as
%se
\begin{eqnarray}
&&\E^{Ij}={\kappa^2\over\Omega}E_I |\chi|X^j_I,\\ &&\A_{Ij} = A_I \chi^I_j,
\end{eqnarray}
the wavefunction $\tilde\Psi(E)$ in the ADM-type representation
corresponding to $\Psi_\Lambda[\A]$ with $\Lambda>0$ is given by
\begin{equation}
e^{-W}\tilde \Psi(E) = C
\int_{-\infty}^\infty{dz\over\sqrt{1+z^2}}\exp\left
[ {\lambda\over2}{-(E_1^2+E_2^2)+2iE_1E_2 z \over 1+z^2} + iE_3 z
-{z^2\over2\lambda}\right],
\end{equation}
where
\begin{eqnarray}
&& W= {1\over2}\left({E_2E_3\over E_1}+ {E_3E_1\over E_2}+
{E_1E_2\over E_3}\right),\\ && \lambda={\kappa^2\Omega\over 3\Lambda}.
\end{eqnarray}
Though $\tilde\Psi(E)$ is singular at the surfaces $E_1=0$ or $E_2=0$
or $E_3=0$ from the structure of $W$, the spatial metric $q_{IJ}$ does
not become degenerate on these surfaces because
$q_{IJ}=q_I\delta_{IJ}$ is expressed as
$q_I=E_JE_K/E_I$($I\not=J\not=K$).  Thus the relation between the
regions $q>0$ and $q<0$ is not simple and the wavefunctions in these
two regions are not connected by a simple analytic continuation.
Outside these singular surfaces the behavior of the wavefunction is
simple.  It rapidly oscillates in the region $q>0$ as $q$ increases
and falls off exponentially in the classically forbidden region $q<0$
as in the isotropic case, as shown in Fig.1.

Finally, in connection with the extension of the solution to region
$q<0$, we comment on the difference of the right and the left chiral
theories.  By inspecting the structure of the chiral Lagrangian
(\ref{ChiralLagrangian:CanonicalForm}), we find that under the
transformation $\E^j \rightarrow -\E^j$ and $\A_j \rightarrow -\A_j$
only the terms such as $V(\Phi)$ which explicitly contain $q$ change
sign if we simultaneously reverse the signs depending on the
chirality.  In particular for the present case the region $q>0$ for
the right chiral theory with $\Lambda>0$ is mapped by this
transformation to the region $q<0$ for the left one with $\Lambda<0$.
Since there exists no symmetry between the two regions $q\Lambda>0$
and $q\Lambda<0$ for a fixed chirality except for the spatially
homogeneous and isotropic sector, this implies that the physical
contents of the left and the right chiral theories are different
in the quantum framework unlike in the classical framework.  For
example it is shown that in the spatially anisotropic cases there
exists no ADM wavefunction corresponding to the solution
$\Psi_\Lambda[\A]$ for the right chiral theory with $\Lambda>0$ though
one can find one for the left chiral theory with
$\Lambda>0$\CITE{Kodama90}.

\section{Summary and Discussion}

To construct a quantum gravity theory one must find a way to reconcile
the general covariance of the classical gravity theory with the
quantum framework. As explained in \S3, the canonical approach which
respects the structure of the conventional classical theories divides
this task to the two problems: the construction of
spatial-diffeomorphism invariant states and observables, and the
formulation of dynamics in terms of them.

In the conventional approach based on the ADM formulation most of the
work done so far is limited to the study of the minisuperspace models
and is concerned mainly with the latter problem.  Since the essential
features of the general covariance appears only in systems with
infinite degrees of freedom, this limitation is severe.  Some aspects
such as the hyperbolicity of the Hamiltonian constraint have been
studied using superspace.  However, the treatment is too formal to be
useful in the explicit construction of the invariant states and
observables.

In contrast, in the approach based on the complex canonical
formulation, we could go beyond the minisuperspace models and directly
attack the generic situations.  The essential points were the
introduction of the additional gauge freedom corresponding to the
tetrad rotation and the chiral decomposition.  They enabled us to
express the fundamental equations by simple differential polynomials
and gave a gauge field theoretical structure to the theory.  In
particular the introduction of loop integral variables based on the
latter feature has reduced the problem of constructing the invariant
states to finding and classifying the knot or the link invariant of
three manifolds.  Further we could find an infinite number of exact
solutions to the quantum Hamiltonian constraint owing to the simple
structure of the equations.

In order to complete the quantization program in this approach,
however, we must solve the following three problems at least.

 First of all, though significant progress has been made in the
construction and the parametrization of the invariant states, our
knowledge about the structure of the invariant operators is still
quite poor.  Since the physical interpretation of the states is found
only with the aid of the operators whose relation to the classical
geometrical variables is known, this situation is quite
unsatisfactory.

 In connection with this problem we comment on the recent work by the
Syracuse group\CITE{Smolin92}.  They succeeded in constructing finite
operators which represent the spatial area of minimal surfaces and the
total volume of the space in the loop space representation, and found
that their eigenvalues are discrete and integer multiples of constants
of order unity in the Planck units when acted on the so-called weave
states which forms a sparse subset of the whole loop-functional space.
Though these results are quite fascinating, the type of the operators
constructed so far is too restrictive
 to be used as the basis of
generic arguments.  The origin of this limitation exists in the fact
that we have no regular covariant tensor densities other than
$\epsilon_{jkl}$ as noted in
\S4.2.4.  The technique developed by the Syracuse group cannot be
applied, for example, to the quantity $Z$ representing the expansion
rate of the space introduced in \S4.1.2.

 Second is the problem of the reality condition.  Since the complex
canonical theory is equivalent to the Einstein theory only under this
condition, one may be studying a theory quite different from the
Einstein theory if one neglects the condition.  In the holomorphic
connection representation this condition can be translated to the
problem of finding a measure which makes the formal expression for the
inner product well-defined.  However, at the present stage, no such
translation is possible in the loop space representation since we do
not yet know how to express the reality condition only in terms of the
loop language.  The study of the relation between the left and the
right chiral variables may shed light on this problem.

Finally we are far from being able to discuss dynamical problems
in realistic situations.
This is partly because of our poor knowledge on the invariant
operators touched upon above.  However, it is not the whole reason.
The main obstacle lies  in the fact that there exists no consensus
on the treatment of the time variables in the quantum framework.
In order to settle this issue of time, detailed studies of realistic
systems beyond the minisuperspace models are needed.

\bigskip

\bigskip

\noindent
{\Large\bf
Acknowledgement}

\medskip

The author would like to thank Prof. Abhay Ashtekar for his valuable
comments on the manuscript.
This work is supported by the Grand-in-Aid for Scientific Research of
the Ministry of Education, Science and Culture of Japan (No.02640228).
The numerical computation and its graphic display in this work is
supported by the Yukawa Institute for Theoretical Physics.

\par 
\begin{thebibliography}{10}

\bibitem{GreenSchwarzWitten87}
{Green, M.~B.}, {Schwarz, J.~H.}, and {Witten, E.}, {\it Superstring Theory}
  (Cambridge Univ. Press, 1987).

\bibitem{Hawking82a}
{Hawking, S.W.}
\newblock in H.~A. {Br\"uk} and G.~V. Coyne, editors, {\it Astrophysical
  Cosmology}, Pontifica Academia Scientarium, Vatican City, 1982.

\bibitem{HartleHawking83}
{Hartle, J.~B.} and {Hawking, S.~W.}, Phys. Rev. {\bf D28}, 2960 (1983).

\bibitem{Hawking84}
{Hawking, S.~W.}, Nucl. Phys. {\bf B239}, 257 (1984).

\bibitem{Vilenkin83}
{Vilenkin, A.}, Phys. Rev. {\bf D27}, 2848 (1983).

\bibitem{Vilenkin84}
{Vilenkin, A.}, Phys. Rev. {\bf D30}, 509 (1984).

\bibitem{Ashtekar86a}
{Ashtekar A.}, Phys. Rev. Lett. {\bf 57}, 2244 (1986).

\bibitem{Ashtekar87}
{Ashtekar, A.}, Phys. Rev. {\bf D36}, 1587 (1987).

\bibitem{Samuel87}
{Samuel, J.}, Pramana J. Phys. {\bf 28}, L429--L432 (1987).

\bibitem{JacobsonSmolin87}
{Jacobson, T.} and {Smolin, L.}, Phys. Lett. {\bf B196}, 39--42 (1987).

\bibitem{JacobsonSmolin88a}
{Jacobson, T.} and {Smolin, L.}, Class. Quantum. Grav. {\bf 5}, 583--594
  (1988).

\bibitem{AshtekarRomanoTate89}
{Ashtekar, A.}, {Romano, J.~D.}, and {Tate, R.~S.}, Phys. Rev. {\bf D40}, 2572
  (1989).

\bibitem{Jacobson88b}
{Jacobson, T.}, Class. Quantum. Grav. {\bf 5}, 923--935 (1988).

\bibitem{GorobeyLukyanenko90}
{Gorobey, N.N.} and {Lukyanenko, A.S.}, Class. Quantum. Grav. {\bf 7}, 67
  (1990).

\bibitem{JacobsonSmolin88b}
{Jacobson, T.} and {Smolin, L.}, Nucl. Phys. {\bf B299}, 295 (1988).

\bibitem{DeWitt67a}
{DeWitt, B.~S.}, Phys. Rev. {\bf 160}, 1113 (1967).

\bibitem{Ashtekar86b}
{Ashtekar, A.}, Self-duality and spinorial techniques in the canonical
approach to quantum gravity,
in {\it Qauntum Concepts in  Space and Time}, {Penrose, R.} and {Isham, C.J.},
eds., p.302 (Clarendon Press, Oxford, 1986).

\bibitem{Husain89}
{Husain, V.}, Nucl. Phys. {\bf B313}, 711 (1989).

\bibitem{BruegmannPullin91}
{Br\"ugmann, B.} and {Pullin, J.}, Nucl. Phys. {\bf B363}, 221 (1991).

\bibitem{RovelliSmolin88}
{Rovelli, C.} and {Smolin, L.}, Phys. Rev. Lett. {\bf 61}, 1155 (1988).

\bibitem{RovelliSmolin90}
{Rovelli, C.} and {Smolin, L.}, Nucl. Phys. {\bf B331}, 80 (1990).

\bibitem{BruegmannGambiniPullin92a}
{Br\"ugmann, B.}, {Gambini, R.}, and {Pullin, J.}, Phys. Rev. Lett. {\bf 68},
  431 (1992).

\bibitem{BruegmannGambiniPullin92b}
{Br\"ugmann, B.}, {Gambini, R.}, and {Pullin, J.}, Syracuse Univ. Preprint,
  SU-GP-92/1-1 (1992).

\bibitem{BruegmannGambiniPullin92c}
{Br\"ugmann, B.}, {Gambini, R.}, and {Pullin, J.}, Syracuse Univ. Preprint,
  SU-GP-92-3/1 (1992).

\bibitem{Kodama90}
{Kodama, H.}, Phys. Rev. {\bf D42}, 2548--2565 (1990).

\bibitem{Witten89}
{Witten, E.}, Comm. Math. Phys. {\bf 121}, 351--399 (1989).

\bibitem{AshtekarRovelliSmolin91}
{Ashtekar, A.}, {Rovelli, C.}, and {Smolin, L.}, Phys. Rev. {\bf D44}, 1740
  (1991).

\bibitem{AshtekarRovelliSmolin92}
{Ashtekar, A.}, {Rovelli, C.}, and {Smolin, L.}, Phys. Rev. Lett. {\bf 69}, 237
  (1992).

\bibitem{Smolin92}
{Smolin, L.}, Syracuse Univ. preprint {\bf }, 1 (1992).

\bibitem{Kodama88a}
{Kodama, H.}, Prog. Theor. Phys. {\bf 80}, 1024 (1988).

\bibitem{AshtekarPullin89}
{Ashtekar, A.} and {Pullin, J.}, Syracuse Univ. Preprint (1990).

\bibitem{AshtekarTateUggla92}
{Ashtekar, A.}, {Tate, R.}, and {Uggla, C.}, Syracuse Univ. Preprint (1992).

\bibitem{HusainSmolin89}
{Husain, V.} and {Smolin, L.}, Nucl. Phys. {\bf B327}, 205 (1989).

\bibitem{HusainPullin90}
{Husain, V.} and {Pullin, J.}, Modern Phys. Lett. {\bf A5}, 733--741 (1990).

\bibitem{Ashtekar88a}
{Ashtekar, A.}, {\it New Perspectives on Canonical Gravity} (Bibliopolous,
  Naples, 1988).

\bibitem{Smolin89}
{Smolin, L.}, in A. Ashtekar and J. Stackel, editors, {\it {\rm Proceedings of
  the Osgood Hill Conference on} Conceptual Problems of Quantum Gravity}
  (Birkhauser, 1989).

\bibitem{Rovelli91a}
{Rovelli, C.}, Class. Quantum. Grav. {\bf 8}, 1613 (1991).

\bibitem{Ashtekar91a}
{Ashtekar, A.}, {\it Lectures on Non-Perturbative Canonical Gravity} (World
  Scientific, Singapore, 1991).

\bibitem{HalliwellHartle91}
{Halliwell, J.~J.} and {Hartle, J.~B.}, Phys. Rev. {\bf D43}, 1170 (1991).

\bibitem{StoneKuchar92}
{Stone, C.L.} and {Kucha\v{r}, K.V.}, Class. Quantum. Grav. {\bf 9}, 757--776
  (1992).

\bibitem{MisnerThorneWheeler73}
{Misner, C.W.}, {Thorne, K.S.}, and {Wheeler, J.A.}, {\it Gravitation}
  (Freeman, San Francisco, 1973).

\bibitem{KobayashiNomizu63}
{Kobayashi, S.} and {Nomizu, K.}, {\it Foundation of Differential Geometry I \&
  II} (Interscience Pub., 1963).

\bibitem{Kuchar81}
{Kucha\v{r}, K.}, Canonical methods of quantization,
in {\it  Quantum Gravity 2: A Second Oxford Symposium}, {Isham, C.J.},
{Penrose, R.}, and {Sciama, D.W.}, eds., p.329 (Clarendon Press,
  Oxford, 1981).

\bibitem{AshtekarStachel91}
{Ashtekar, A.} and {Stachel, J.}, editors, {\it Conceptual Problems of Quantum
  Gravity} ({Birkh\"auser}, Boston, 1991).

\bibitem{NewmanRovelli92}
{Newman, E.T.} and {Rovelli, C.}, Univ. Pittsburgh preprint (1992).

\bibitem{Vilenkin82}
{Vilenkin, A.}, Phys. Lett. {\bf B117}, 25 (1982).

\bibitem{Hawking79}
{Hawking, S.W.}, The path-integral approach to quantum gravity,
 in {\it General Relativity: an Einstein Centenary Survey}, {Hawking, S.W.} and
{Israel, W.}, eds., p.746 (Cambridge Univ. Press,   Cambridge, 1979).

\bibitem{HalliwellLouko89a}
{Halliwell, J.~J.} and {Louko, J.}, Phys. Rev. {\bf D39}, 2206 (1989).

\bibitem{HalliwellLouko89b}
{Halliwell, J.~J.} and {Louko, J.}, Phys. Rev. {\bf D40}, 1868 (1989).

\bibitem{HalliwellHartle90}
{Halliwell, J.~J.} and {Hartle, J.~B.}, Phys. Rev. {\bf D41}, 1815 (1990).

\bibitem{GarayHalliwellMarugan91}
{Garay, L.J.}, {Halliwell, J.J.}, and {Marugan, G.A.M.}, Phys. Rev. {\bf D43},
  2572 (1991).

\bibitem{Witten88c}
{Witten, E.}, Nucl. Phys. {\bf B311}, 46--78 (1988).

\bibitem{FujiwaraEtAl91a}
{Fujiwara, Y.}, {Higuchi, S.}, {Hosoya, A.}, {Mishima, T.}, and {Siino, M.}, PR
  {\bf D44}, 1756 (1991).

\bibitem{FujiwaraEtAl91b}
{Fujiwara, Y.}, {Higuchi, S.}, {Hosoya, A.}, {Mishima, T.}, and {Siino, M.}, PR
  {\bf D44}, 1763 (1991).

\bibitem{FujiwaraEtAl92a}
{Fujiwara, Y.}, {Higuchi, S.}, {Hosoya, A.}, {Mishima, T.}, and {Siino, M.},
  Tokyo Inst. Tech. preprint, TIT/HEP-184/COSMO-13 (1992).

\bibitem{Banks85}
{Banks, T.}, Nucl. Phys. {\bf B249}, 332 (1985).

\bibitem{HalliwellHawking85}
{Halliwell, J.~J.} and {Hawking, S.~W.}, Phys. Rev. {\bf D31}, 1777 (1985).

\bibitem{Hartle91}
{Hartle, J.B.}, The quantum mechanics of cosmology,
in {\it Quantum Cosmology and Baby Universe, \rm Jerusalem
  Winter School}, C.Coleman et al, eds. (1991).

\bibitem{AshtekarIsham92}
{Ashtekar, A.} and {Isham, C.J.}, Syracuse Univ. Preprint, SU-GP-91-12-2
(1992).

\bibitem{GoldbergLewandowskiStornaiolo92}
{Goldberg, J.N.}, {Lewandowski, L.}, and {Stornaiolo, C.}, Syracuse Univ.
(1992).

\bibitem{Samuel88}
{Samuel, J.}, Class. Quantum. Grav. {\bf 5}, L123--L125 (1988).

\bibitem{CapovillaJacobsonDell90}
{Capovilla, R.}, {Jacobson, T.}, and {Dell, J.}, Class. Quantum. Grav. {\bf 7},
  L1--L3 (1990).

\bibitem{Uehara91}
{Uehara, S.}, Class. Quantum. Grav. {\bf 8}, L229--L234 (1991).

\end{thebibliography}
\end{document}